\documentclass[structabstract]{aa}

\usepackage{graphics,epsfig,txfonts}
\usepackage{natbib}
\usepackage{longtable}
\usepackage{lscape}
\usepackage{textcomp}

\bibpunct{(}{)}{;}{a}{}{,}

\newcommand{\ergcm}[1]{$\times 10^{#1}$ erg cm$^{-2}$ s$^{-1}$}
\newcommand{\oergcm}[1]{$10^{#1}$ erg cm$^{-2}$ s$^{-1}$}
\newcommand{\ergs}[1]{$\times 10^{#1}$ erg s$^{-1}$}
\newcommand{\oergs}[1]{$10^{#1}$ erg s$^{-1}$}

\newcommand{\nh}{N$_{\rm H}$}

\newcommand{\Hone}{\ion{H}{i}}

\newcommand{\Halp}{H${\alpha}$}
\newcommand{\ltsima}{$\buildrel < \over \sim$}
\newcommand{\lsim}{\lower.5ex\hbox{\ltsima}}
\newcommand{\gtsima}{$\buildrel > \over \sim$}
\newcommand{\gsim}{\lower.5ex\hbox{\gtsima}}

\newcommand{\xmm}{{\it XMM-Newton}}
\newcommand{\cxo}{{\it Chandra}}
\newcommand{\calsrc}{1E0102.2-7219}

\def\rahour{\hbox{\ensuremath{^{\rm h}}}}
\def\ramin{\hbox{\ensuremath{^{\rm m}}}}
\usepackage{color}

\newcommand{\no}[1]{\textnumero\,#1}

\begin{document}

\title{The \emph{XMM-Newton} survey of the Small Magellanic Cloud:\\
       The X-ray point-source catalogue
       \thanks{Based on observations obtained with XMM-Newton, an ESA science mission with instruments and contributions directly funded by ESA Member States and NASA}
      }

\author{R.~Sturm\inst{1}
   \and F.~Haberl\inst{1}
   \and W.~Pietsch\inst{1}
   \and J.~Ballet\inst{2}
   \and D.~Hatzidimitriou\inst{3}
   \and D.~A.~H.~Buckley \inst{4}
   \and M.~Coe\inst{5}
   \and M.~Ehle \inst{6}
   \and M.~D.~Filipovi{\'c}\inst{7}
   \and N.~La~Palombara\inst{8}
   \and A.~Tiengo\inst{8,9,10}
      }

\titlerunning{The \xmm\ SMC-survey point-source catalogue.}
\authorrunning{Sturm et al.}

\institute{Max-Planck-Institut f\"ur extraterrestrische Physik, Giessenbachstra{\ss}e, 85748 Garching, Germany
	   \and
           Laboratoire AIM, CEA/DSM - CNRS - Universit\'{e} Paris Diderot, IRFU/SAp, CEA-Saclay, 91191 Gif-sur-Yvette, France
	   \and
           Department of Physics, University of Athens, Panepistimiopolis, Zografos, Athens, Greece
	   \and
           South African Astronomical Observatory, PO Box 9, Observatory 7935, Cape Town, South Africa
	   \and
           School of Physics and Astronomy, University of Southampton, Highfield, Southampton SO17 1BJ, United Kingdom
	   \and
           XMM-Newton Science Operations Centre, ESAC, ESA, PO Box 78, 28691 Villanueva de la Ca\~{n}ada, Madrid, Spain
	   \and
           University of Western Sydney, Locked Bag 1797, Penrith South DC, NSW1797, Australia
	   \and
           INAF, Istituto di Astrofisica Spaziale e Fisica Cosmica Milano, via E. Bassini 15, 20133 Milano, Italy
	   \and
           Istituto Universitario di Studi Superiori di Pavia, Piazza della Vittoria 15, I-27100 Pavia, Italy
	   \and
           INFN, Istituto Nazionale di Fisica Nucleare, Sezione di Pavia, via A. Bassi 6, I-27100 Pavia, Italy
	   }

\date{Received 2 July 2012 / Accepted 3 July 2013}

\abstract{Local-Group galaxies provide access to samples of X-ray source populations of whole galaxies.
          The XMM-Newton survey of the Small Magellanic Cloud (SMC) completely covers the bar and eastern wing with a 5.6 deg$^2$ area in the (0.2$-$12.0) keV band.
        }{
          To characterise the X-ray sources in the SMC field, we created a catalogue of point sources and sources with moderate extent.
          Sources with high extent ($\geq$40\arcsec) have been presented in a companion paper.
        }{
          We searched for point sources in the EPIC images using sliding-box and maximum-likelihood techniques
          and classified the sources using hardness ratios, X-ray variability, and their multi-wavelength properties.
        }{
          The catalogue comprises 3053 unique X-ray sources with a median position uncertainty of 1.3\arcsec
          down to a flux limit for point sources of $\sim$10$^{-14}$ erg cm$^{-2}$ s$^{-1}$ in the (0.2$-$4.5) keV band,
          corresponding to $5\times 10^{33}$ erg s$^{-1}$ for sources in the SMC.
          We discuss statistical properties, like the spatial distribution, X-ray colour diagrams, luminosity functions, and time variability.
          We identified
                        49 SMC high-mass X-ray binaries (HMXB),
                        four super-soft X-ray sources (SSS),
                        34 foreground stars, and
                        72 active galactic nuclei (AGN) behind the SMC.
          In addition, we found candidates for
                        SMC HMXBs (45)
                        and faint SSSs (8)
                        as well as AGN (2092)
                        and galaxy clusters (13).
        }{We present the most up-to-date catalogue of the X-ray source population in the SMC field. In particular, the known population of X-ray binaries is greatly increased.
          We find that the bright-end slope of the luminosity function of Be/X-ray binaries significantly deviates from the expected universal high-mass X-ray binary luminosity function.
        }


\maketitle

\section{Introduction}
  \label{sec:intro}

In contrast to the Milky Way, nearby galaxies are well
suited to investigate the X-ray source populations of a galaxy as a whole. This is because
most X-ray sources in the Galactic plane are obscured by large amounts of absorbing gas and dust
and uncertainties in distances complicate the determination of luminosities.

The Small Magellanic Cloud (SMC) is a gas-rich dwarf irregular galaxy orbiting the Milky Way
and is the second nearest star-forming galaxy after the Large Magellanic Cloud (LMC).
Gravitational interactions with the LMC and the Galaxy are believed to have tidally triggered recent bursts of star formation \citep{2004ApJ...604..167Z}.
In the SMC this has resulted in a remarkably large population of high-mass X-ray binaries (HMXBs)
that formed $\sim$40 Myr ago \citep{2010ApJ...716L.140A}.
The relatively close distance of $\sim$60 kpc \citep[assumed throughout the paper, e.g. ][]{2005MNRAS.357..304H,2011MNRAS.415.1366K}
and the moderate Galactic foreground absorption of \nh$\approx 6\times10^{20}$ cm$^{-2}$ \citep[][]{1990ARA&A..28..215D}
enable us to study complete X-ray source populations in the SMC, like supernova remnants (SNRs), HMXBs or
super-soft X-ray sources (SSSs) in a low metallicity \citep[$Z \approx 0.2Z_{\sun}$, ][]{1992ApJ...384..508R} environment.
The \xmm\ large-programme survey of the SMC \citep[][]{2012A&A...545A.128H} allows to continue 
the exploration of this neighbouring galaxy in the (0.2--12.0)~keV band.

In this study we present the \xmm\ SMC-survey point-source catalogue and describe the classification
of X-ray sources, with the main purpose of discriminating between sources within the SMC and fore- or background sources.
The detailed investigation of individual source classes, such as Be/X-ray binaries (BeXRBs) or SSSs, will be the subject of subsequent studies.
Extended sources, with angular diameters of $\geq$40\arcsec\, are not appropriate for the \xmm\ point-source detection software.
For example, substructures in SNRs can result in the detection of several spurious point sources.
Highly extended sources in and behind the SMC have been identified on a mosaic image and are discussed in \citet{2012A&A...545A.128H}.
These are all SNR in the SMC or clusters of galaxies with large angular diameter.
Clusters with smaller angular diameter cannot be easily discriminated 
from point sources and are therefore included 
in the present study.

The paper is organised as follows:
In Section~\ref{sec:observations}, we briefly review the \xmm\ observations of the SMC.
Section~\ref{sec:catalogue} describes the creation of the point-source catalogue,
which is characterised in Section~\ref{sec:characterisation}.
After reporting the results of the cross-correlation with other catalogues in Section~\ref{sec:correlations},
we present our source classification in Section~\ref{sec:class}.
Finally, statistical properties of the dataset are discussed in Section~\ref{sec:discussion}.
A summary is given in Section~\ref{sec:conclusions}.

\section{XMM-Newton observations of the SMC}
  \label{sec:observations}

\begin{figure}
  \resizebox{\hsize}{!}{\includegraphics[angle=0,clip=]{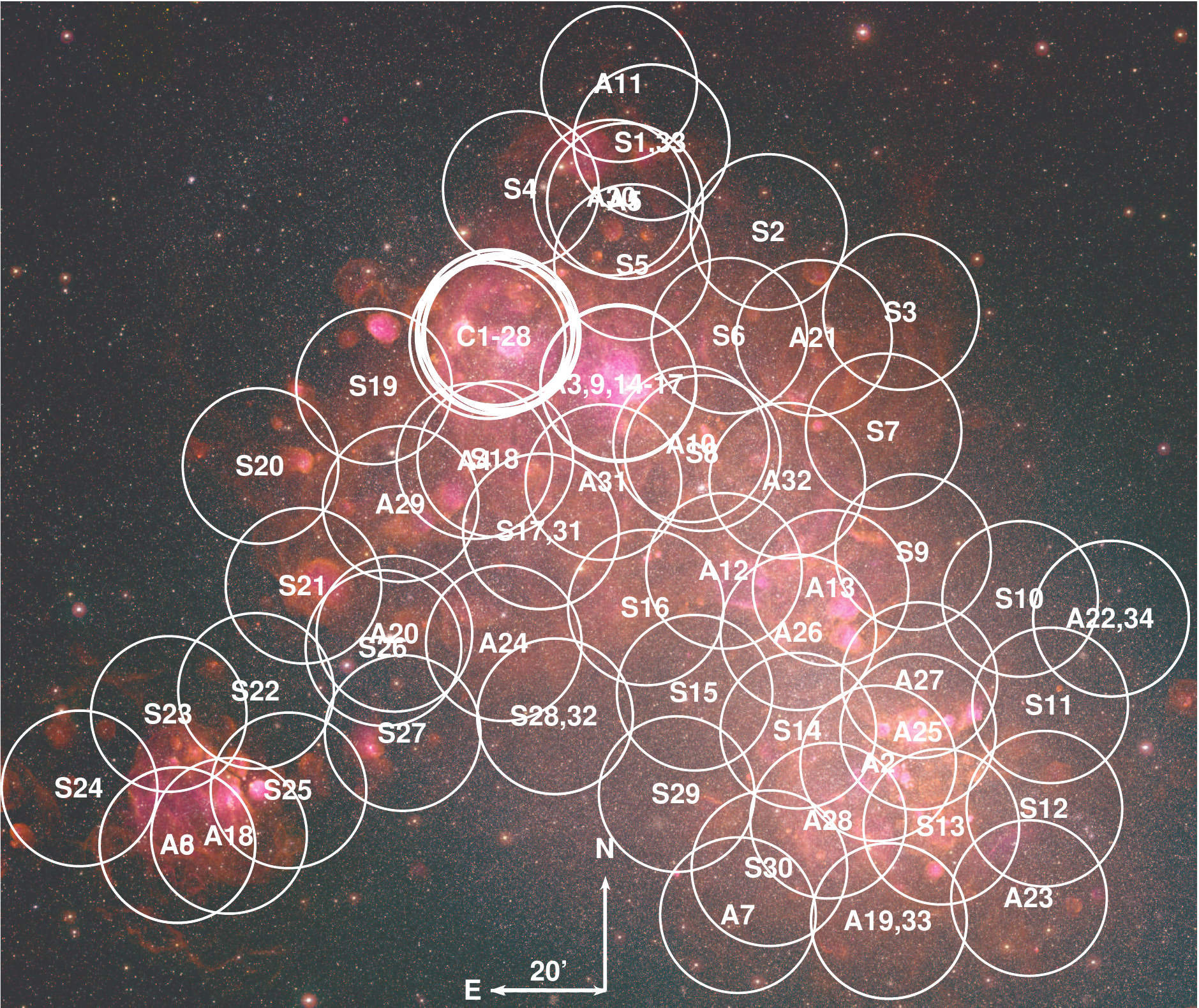}}

  \caption{
    Colour image of the SMC from MCELS \citep[e.g.][]{2000ASPC..221...83S,2005AAS...20713203W}
    with \Halp/[\ion{S}{ii}]/[\ion{O}{iii}] in red, green, and blue.
    The overlaid circles indicate the distribution of \xmm\ observations and have radii of 800\arcsec.
    Labels correspond to column 1 of Table~\ref{tab:observations}.
  }
  \label{fig:fields}
\end{figure}

\begin{figure}
  \resizebox{\hsize}{!}{\includegraphics[angle=0,clip=]{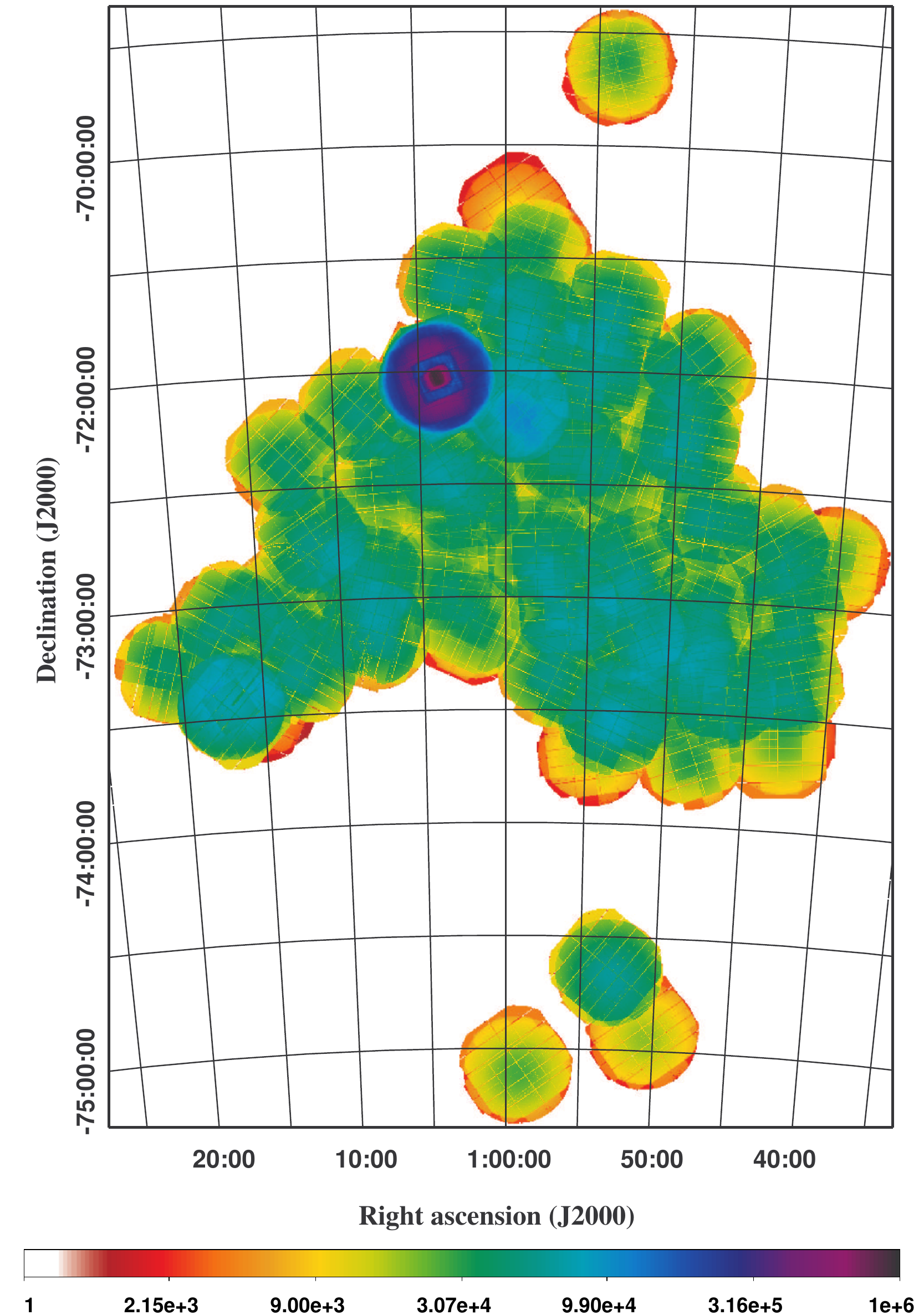}}

  \caption{
  Combined vignetting-corrected EPIC exposure map of the SMC main field and additional outer fields, as used for the catalogue.
  The survey observation provide a homogeneous coverage in addition to the deep field around \calsrc.
  EPIC-MOS1 and -MOS2 exposure is weighted by a factor of 0.4 relative to EPIC-pn to account for the lower effective area.
  }
  \label{fig:exp}
\end{figure}

The \xmm\ observatory \citep{2001A&A...365L...1J} comprises three X-ray telescopes \citep{2002SPIE.4496....8A} each equipped
with a European Photon Imaging Camera (EPIC) at their focal planes, one of pn type \citep{2001A&A...365L..18S} and two of MOS type \citep{2001A&A...365L..27T}.
This enables observations in the (0.2--12.0) keV band with an angular resolution of 5\arcsec--6\arcsec\ (FWHM),
which corresponds to a spatial resolution of $\sim$1.5 pc at the distance of the SMC.

In combination with archival observations, our large-programme SMC-survey provides 
complete coverage of the  main body of the SMC (see Fig.~\ref{fig:fields}).
The survey was executed with EPIC in {\it full-frame} imaging mode, using the thin and medium filter for EPIC-pn and EPIC-MOS, respectively.
Archival observations were partly performed in other modes.

To construct the \xmm\ point-source catalogue for the SMC, 
we combined the data of the large-programme SMC survey (33 observations of 30 different fields, 1.1 Ms total exposure),
with all publicly available archival data up to April 2010 (62 observations, 1.6 Ms exposure).
Some 28 archival observations (850 ks in total) are calibration observations of the SNR \calsrc.
These calibration observations are performed 
every 6 months and constitute the deepest exposure \xmm\ field in the
SMC (see Fig.~\ref{fig:exp}). All the observations form a continuous field, which we will refer to the {\it main field} (5.58 deg$^2$).
In addition, we included two observations of a field to the north and three fields to the south of the SMC main field (98 ks exposure in total).
These are somewhat further away from the SMC bar and wing but 
contain SSSs found by ROSAT.
The total area covered by the catalogue is 6.32 deg$^2$.
A list of all exposures is given in Table~\ref{tab:observations}.
The table columns are described in detail in Sec.~\ref{sec:description}.

\section{Creation of the source catalogue}
\label{sec:catalogue}
To create a source catalogue for the SMC, we used a similar procedure as for the \xmm\ Serendipitous Source Catalogue \citep[2XMM, ][]{2009A&A...493..339W}.
We built on similar studies of \xmm\ observations of the Local-Group galaxies M\,31 \citep{2005A&A...434..483P,2011A&A...534A..55S} and M\,33 \citep{2004A&A...426...11P,2006A&A...448.1247M}.
Compared to the standard \xmm\ source catalogue, these catalogues have an improved positional accuracy by using boresight corrections from identified sources, plus a comprehensive source screening.
The creation of our source catalogue is described in the following subsections.
Further details are given in Appendix~\ref{sec:cat:details}.

\subsection{Maximum-likelihood source detection}
\label{sec:cat:detection}

We first reprocessed all observations homogeneously with SAS 10.0.0\footnote{Science Analysis Software (SAS), http://xmm.esac.esa.int/sas/} 
and created event lists using {\tt epchain} or {\tt emchain}, respectively.
We used {\tt epreject} to correct for artefacts in the EPIC-pn offset map.
This avoids the detection of spurious apparently very-soft sources later on,
but has the disadvantage of enhancing the effect of optical loading due to optically bright stars.
Therefore we screened for bright stars  as described in Sec.~\ref{sec:cat:screening}.

To exclude time intervals of high background at the beginning or end of the satellite orbit, or during soft-proton flares,
we applied a screening to remove time intervals with background rates in the (7.0$-$15.0) keV band,
above 8 and 2.5 cts ks$^{-1}$ arcmin$^{-2}$ for EPIC-pn and EPIC-MOS, respectively.
Since soft-proton flares affect all EPIC detectors, EPIC-pn and EPIC-MOS were allowed to veto each other,
except for the observations 0503000301 and 0011450201, 
where the count rate in the high-energy band was significantly increased for EPIC-pn only.
Therefore we used the EPIC-MOS data in these cases.
For observations 0112780601, 0164560401, 0301170501 and 0135722201, the good time exposure was below 1 ks and we therefore rejected these observations.
The resulting net exposures are given in Table~\ref{tab:observations}.
This good time selection procedure  removed about 16\% of exposures from the survey data, 22\% from the calibration observations,
34\% from other archival data and 10\% of the outer fields.
We discarded EPIC-pn events between 7.2 keV and 9.2 keV, since these are affected by
background fluorescent emission lines, inhomogeneously distributed over the detector area \citep{2004SPIE.5165..112F}.
In the lowest energy band of EPIC-pn, we used an additional screening of recurrent hot pixels and for a few columns with increased noise where
we rejected events below individual energy-offsets between 220 and 300 eV.

We produced images and exposure maps corrected for vignetting
with an image pixel binning of $2\arcsec\times2$\arcsec\
in the five \xmm\ standard energy bands:
1 $\rightarrow$ (0.2--0.5) keV,
2 $\rightarrow$ (0.5--1.0) keV,
3 $\rightarrow$ (1.0--2.0) keV,
4 $\rightarrow$ (2.0--4.5) keV, and
5 $\rightarrow$ (4.5--12.0) keV.
We used single-pixel events for EPIC-pn in the (0.2--0.5) keV band,
single- and double-pixel events for the other EPIC-pn bands,
and single- to quadruple-pixel events for all EPIC-MOS bands.
EPIC-MOS events were required to have {\tt FLAG=0}.
For EPIC-pn we selected {\tt (FLAG \& 0xfa0000) = 0}, which, as opposed to {\tt FLAG=0}, allows events in pixels, when next to bad pixels or bad columns.
This increases the sky coverage, but can also cause additional spurious detections, which need to be taken into consideration (see Sec.~\ref{sec:cat:screening}).

We accomplished source detection on the images with the SAS task {\tt edetect\_chain} in all energy bands and instruments (up to $3\times5$ images) simultaneously.
We allowed two sources with overlapping point-spread function (PSF) to be fitted simultaneously.
Possible source extension was investigated by using a $\beta$-model that approximates the brightness profile of galaxy clusters \citep{1976A&A....49..137C}.
This results in a source extent, with corresponding uncertainty, and a likelihood of source extent $ML_{\rm ext}$.
A detailed description of the detection procedure can be found in \citet{2009A&A...493..339W}.
As in the case of their catalogue, we accepted detections with $ML_{\rm det} = -\ln (P) \geq 6$,
where $ML_{\rm det}$ is the detection likelihood, normalised to two degrees of freedom,
and $P$ is the chance detection probability due to Poissonian background fluctuations.

\subsection{Compilation of the point-source catalogue}
\label{sec:cat:compilation}

Astrometric corrections of the positions of detections were applied, as described in Sec.~\ref{sec:astcor}.
We uniformly assume a systematic positional uncertainty of $\sigma_{\rm sys}=0.5$\arcsec\ \citep{2005A&A...434..483P}.
The total positional uncertainty was estimated by $ \sigma = (\sigma_{\rm sys}^2 + \sigma_{\rm stat}^2)^{1/2}$,
where $\sigma_{\rm stat}$ is the statistical uncertainty, determined by {\tt emldetect}.
After a screening of the catalogue (see Sec.~\ref{sec:cat:screening}), 
all 5236 non-spurious detections of point, or moderately extended, sources were auto-correlated
to identify detections originating from the same source in a field which was observed several times.
For the auto-correlation, we accepted correlations with a maximal angular separation of $d_{\rm sep}<7$\arcsec and $d_{\rm sep} < 3 (\sigma_1 + \sigma_2$),
where $\sigma_{1,2}$ is the total positional uncertainty of the two detections \citep[see][]{2009A&A...493..339W}.
This resulted in 3053 unique X-ray sources.
Master source positions and source extent were calculated from the error-weighted average of the individual detection values.
Detection likelihoods were combined and renormalised for two degrees of freedom.

To investigate the spectral behaviour of all sources, we used hardness ratios $HR_i$ ($i=1,2,3,4$), defined by
$$ HR_i=\frac{R_{i+1}-R_{i}}{R_{i+1}+R_{i}}
$$
where $R_{i}$ is the count rate in energy band $i$ as defined in Sec.\ref{sec:cat:detection}.
To increase statistics, we also calculated average HRs, combining all available instruments and observations.
$HR_i$ is not given, if both rates $R_i$ and $R_{i+1}$ are null
or if the 1$\sigma$ uncertainty of $\Delta HR_i$ covers the complete $HR$ interval from -1 to 1.

To convert an individual count rate $R_i$ of each energy band $i$ into an observation setup-independent, observed flux $F_i$,
we calculated energy conversion factors (ECFs) $f_i=R_i / F_i$ as described in Sec.~\ref{sec:ecf}.
For the calculation, we assumed a universal spectrum for all sources, described by a power-law model with a photon index of $\Gamma=1.7$
and a photo-electric foreground absorption by the Galaxy of $N_{\rm H, Gal} = 6\times 10^{20}$ cm$^{-2}$ \citep[average for SMC main field in \Hone\ map of ][]{1990ARA&A..28..215D}.
For sources with several detections, we give the minimum, maximum and error-weighted average values for the flux.

In addition to the fluxes for each detection,
we calculated flux upper limits $F_{\rm UL}$ for each observation and source, if the source was 
observed but not detected in the individual observation.
As for the initial source detection, we used {\tt emldetect}, with the same parameters as above, to fit sources,
but kept the source positions fixed ({\tt xidfixed=yes}) at the master positions 
and accepted all detection likelihoods in order to get an upper limit for the flux.

Following \citet{1993ApJ...410..615P}, \citet{2006A&A...448.1247M} and \citet{2008A&A...480..599S},
for the characterisation of the observed variability of sources covered by various \xmm\ observations,
we calculated the variability $V$ and its significance $S$ from

$$V=\frac{F_{\rm max}}{F_{\rm min}}
\hspace{1cm}
S=\frac{F_{\rm max} - F_{\rm min}}{\sqrt{\sigma_{\rm max}^2+\sigma_{\rm min}^2}}$$
where $F_{\rm max}$ and $\sigma_{\rm max}$ are the source flux and 1$\sigma$ uncertainty in the (0.2--4.5) keV band of the detection, for which $F-\sigma$ is maximal among
all detections with a significance of $F>2\sigma$.
In a similar way, $F_{\rm min}$ and $\sigma_{\rm min}$ were chosen from the detection, for which $F+\sigma$ is minimal among all detections, with $F>2\sigma$.
In cases of $F<2\sigma$, we also considered $F_{\rm min} = 3\sigma$ as a possible lower limit.
Analogously, the minimum upper limit flux $F_{\rm UL}$ was selected from the observations, where the source was not detected.
If the minimum $F_{\rm UL}$ was smaller than $F_{\rm min}$ defined above, we used it instead.

To investigate the flux variability within the individual observations,
we used a Kolmogorow-Smirnow (KS) test to compare the photon arrival time distribution with the expected distribution from a constant source.
This method is also applicable to sources with poor statistics, where background subtracted binned light curves cannot be obtained.
A detailed description is given in Sec.~\ref{sec:sttimevar}.

\subsection{Estimation of sensitivity }
\label{sec:sensmaps}
To have an estimate of the completeness of the catalogue,
we calculated sensitivity maps with {\tt esensmap} for the individual energy bands and instruments,
as well as for combinations of them, for each observation.
Assuming Poisson statistics, detection limits for each position were calculated from the exposure and background maps.
In the case of combined energy bands or instruments,
the background images were added and the exposure maps were averaged,
weighted by the expected count rate for the adopted universal spectrum of Sec.~\ref{sec:cat:compilation}.
The individual observations were combined, by selecting the observation with highest sensitivity at each position.
We note that, depending on the individual source spectra, the detection limits deviate from this estimated value,
but a detailed simulation of the detection limit goes beyond the scope of this study.

\begin{figure}
  \resizebox{\hsize}{!}{\includegraphics[angle=0,clip=]{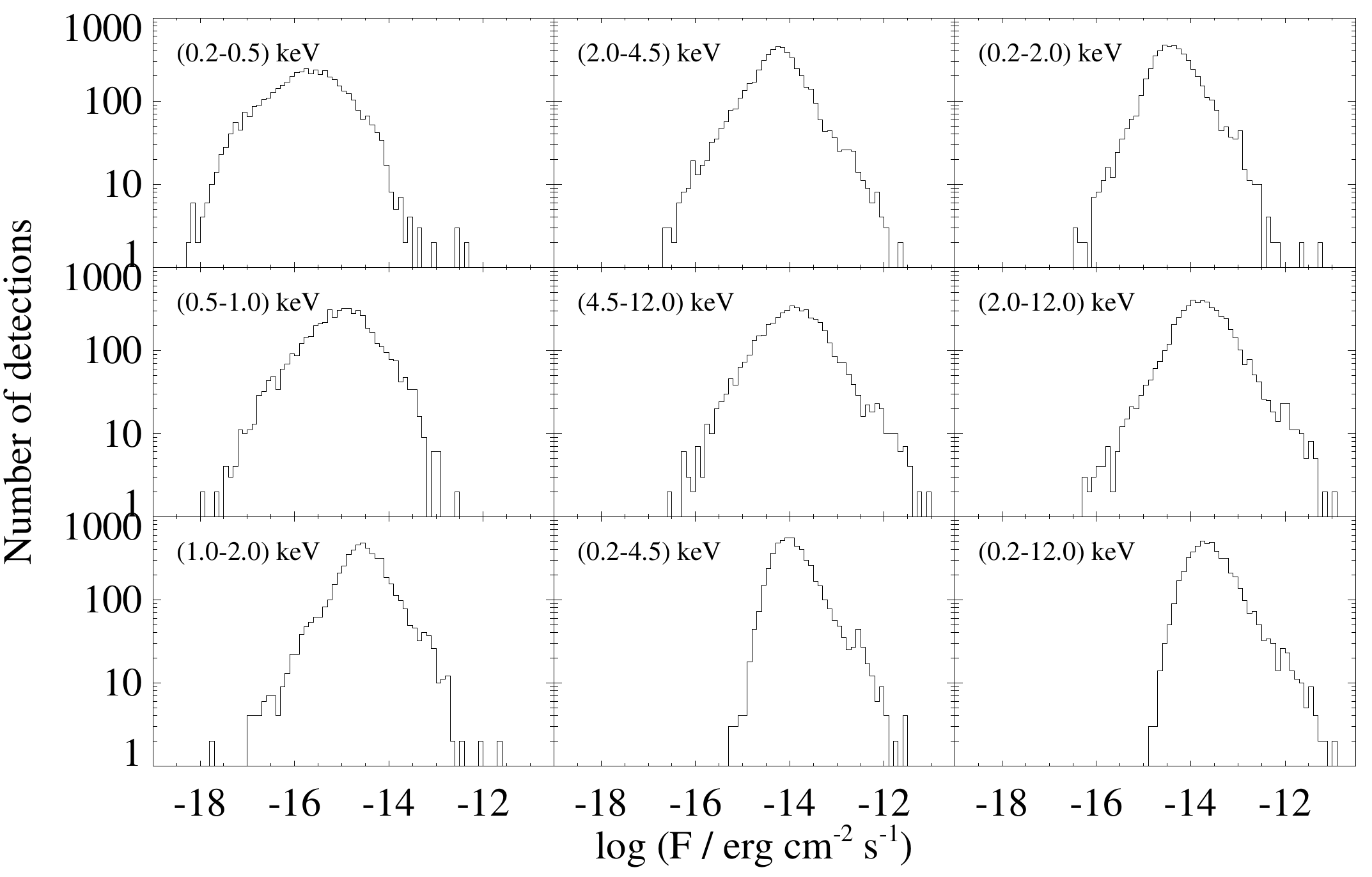}}
  \caption{
    Histograms of the flux distribution of the individual detections of the \xmm\ catalogue in various energy bands.
  }
  \label{fig:histflux}
\end{figure}

\section{Catalogue description and characterisation}
\label{sec:characterisation}

\begin{figure*}
  \resizebox{\hsize}{!}{\includegraphics[angle=0,clip=]{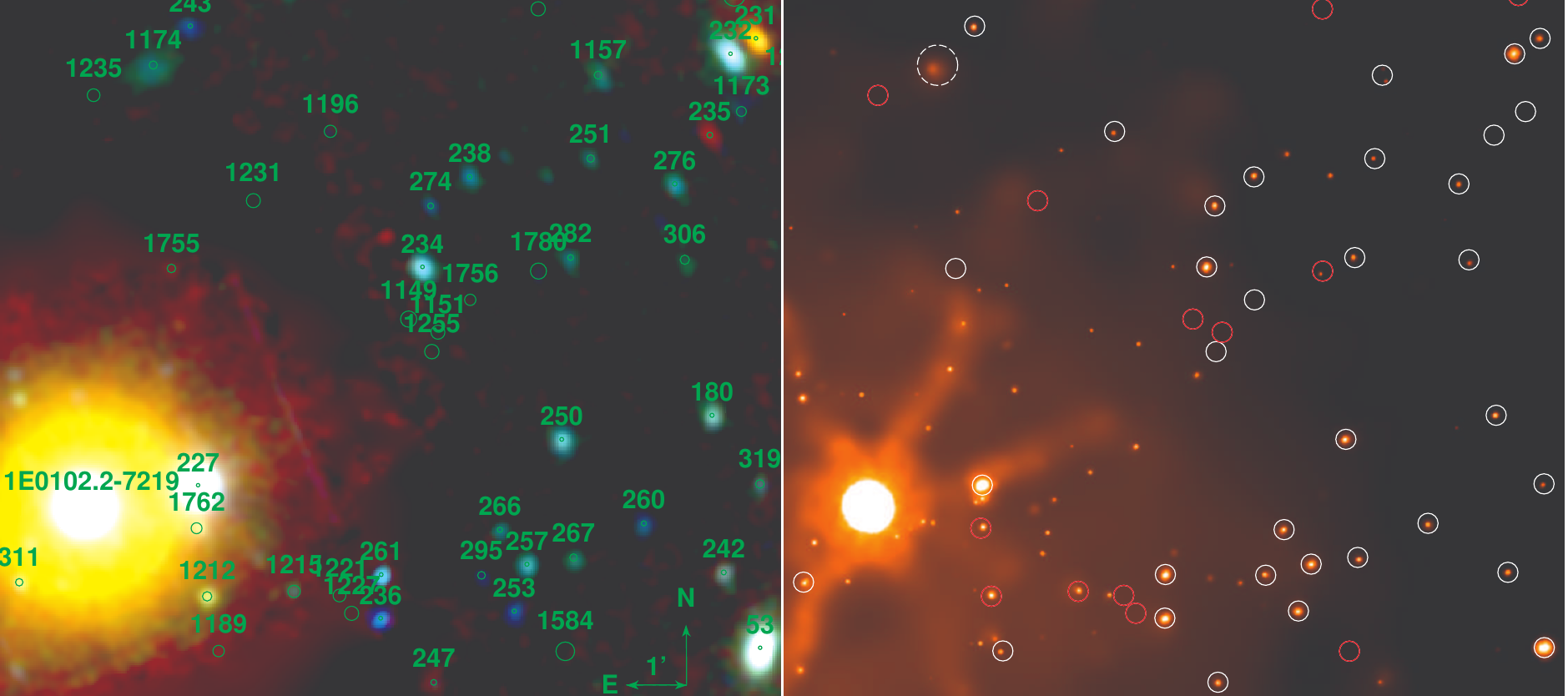}}
  \caption{
    {\it Left:} \xmm\ mosaic image of the deepest field in the SMC overplotted with the point-source catalogue.
                 The image is background subtracted. Colours (red/green/blue) give logarithmically scaled intensities in the (0.2--1.0)/(1.0--2.0)/(2.0--4.5) keV bands.
                 The brightest source is the oxygen-rich SNR \calsrc.
    {\it Right:} Comparison with a deep \cxo\ image of the same region. The false-colour image gives logarithmically scaled intensities in the (0.2--10.0) keV band.
                 \xmm\ sources with low detection likelihood ($ML_{\rm det}<8$) are plotted in red, others in white.
  }
  \label{fig:deepxmmcxo}
\end{figure*}

\begin{figure*}
  \resizebox{\hsize}{!}{\includegraphics[angle=0,clip=]{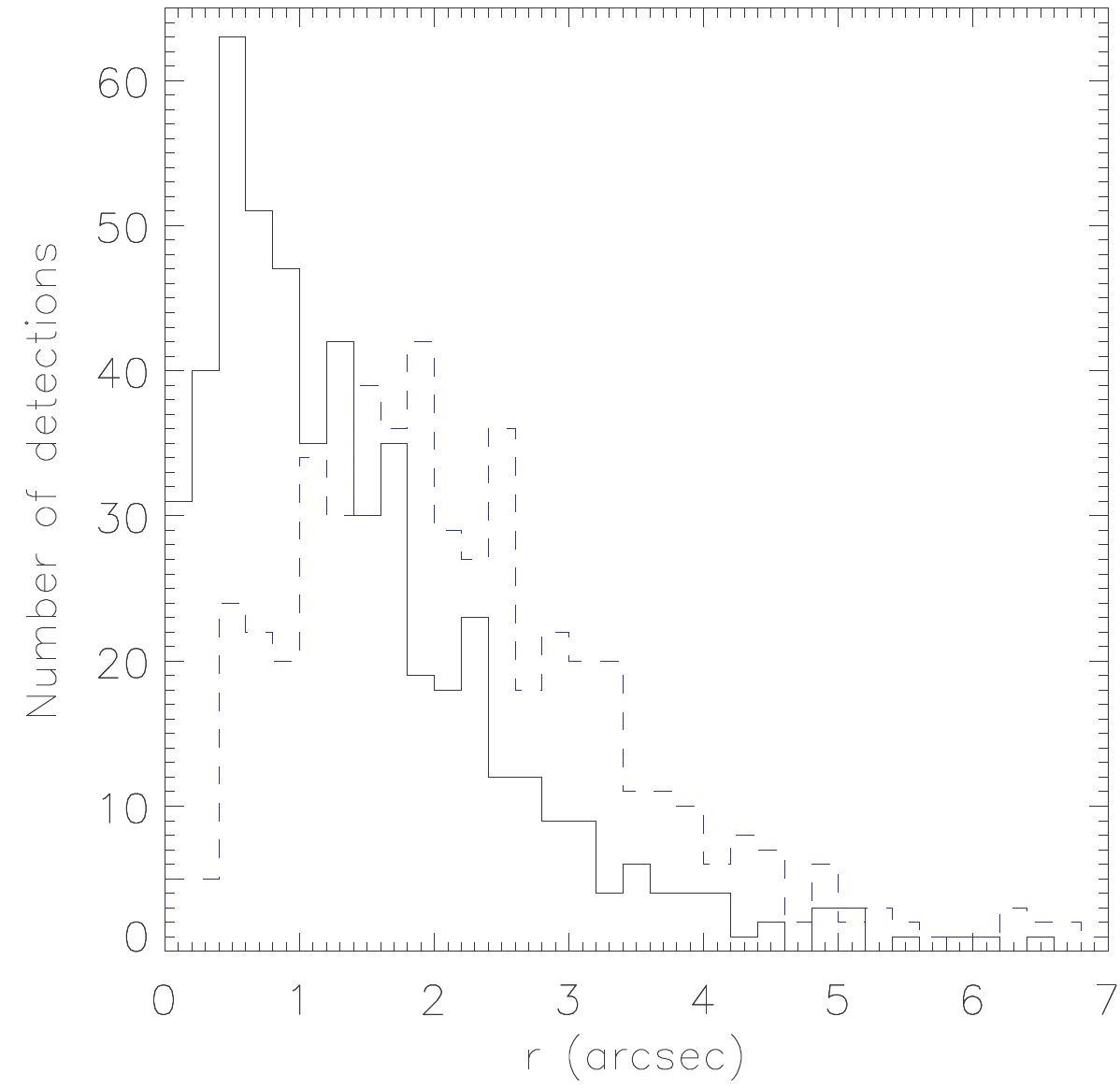}
                        \includegraphics[angle=0,clip=]{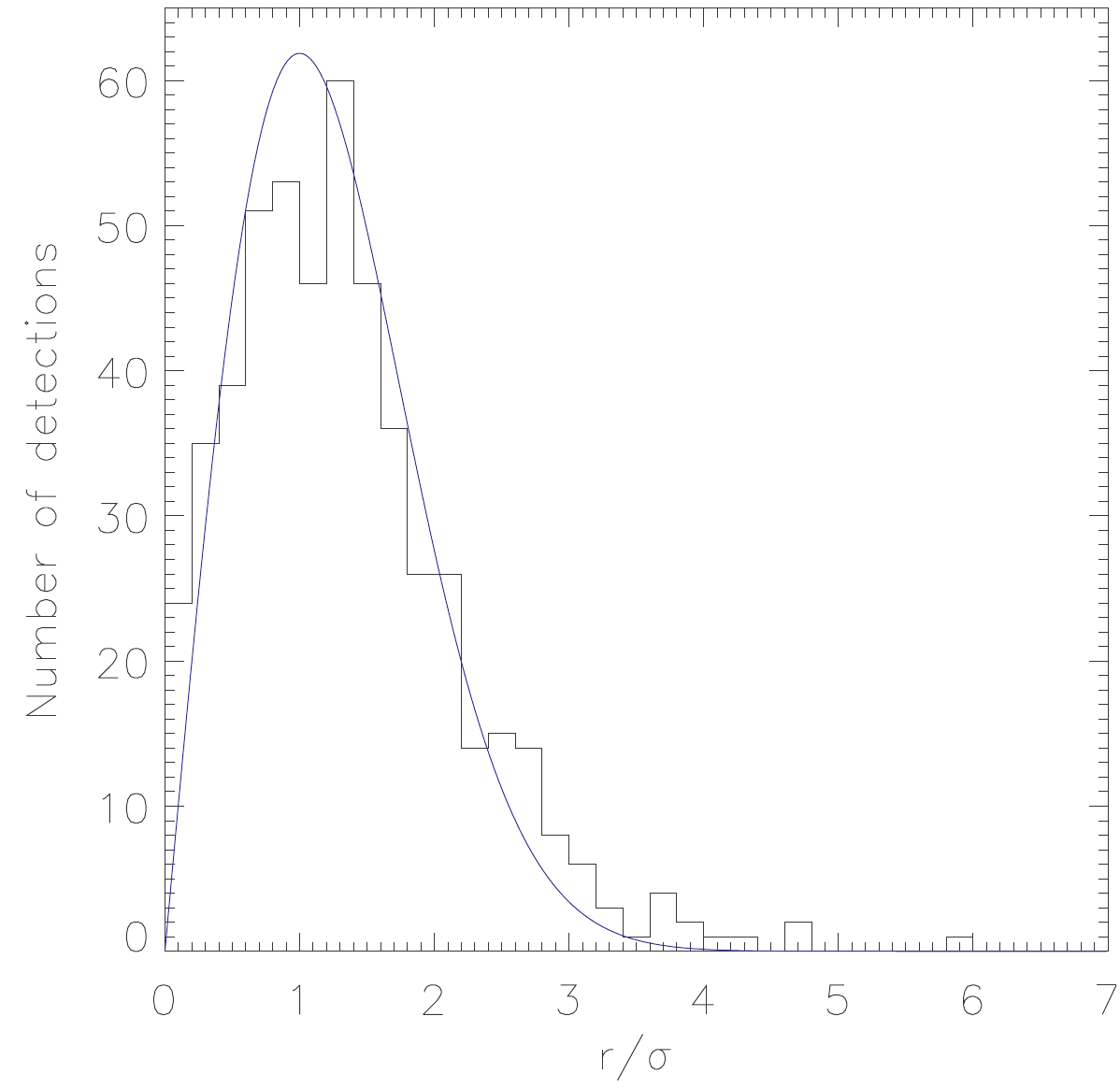}
                        \includegraphics[angle=0,clip=]{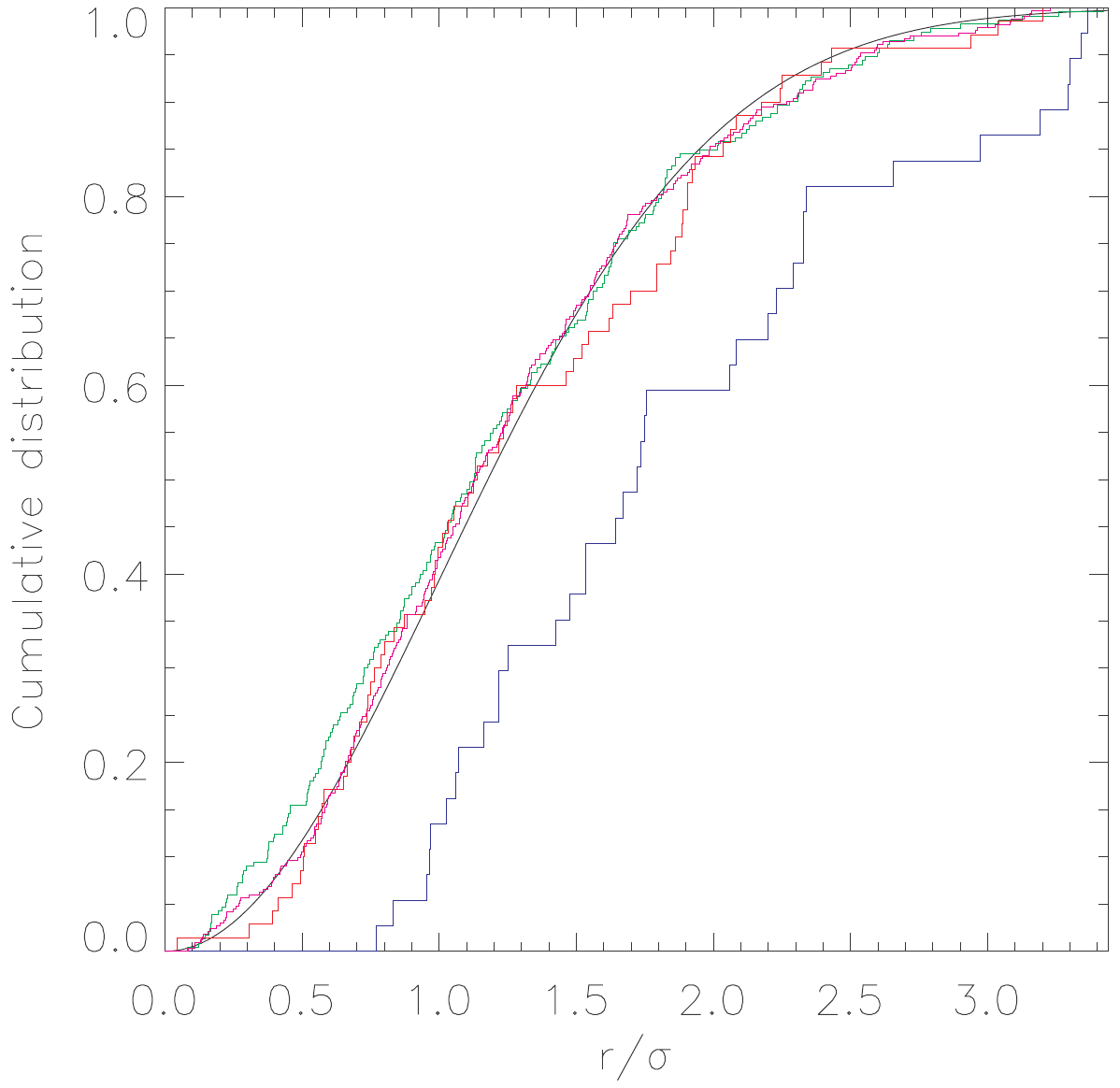}
                       }
  \caption{
    {\it Left:}    Angular separation of X-ray and reference position of identified sources before (dashed blue line) and after (solid black line) boresight correction.
    {\it Middle:}  Distribution of $r/\sigma$, compared with a Rayleigh distribution (blue line).
    {\it Right:}   Cumulative distribution of angular separation between the \xmm\ SMC catalogue and the \cxo\ catalogues of
                       \citet{2008MNRAS.383..330M} in green,
                       \citet{2010ApJ...716.1217L} in red,
                       \citet{2003ApJ...586..983N} in blue, and
                       \citet{2010ApJS..189...37E} in orange.
                       The cumulative Rayleigh distribution is shown by the black line.
  }
  \label{fig:poserr}
\end{figure*}

\begin{figure*}
  \resizebox{\hsize}{!}{\includegraphics[angle=0,clip=]{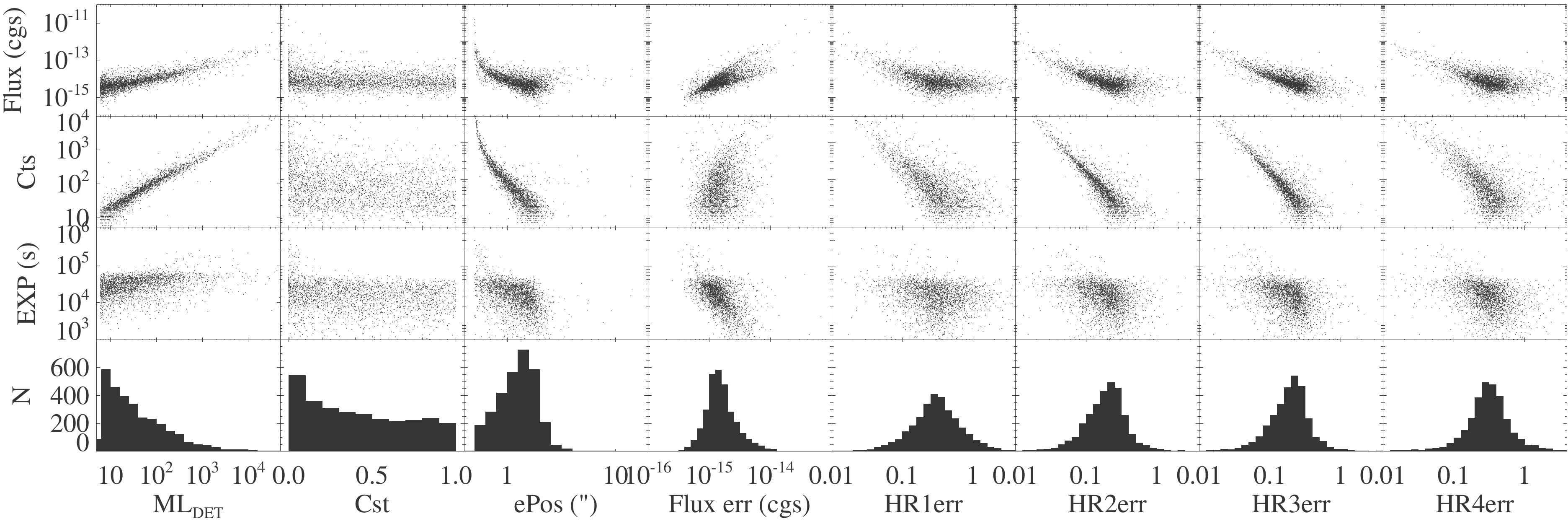}}
  \caption{
    Dependence of various source parameters of the catalogue (maximum likelihood $ML_{\rm det}$, probability for time constancy $Cst$
    and uncertainty of  position $ePos$, flux in the (0.2--4.5) keV band and  hardness ratios $HR$)
    on source flux, detected counts, and observation exposure.
    The lower panels show histograms of the distribution of these source parameters.
  }
  \label{fig:errdep}
\end{figure*}

The catalogue contains a total of 3053 X-ray sources from a total of 5236 individual detections, either from the large-programme SMC survey between May 2009 and March 2010,
or from re-analysed public archival observations between April 2000 and April 2010.
For 927 sources, there were detections at multiple epochs, 
with some SMC fields observed up to 36 times.

\subsection{Description}
\label{sec:description}

The parameters and instrumental setup of the individual observations are summarised in Table~\ref{tab:observations}.
The columns give the following parameters:
\\ (1) = ID of the observation, where S,A,C, and O denote observations from the large-programme SMC survey, archival data, calibration observations and outer fields;
\\ (2) = \xmm\ Observation Id;
\\ (3) = name of the observation target;
\\ (4) = date of the beginning of the observation;
\\ (5-6) = pointing direction;
\\ (7-8) = boresight correction;
\\ (9) = exposure Id;
\\ (10) = start time of the exposure;
\\ (11) = instrument filter;
\\ (12) = instrument mode;
\\ (13) = total exposure time;
\\ (14) = exposure time after GTI screening, not considering the instrumental death time.

The source catalogue will be available at the
Strasbourg Astronomical Data Center (CDS) and contains the following information:
\\ (1) = unique source id (not continuous or ordered by coordinates);
\\ (2) = XMM name;
\\ (3) = number of detections of the source;
\\ (4) = number of observations of the source;
\\ (5) = combined maximum detection likelihood normalised for two degrees of freedom;
\\ (6-7) = J2000 coordinates in degrees;
\\ (8) = position uncertainty for 1$\sigma$ confidence (99.7\% of all true sources positions are expected within a radius of $3.439\sigma$);
\\ (9-22) = averaged fluxes and uncertainties in the five standard bands, in the combined band (0.2$-$12.0) keV and the XID band (0.2$-$4.5) keV, all in erg  cm$^{-2}$ s$^{-1}$;
\\ (23-24) = maximum of all detected fluxes of this source in the XID band in erg cm$^{-2}$ s$^{-1}$ and the corresponding uncertainties;
\\ (25-26) = minimum or upper limit of all detected fluxes (as described in Sec.~\ref{sec:cat:compilation})  in the XID band in erg cm$^{-2}$ s$^{-1}$;
\\ (27-34) = hardness ratios between the standard bands and corresponding uncertainties;
\\ (35-37) = averaged source extension, corresponding uncertainty and likelihood of extent;
\\ (38) = KS-test probability, $Cst$, that the source was constant in all observations (the minimum value of all detections is taken, corresponding to the highest observed variability);
\\ (39-40) = source variability $V$ between individual observations and corresponding significance $S$;
\\ (41) = source classification;
\\ (42) = name of identified sources.

\subsection{Completeness,  confusion, and spurious detections}

In Fig.~\ref{fig:histflux}, the flux distribution of all individual source detections in several energy bands is presented.
In the (0.2$-$4.5) keV and (0.2$-$12.0) keV band, we see a decreasing number density for fluxes lower than \oergcm{-14} and $\sim$2\ergcm{-14}.
Thus we estimate the average detection threshold of our catalogue for sources in the SMC to be $\sim$5\ergs{33} and \oergs{34}, respectively.
However, the inhomogeneous exposure time of the individual observations has to be taken into consideration.

In the field containing the calibration source \calsrc, we can compare our catalogue with a deep \xmm\ mosaic image (Fig.~\ref{fig:deepxmmcxo}, left).
Sources from our catalogue are overplotted with circles of 3.4$\sigma$ radii.
Detections of \calsrc\ were screened, due to the high extent of this SNR. It is usually fitted with $\sim$5 sources.
Other examples of identified sources (see Sec.~\ref{sec:class}) in the field are
an active galactic nucleus (AGN) (\no{53}),
a HMXB (\no{227}),
a Wolf-Rayet star in the SMC (\no{1212}),
a Galactic star (\no{231}),
a SSS candidate (\no{235}), and
a cluster of galaxies (ClG, \no{1174}).
Sources that are not clearly visible in the mosaic image can be due to them being weak and variable.
Also, we stress that spurious detections in this field accumulate from 28 observations,
because spurious detections are determined from the result of independent source detections performed on all 28 observations comprising the \calsrc\ calibration field.
From the estimated number of spurious detections (see below) we expect $\sim$7 spurious sources in this image.
A few additional sources appear in the deep mosaic image that are not listed in our catalogue, e.g. two sources left of \no{251}.
The flux of these sources is below the detection limit of individual observations.

Since \cxo\ performs similar calibration observations of \calsrc, we compare our results with a deep \cxo\ ACIS image (Fig.~\ref{fig:deepxmmcxo}, right).
It was created by merging 107 observations with the CIAO (version 4.3) task {\tt merge\_all} and adaptively smoothed.
The exposure time is $\sim$920 ks decreasing with distance from \calsrc, as the outer fields are not covered in all calibration observations.
Our sources are overplotted with radii of 10\arcsec\ in white and red for detection likelihoods of $ML_{\rm det}\geq8$ and $ML_{\rm det}<8$, respectively.
We see that source confusion is only relevant near the brightest sources (c.f. the surrounding of \calsrc) and in some rare cases of close-by sources
(e.g. source \no{1157} might consist of two weak sources seen by \cxo).
Source \no{1174} is extended in the \cxo\ image, further supporting our ClG classification.
We see that most sources with $ML_{\rm det}\geq8$ are clearly visible in the \cxo\ image.
\no{235} is not found, due to the very soft spectrum and time variability.
For sources with  $ML_{\rm det}<8$, a corresponding source in the \cxo\ image is not always obvious.

To quantify spurious detections, we compared our catalogue with two deep \cxo\ SMC fields, where source lists are available \citep{2010ApJ...716.1217L}.
All \xmm\ sources, which were detected more than once or have a detection likelihood of $ML\gtrsim 8$,
are also listed in the \cxo\ catalogue.
Only 3 of 12 \xmm\ sources with $ML_{\rm det}<8$ and one source with $ML_{\rm det}=8.2$ were not detected by \cxo.
Some non-detections might be due to variability and the lower effective area of \cxo\ at the highest and lowest energies, but in general, as for the 2XMM catalogue,
a fraction of detections with $ML_{\rm det}<8$ is expected to be spurious and should be regarded with care.
In total, our catalogue contains 418 sources with $ML_{\rm det}<8$. From the former comparison, we roughly estimate around one hundred spurious detections among those, i.e. about one per observation.

\subsection{Accuracy of source parameters}

In Fig.~\ref{fig:poserr} left, we show the angular separation of identified sources before and after the astrometric correction (Sec.~\ref{sec:astcor}).
The median of the total position uncertainty of all source positions is 1.3\arcsec.
In Fig.~\ref{fig:poserr} middle, the distribution of angular separation $r$ scaled by the total uncertainty $\sigma = (\sigma_{\rm sys}^2+\sigma_{\rm sta}^2+\sigma_{\rm ref}^2)^{1/2}$
is shown, where $\sigma_{\rm ref}$ is the position uncertainty of the reference source.
The distribution is similar to a Rayleigh distribution (blue line), 
which justifies our estimation of the systematic error of $\sigma_{\rm sys} = 0.5$\arcsec.
Since the same sample was used to determine the boresight corrections,
some deviations from the Rayleigh distribution are expected.
For example, for all observations containing only one identified source, the angular separation will be reduced to 0 due to the boresight shift.

To further test our positional accuracies with a statistically independent sample,
we compared the final catalogue with available \cxo\ catalogues.
In Sec.~\ref{sec:correlations:X-ray}, we show that the correlation with these catalogues is close to a one-to-one, 
with a negligible number of chance coincidences.
The catalogues are listed in Table~\ref{tab:cat}.
Sources, that have been used for boresight correction were excluded from this comparison.
In Fig.~\ref{fig:poserr} right, the cumulative distribution yields a good agreement with the
catalogues of \citet{2008MNRAS.383..330M}, \citet{2010ApJ...716.1217L}, and \citet{2010ApJS..189...37E}
with KS-test statistics of 22\%, 47\%, and 77\%, respectively.
Only for sources of \citet{2003ApJ...586..983N}, an unexpected distribution of angular separations is found, with a KS-test statistic of 0.097\%. 
In a further investigation, we found a systematic offset of the \cxo\ positions relative to the \xmm\ positions by $\sim$1.7\arcsec.
The offset is also evident when we compare the \cxo\ coordinates to the Tycho-2 position of HD\,5980 and the Magellanic Clouds Photometric Survey catalogue \citep[MCPS, ][]{2002AJ....123..855Z} positions of SXP\,152 and SXP\,304.
Therefore, we conclude that the coordinates of these \cxo\ sources are wrong by a systematic offset.

An overview of the distribution of source parameter uncertainties and probabilities for existence $ML_{\rm det}$ and constancy $Cst$,
as well as their dependence on observational parameters, is shown in Fig.~\ref{fig:errdep}.
The number of counts is the main quantity on which they depend.
For 2378 and 2635 sources, the detection maximum likelihood is $ML_{\rm det} > 10$ and $>$8, respectively.
The relative uncertainties of fluxes in the (0.2$-$4.5) keV band have a median of  22\%.
For the uncertainties of the hardness ratios 1 to 4, we obtain the medians of 0.30, 0.20, 0.20, and 0.31, respectively.

\section{Cross-correlation with other catalogues}
\label{sec:correlations}

\begin{table*}
\caption[]{Reference catalogues used for cross-correlation.}
\begin{center}
\begin{tabular}{lcccrrrrr}
\hline\hline\noalign{\smallskip}
\multicolumn{1}{l}{Catalogue} &
\multicolumn{1}{c}{Type} &
\multicolumn{1}{l}{Reference} &
\multicolumn{1}{c}{$\sigma_{\rm ref}(\arcsec)$} &
\multicolumn{1}{c}{$N_{\rm cat}$\tablefootmark{a}} &
\multicolumn{1}{c}{$N_{\rm ref}$\tablefootmark{b}} &
\multicolumn{1}{c}{$N_{\rm XMM}$\tablefootmark{c}} &
\multicolumn{1}{c}{$C_{\rm ref}$\tablefootmark{d}} &
\multicolumn{1}{c}{$C_{\rm XMM}$\tablefootmark{e}} \\
\noalign{\smallskip}\hline\noalign{\smallskip}
{\it Einstein}                   & X-ray                                    &     1 &  40\tablefootmark{f}     & 50      & 48   & 154  & 45.1$\pm$1.6 & 131$\pm$13   \\
{\it Einstein}\tablefootmark{g}  & X-ray                                    &     1 &  40\tablefootmark{f}     & 50      & 26   & 27   &  6.1$\pm$2.1 & 6.5$\pm$2.4  \\
ROSAT PSPC                       & X-ray                                    &     2 &  10\tablefootmark{i}     & 353     & 282  & 353  &  40.9$\pm$4.5 & 55.5$\pm$7.4   \\
ROSAT PSPC\tablefootmark{h}      & X-ray                                    &     2 &  10\tablefootmark{i}     & 353     & 236  & 264  &  15.3$\pm$4.0&17.9$\pm$4.6  \\
ROSAT HRI                        & X-ray                                    &     3 &  2.6\tablefootmark{i}    & 109     & 76   & 78   &  2.4$\pm$1.8&2.4$\pm$1.8  \\
ASCA                             & X-ray                                    &     4 &  18.6\tablefootmark{f}   & 83      & 69   & 111  &  33.1$\pm$4.2 & 42.5$\pm$6.1  \\
\cxo\ Wing Survey                & X-ray                                    &     5 &  1.02\tablefootmark{i}   & 393     & 242  & 240  &  2.3$\pm$1.6&2.3$\pm$1.5  \\
\cxo\ deep fields                & X-ray                                    &     6 &  0.66\tablefootmark{i}   & 394     & 85   & 85   & 1.8$\pm$1.4&1.7$\pm$1.3  \\
\cxo\ NGC\,346                   & X-ray                                    &     7 &  0.30\tablefootmark{i}   & 75      & 41   & 41   & 0.58$\pm$0.64&0.63$\pm$0.70 \\
CSC (release 1.1)                & X-ray                                    &     8 &  1.30\tablefootmark{i}   & 496     & 368  & 373  &  8.2$\pm$2.4&9.4$\pm$3.4  \\
\noalign{\smallskip}
MCPS	                         & opt.                                     &     9 &  0.3                     & 2872224 & 10484& 2604 &  10082$\pm$75 & 2431$\pm$21   \\
Tycho-2                          & opt.                                     &    10 &  0.078\tablefootmark{i}  &   321   & 41   & 41   & 1.5$\pm$1.1&1.5$\pm$1.2   \\
GSC (version 2.3.2)              & opt.                                     &    11 &  0.43\tablefootmark{i}   &  855524 & 3476 & 2099 &  3045$\pm$42&1752$\pm$20    \\
\noalign{\smallskip}
2MASS                            & NIR                                      &    12 &  0.15\tablefootmark{i}   & 159491   &  923 & 743  & 565$\pm$27&427$\pm$17    \\
2MASX                            & NIR                                      &    12 &  4.4\tablefootmark{i}    & 223     & 26   & 26   & 8.5$\pm$2.4 & 9.0$\pm$2.7   \\
DENIS MC                         & NIR                                      &    13 &  0.47\tablefootmark{i}   & 94357   & 609  & 540  &  364$\pm$19 & 303$\pm$15    \\
DENIS (3rd release)              & NIR                                      &    14 &  0.3                     & 438517  & 2058 & 1043 & 1477$\pm$55&737$\pm$18    \\
IRSF Sirius                      & NIR                                      &    15 &  0.1                     & 1855973 & 8426 & 2407 &  6500$\pm$110&1914$\pm$22   \\
\noalign{\smallskip}
S$^{3}$MC                         & IR                                       &    16 & 1, 3, 6\tablefootmark{j} & 400735  & 3403 & 1711 & 2193$\pm$40&1108$\pm$17    \\
ATCA RCS                         & radio                                    & 17,18 &  1.0                     & 301     & 31   & 31   &  1.6$\pm$1.2&1.6$\pm$1.2     \\
SUMSS (version 2.1)              & radio                                    & 19,20 &  3.0\tablefootmark{i}    & 246     & 46   & 47   &  5.3$\pm$2.2&5.5$\pm$2.3   \\
\noalign{\smallskip}
MA93                             & \Halp                                    &  21   &  2.0\tablefootmark{f}    & 1805    & 63   & 62   & 18.6$\pm$3.0&18.2$\pm$3.4   \\
Murphy2000                       & \Halp, [\ion{O}{iii}]                    &  22   &  3.5, 4.4                &  286    & 12   & 12   & 7.4$\pm$2.6&7.3$\pm$2.7     \\
2dF SMC                          & stellar classification                   &  23   &  0.5\tablefootmark{f}    & 2874    & 31   & 31   & 8.8$\pm$3.4 &8.7$\pm$3.4     \\
6dF GS                           & galaxy redshifts                         &  24   &  1.0\tablefootmark{f}    & 16      & 6    & 6    & 0.04$\pm$0.20&0.04$\pm$0.20  \\
Kozlowski2009                    & AGN candidates                           &  25   &  1.0\tablefootmark{f}    & 655     & 146  & 148  & 3.9$\pm$2.2 &3.8$\pm$2.1     \\
Bica2008                         & star cluster                             &  26   &  26.6\tablefootmark{k}   & 409     & 41   & 45  & 29.0$\pm$6.1&32.1$\pm$6.6      \\
Bonatto2010                      & star cluster                             &  27   &  33.7\tablefootmark{k}   & 75      & 11   & 14   & 7.8$\pm$2.9&9.4$\pm$3.2       \\

\noalign{\smallskip}\hline

\end{tabular}

\tablefoot{
\tablefoottext{a}{Number of reference-catalogue sources in the \xmm\ survey field.}
\tablefoottext{b}{Number of reference-catalogue sources matching at least one \xmm\ source.}
\tablefoottext{c}{Number of \xmm\ sources matching at least one reference-catalogue source.}
\tablefoottext{d}{Expected number of reference sources matched by chance.}
\tablefoottext{e}{Expected number of \xmm\ sources matched by chance.}
\tablefoottext{f}{Value estimated.}
\tablefoottext{g}{Compared with a subset of \xmm\ sources brighter than $5 \times 10^{-14}$ erg cm$^{-2}$ s$^{-1}$.}
\tablefoottext{h}{Compared with a subset of \xmm\ sources brighter than $10^{-14}$ erg cm$^{-2}$ s$^{-1}$.}
\tablefoottext{i}{Catalogue contains individual position uncertainties for each source, value gives the average of the used sample.}
\tablefoottext{j}{Uncertainty is 3\arcsec\ for sources detected at 24~$\mu$m or higher, 6\arcsec\ for sources detected at 70~$\mu$m only, 1\arcsec\ otherwise.}
\tablefoottext{k}{This is the average of the semi-mayor axis extent.}
}
\tablebib{
(1)~\citet{1992ApJS...78..391W};
(2) \citet{2000A&AS..142...41H};
(3) \citet{2000A&AS..147...75S};
(4) \citet{2003PASJ...55..161Y};
(5) \citet{2008MNRAS.383..330M};
(6) \citet{2010ApJ...716.1217L};
(7) \citet{2003ApJ...586..983N};
(8) \citet{2010ApJS..189...37E};
(9) \citet{2002AJ....123..855Z};
(10) \citet{2000A&A...355L..27H};
(11) \citet{2008AJ....136..735L};
(12) \citet{2006AJ....131.1163S};
(13) \citet{2000A&AS..144..235C};
(14) \citet{2005yCat.2263....0D};
(15) \citet{2007PASJ...59..615K};
(16) \citet{2007ApJ...655..212B};
(17) \citet{2002MNRAS.335.1085F};
(18) \citet{2004MNRAS.355...44P};
(19) \citet{1999AJ....117.1578B};
(20) \citet{2003MNRAS.342.1117M};
(21) \citet{1993A&AS..102..451M};
(22) \citet{2000MNRAS.311..741M};
(23) \citet{2004MNRAS.353..601E};
(24) \citet{2009MNRAS.399..683J};
(25) \citet{2009ApJ...701..508K};
(26) Table 3 of \citet{2008MNRAS.389..678B};
(27) \citet{2010MNRAS.403..996B}.
}
\end{center}
\label{tab:cat}
\end{table*}

To classify and identify individual sources, we cross-correlated the boresight-corrected positions of our \xmm\ SMC point-source catalogue with publicly available catalogues.
The correlations with X-ray catalogues from previous studies allows us to study the evolution of X-ray sources with time.
Other wavelength catalogues add ancillary information, enabling a multi-wavelength analysis.
The catalogues used are listed in Table~\ref{tab:cat} together with statistical properties of the correlations.

\subsection{Selection of counterparts}
The uncertainties of the \xmm\ source coordinates are radially symmetric, as is the case for most of the other catalogues.
For some catalogues with higher positional accuracy, elliptical errors are given (e.g. 2MASS).
Since the \xmm\ positional uncertainty is dominant, for simplicity we assumed radial symmetric uncertainties for all catalogues
and used the semi-major axis as the radius if elliptical errors are given.
When confidence levels for the  positional uncertainty are given,
we recalculated  the positional uncertainty of the reference catalogue $\sigma_{\rm ref}$ for 1$\sigma$ confidence.
In some cases, the uncertainties had to be estimated.
Following \citet{2009A&A...493..339W}, we consider all correlations
having an angular separation of $d \leq 3.439\times (\sigma_{\rm sys}^2+\sigma_{\rm sta}^2+\sigma_{\rm ref}^2)^{1/2}$ as counterpart candidates.
This corresponds to a 3$\sigma$ (99.73\%) completeness when we assume a Rayleigh distribution.
The resulting number of matched \xmm\ and reference sources, $N_{\rm XMM}$ and $N_{\rm ref}$, is given in Table~\ref{tab:cat}.

\subsection{Estimation of chance correlations}
\label{sec:correlations:chance}

Depending on the source density and positional uncertainty, the number of chance coincidences, $C_{\rm XMM}$ and $C_{\rm ref}$, has to be considered.
These were estimated by shifting our catalogue in right ascension and declination
by multiples of the maximal possible correlation distance between two sources in both catalogues and using the same correlation criterion as above.
We performed several of these correlation runs to investigate variations of chance coincidences.

\begin{figure}
  \resizebox{\hsize}{!}{\includegraphics[angle=0,clip=]{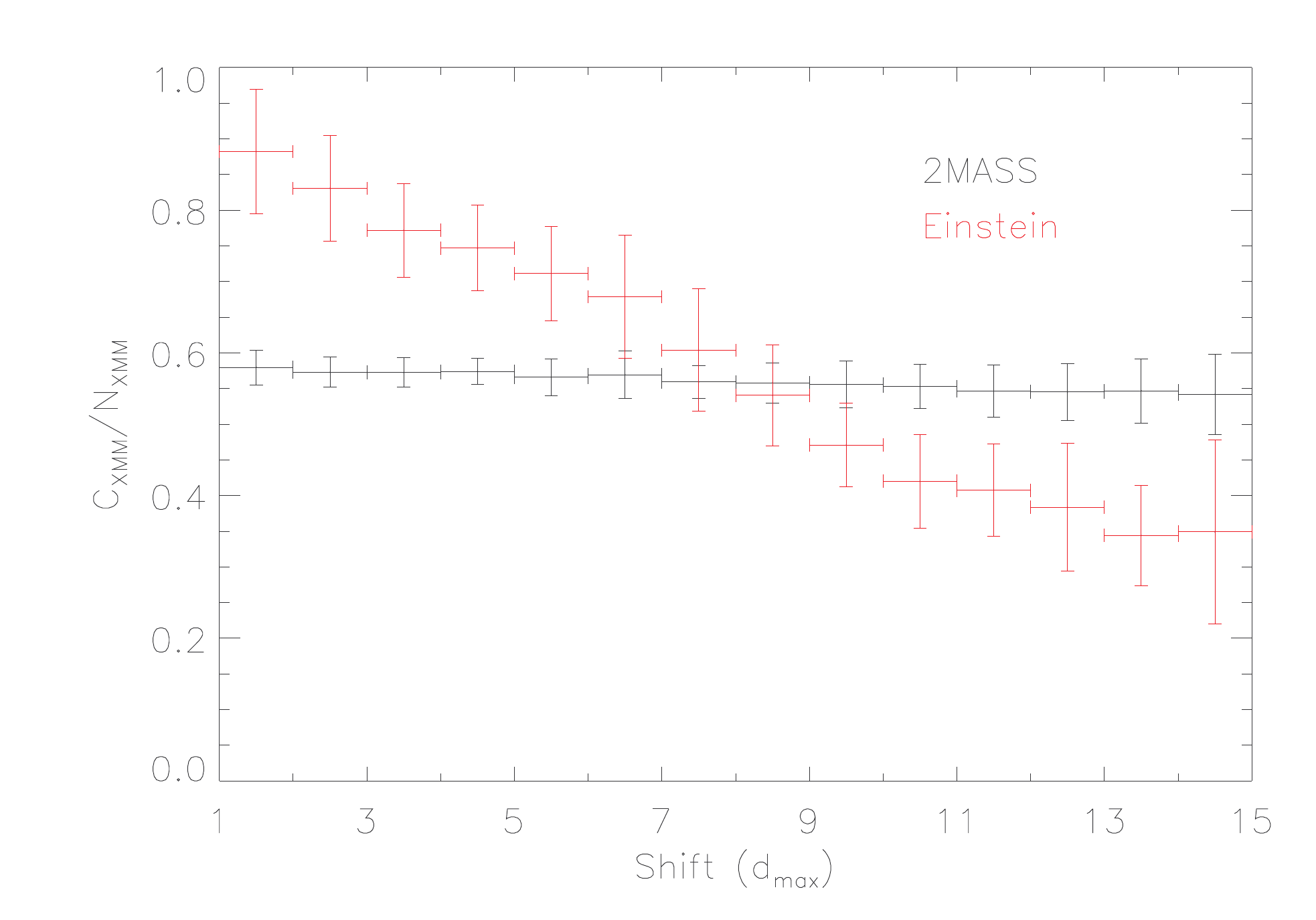}}
  \caption{
Examples for the chance-correlation dependence on offset.
The data points are binned in units of maximal correlation distance ($d_{\rm max}$).
For each bin, the average of chance-correlation $C_{\rm XMM}$ is given, normalised to the number of correlations $N_{\rm XMM}$ with unshifted coordinates.
The error bars give the standard deviation for each bin.
  }
  \label{fig:corstat}
\end{figure}

In Fig.~\ref{fig:corstat} we give examples of the dependence of the number of chance-correlations $C_{\rm XMM}$ on the shifting distance.
In the case of the 2MASS catalogue, we see only a small systematic decrease with increasing offset that is negligible, compared to the standard deviations.
If the coordinate shift becomes too large, a variable source-density can affect the number of correlations.
This is the case for catalogues with inhomogeneously distributed sources, 
e.g. due to the SMC morphology or a limited SMC-specific field of the catalogue.
When investigating the dependence of the number of correlations on shifting distance,
we found no significant variations on a scale of a few shifts,
with the exception of the correlation with the {\it Einstein} catalogue.
The variations found for the {\it Einstein} catalogue are caused by relatively large positional uncertainties that require a large coordinate shift.

In order to estimate the variation of the number of chance coincidences,
we used 24 shifted correlations of a $5\times5$ grid. All these samples result from coordinate shifts between  $d_{\rm max}$ and $\sqrt{8} d_{\rm max}$.
The comparison with the {\it Einstein} catalogue was done with a  $3\times3$ grid.
The averaged numbers of chance coincidences for our catalogue $C_{\rm XMM}$ and the reference catalogue $C_{\rm ref}$ are given in Table~\ref{tab:cat}.
Their uncertainties are estimated using their standard deviation.
By comparing these values with the number of real correlations, we can estimate the contribution of chance coincidences.
In general, we find that correlations with \cxo\ X-ray sources, radio sources and IR-selected AGN candidates are quite robust,
whereas correlations with optical to infrared catalogues are dominated by the contribution of chance coincidences.
The number of multiple coincidences can be estimated by comparing the number of matched sources in our catalogue $N_{\rm XMM}$ to the number of matched sources of the reference catalogues $N_{\rm ref}$.
Correlations with radio and \cxo\ X-ray catalogues are close to a one-to-one correlation, whereas for dense optical catalogues four times more reference sources are found than X-ray sources.
In the case of the MCPS, 74\% of the matched \xmm\ sources have more than one, and 53\% have more than two counterpart candidates.

\subsection{Correlation with other X-ray catalogues}
\label{sec:correlations:X-ray}

We correlated our catalogue with X-ray catalogues from previous studies.
From earlier epochs we used X-ray sources detected with the {\it Einstein} observatory between 1979 and 1980 \citep{1992ApJS...78..391W},
ROSAT sources from \citet{2000A&AS..142...41H} and \citet{2000A&AS..147...75S} detected between 1990 and 1998,
and ASCA sources from observations between 1993 and 2000 \citep{2003PASJ...55..161Y}.
Due to the aforementioned high positional uncertainties of the {\it Einstein} catalogue and the higher sensitivity of \xmm,
the correlation is dominated by chance coincidences, so
most {\it Einstein} sources cannot be assigned uniquely to an \xmm\ source.
A more unambiguous correlation can be achieved if a set of the brightest \xmm\ sources, with fluxes  $> 5 \times 10^{-14}$ erg cm$^{-2}$ s$^{-1}$, is used.
Similarly, we find an improvement for the correlation with the ROSAT PSPC catalogue, if we impose a limit of sources with fluxes $> 10^{-14}$ erg cm$^{-2}$ s$^{-1}$.
These results are also listed in Table~\ref{tab:cat}.
By comparing the catalogues and mosaic images, we found about 30 ROSAT sources, without a corresponding \xmm\ source.
Eight can be associated with variable sources (HMXB or SSS), while others are faint and might be spurious or affected by confusion of multiple sources or diffuse emission.

Based on \cxo\ observations since 1999, there are several catalogues from the same era as the \xmm\ data but covering only some part of the SMC main field:
fields in the SMC wing \citep{2008MNRAS.383..330M},
deep fields in the SMC bar \citep{2010ApJ...716.1217L}, and
sources around NGC\,346 \citep{2003ApJ...586..983N}.
Additional sources were taken from the \cxo\ Source Catalogue \citep[CSC, ][]{2010ApJS..189...37E}.
In general for comparable exposures, these catalogues offer more precise positions but have fewer counts per detection compared to \xmm\ detections.
The correlation between the \xmm\ and \cxo\ sources is close to a one-to-one correlation with less then 2\% of chance coincidences.

\subsection{Correlation with catalogues at other wavelength}
To identify the optical counterparts we used the Magellanic Clouds Photometric Survey catalogue \citep[MCPS, ][]{2002AJ....123..855Z},
providing stellar photometry in $U$, $B$, $V$ and $I$ down to magnitudes of $\sim$20--22 mag.
Due to the high source density compared to the \xmm\ resolution, the cross correlation is dominated by chance coincidences.
To identify bright foreground stars, which are not listed in the MCPS, we used the Tycho-2 catalogue \citep{2000A&A...355L..27H}, which has a completeness of 99\% for $V\sim11.0$ mag
and provides proper motions and $B_{\rm T}$ and $V_{\rm T}$ magnitudes.
Since the MCPS does not cover all parts of the \xmm\ field and some stars around $V \sim 12$ mag are too faint for the Tycho-2 catalogue but too bright for the MCPS,
we used the Guide Star Catalogue \citep[GSC, ][]{2008AJ....136..735L} in these cases, which gives $B_{\rm J}$ and $R_{\rm F}$ magnitudes down to $\sim$21 mag.
For 129 X-ray sources which do not have a counterpart in either of the MCPS and Tycho-2 catalogues, we found a possible counterpart in the GSC.

Near-infrared sources in $J$, $H$, and $K_{\rm S}$ were taken from the Two Micron All Sky Survey \citep[2MASS, ][]{2006AJ....131.1163S},
the Deep Near Infrared Survey \citep[DENIS, ][]{2000A&AS..144..235C,2005yCat.2263....0D},
and the InfraRed Survey Facility (IRSF) Sirius catalogue of \citet{2007PASJ...59..615K}.
Since these catalogues contain measurements from different epochs,
they allow us to estimate the NIR variability of X-ray sources, which is especially interesting for HMXBs.

Infrared fluxes at 3.6, 4.5, 5.8, 8.0, 24, and 70 $\mu$m are taken from the {\it Spitzer} Survey of the SMC \citep[S$^3$MC, ][]{2007ApJ...655..212B}.
Radio sources were taken from the ATCA radio-continuum study \citep{2004MNRAS.355...44P,2002MNRAS.335.1085F},
with ATCA radio point-source flux densities at 1.42, 2.37, 4.80, and 8.64 GHz, 
and from the Sydney University Molonglo Sky Survey at 843 MHz \citep[SUMSS, ][]{2003MNRAS.342.1117M}.
These correlations enable a classification of background sources.

Furthermore, we compared our sources with some individual catalogues providing
emission-line sources \citep{1993A&AS..102..451M,2000MNRAS.311..741M},
stellar classification \citep[][]{2004MNRAS.353..601E},
galaxies confirmed by redshift measurements \citep{2009MNRAS.399..683J},
and IR selected AGN candidates \citep[][]{2009ApJ...701..508K}.
For the correlation with the catalogues of star clusters \citep{2008MNRAS.389..678B,2010MNRAS.403..996B},
we used the semi-major axis of the cluster extent as a $3\sigma$ uncertainty for the reference position.

\section{Source identification and classification}
\label{sec:class}

\begin{figure*}
  \resizebox{\hsize}{!}{\includegraphics[angle=-90,clip=]{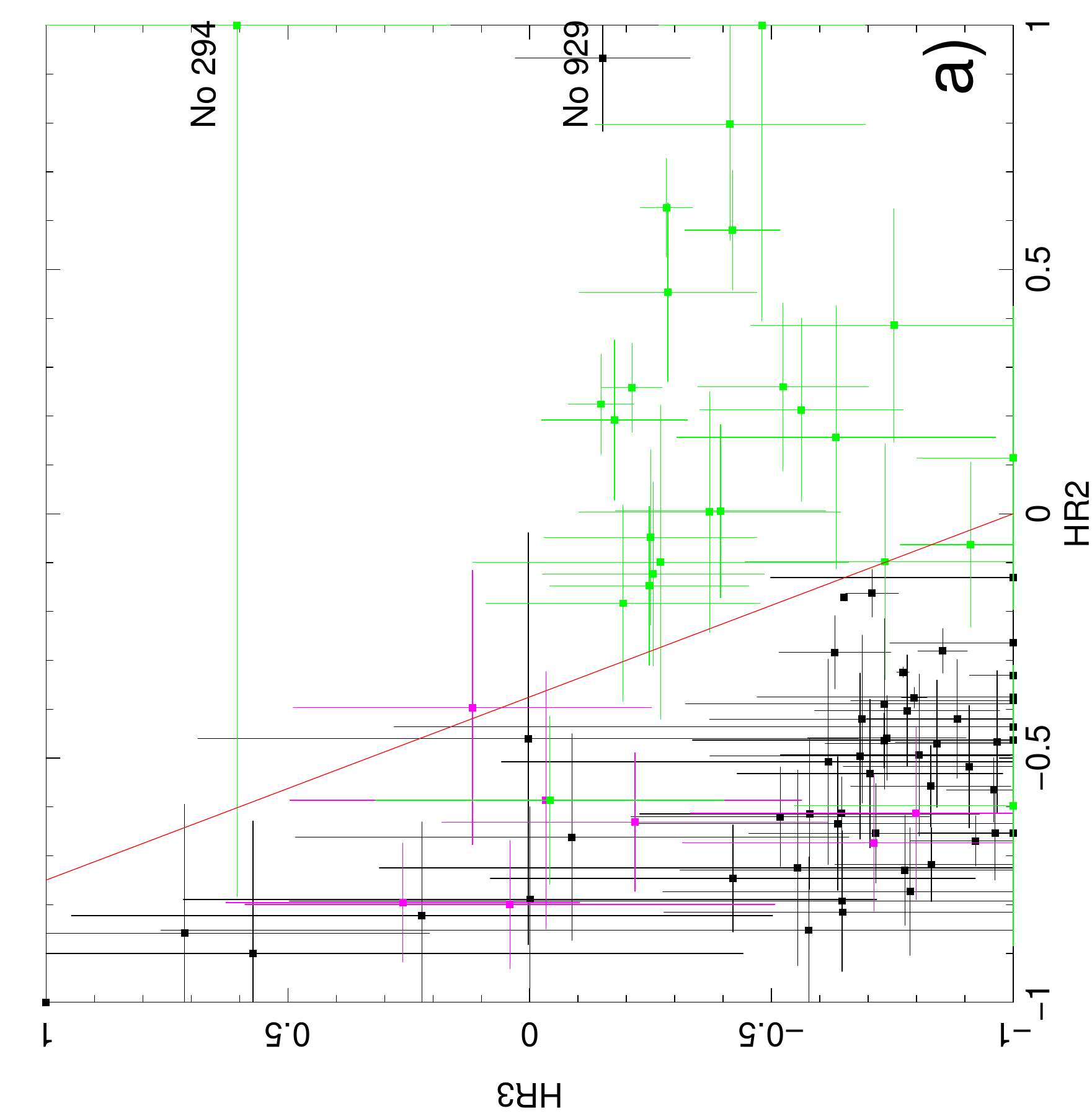}\includegraphics[angle=-90,clip=]{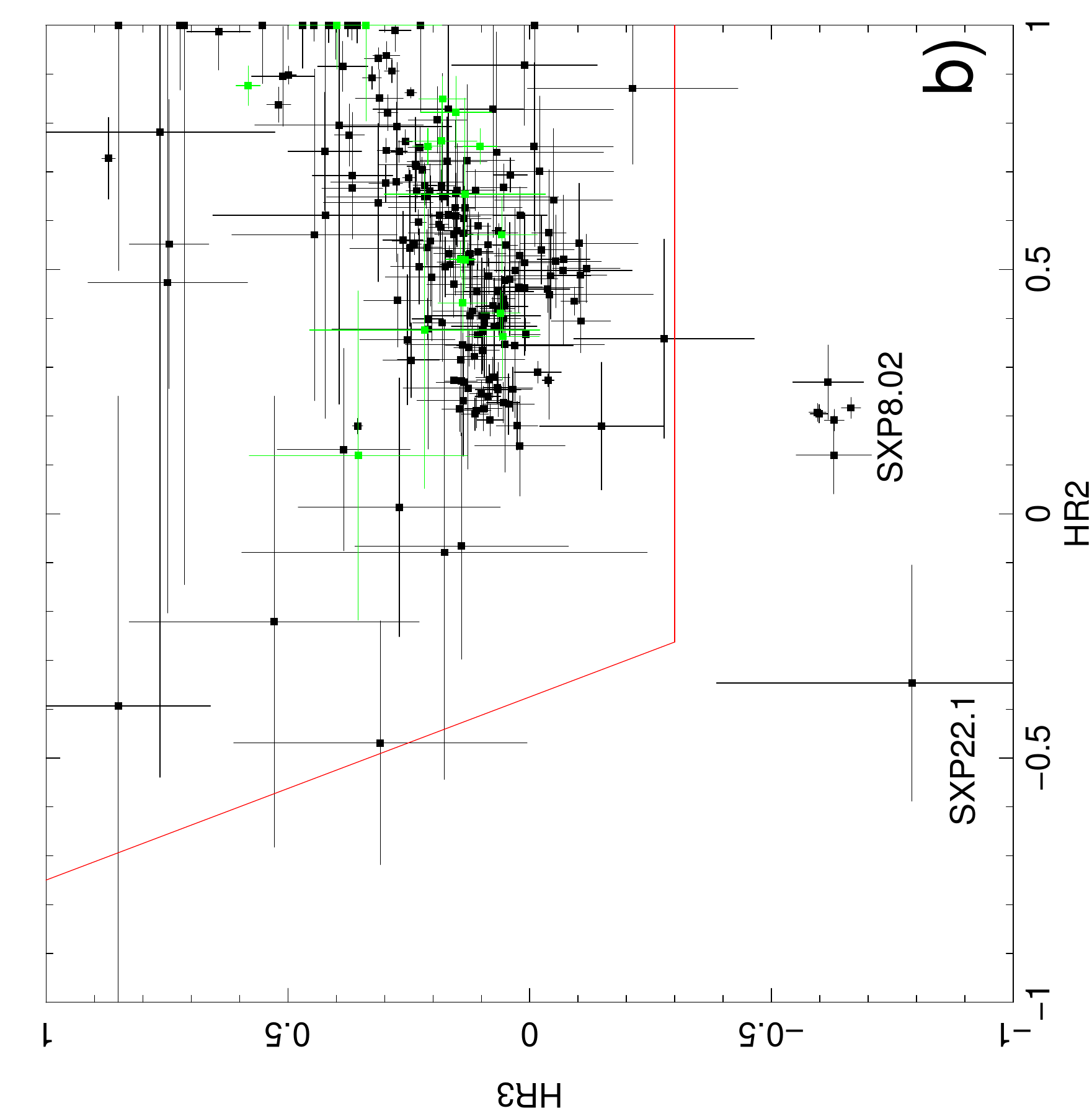}\includegraphics[angle=-90,clip=]{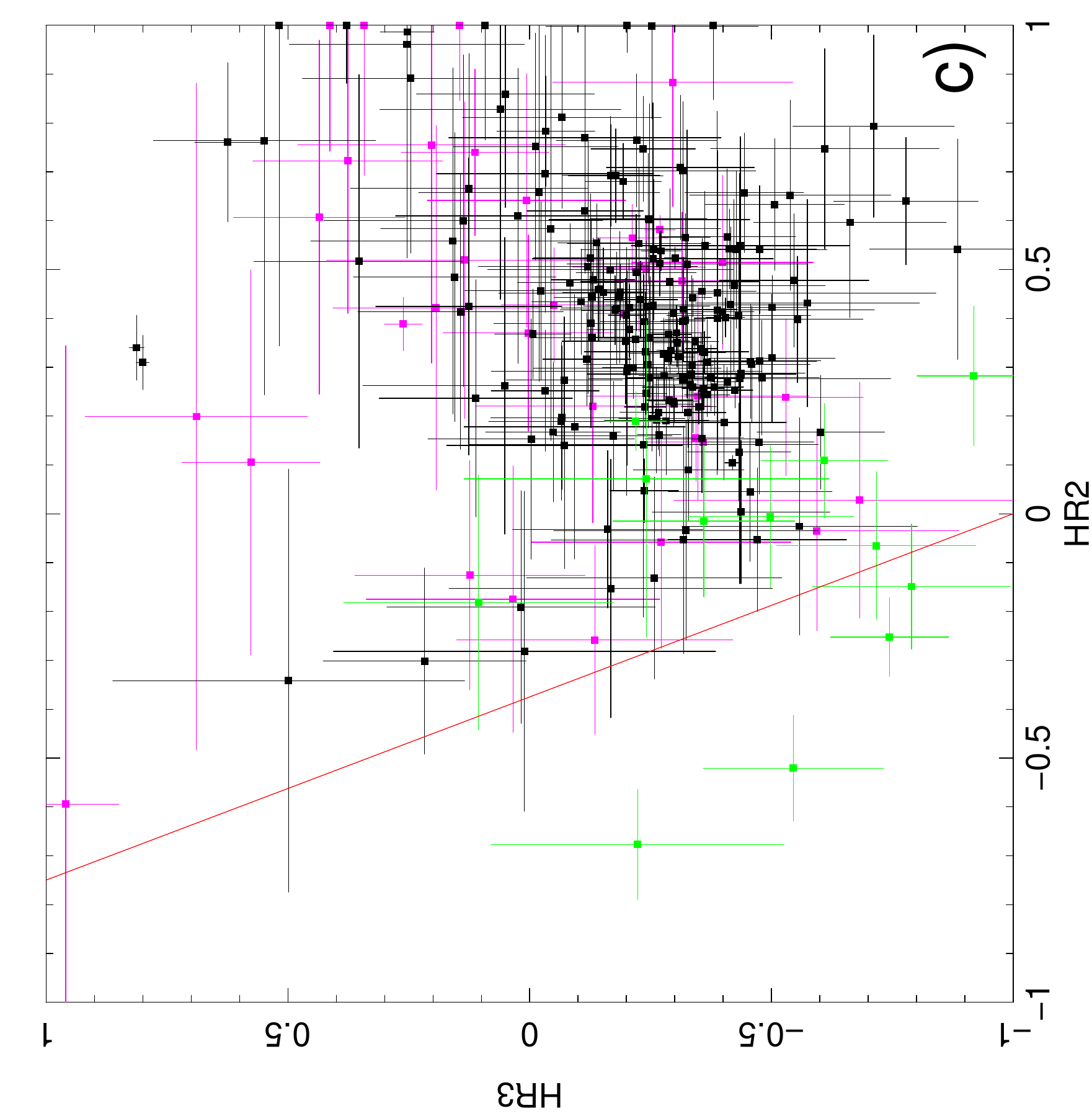}}
  \resizebox{\hsize}{!}{\includegraphics[angle=-90,clip=]{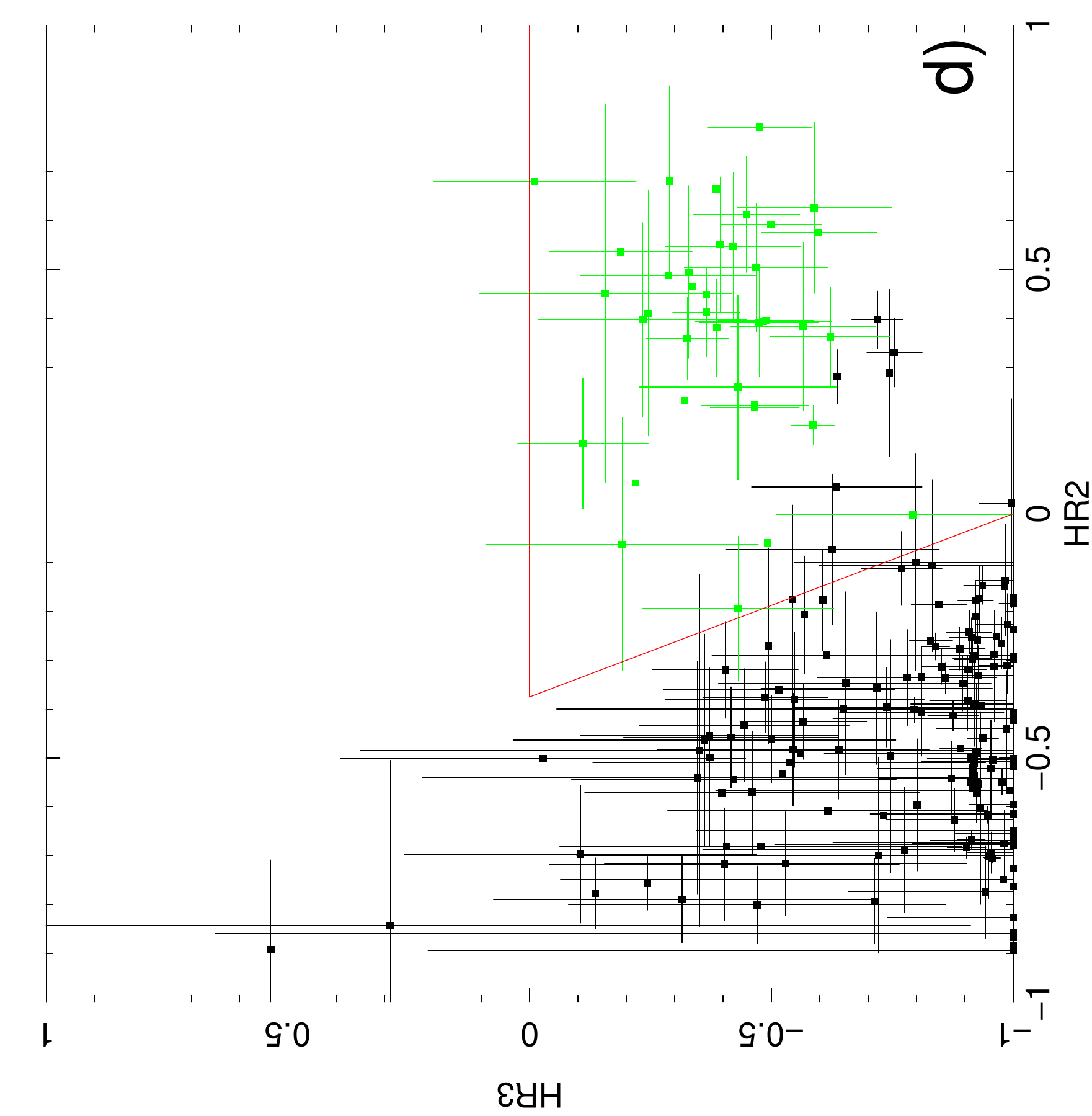}\includegraphics[angle=-90,clip=]{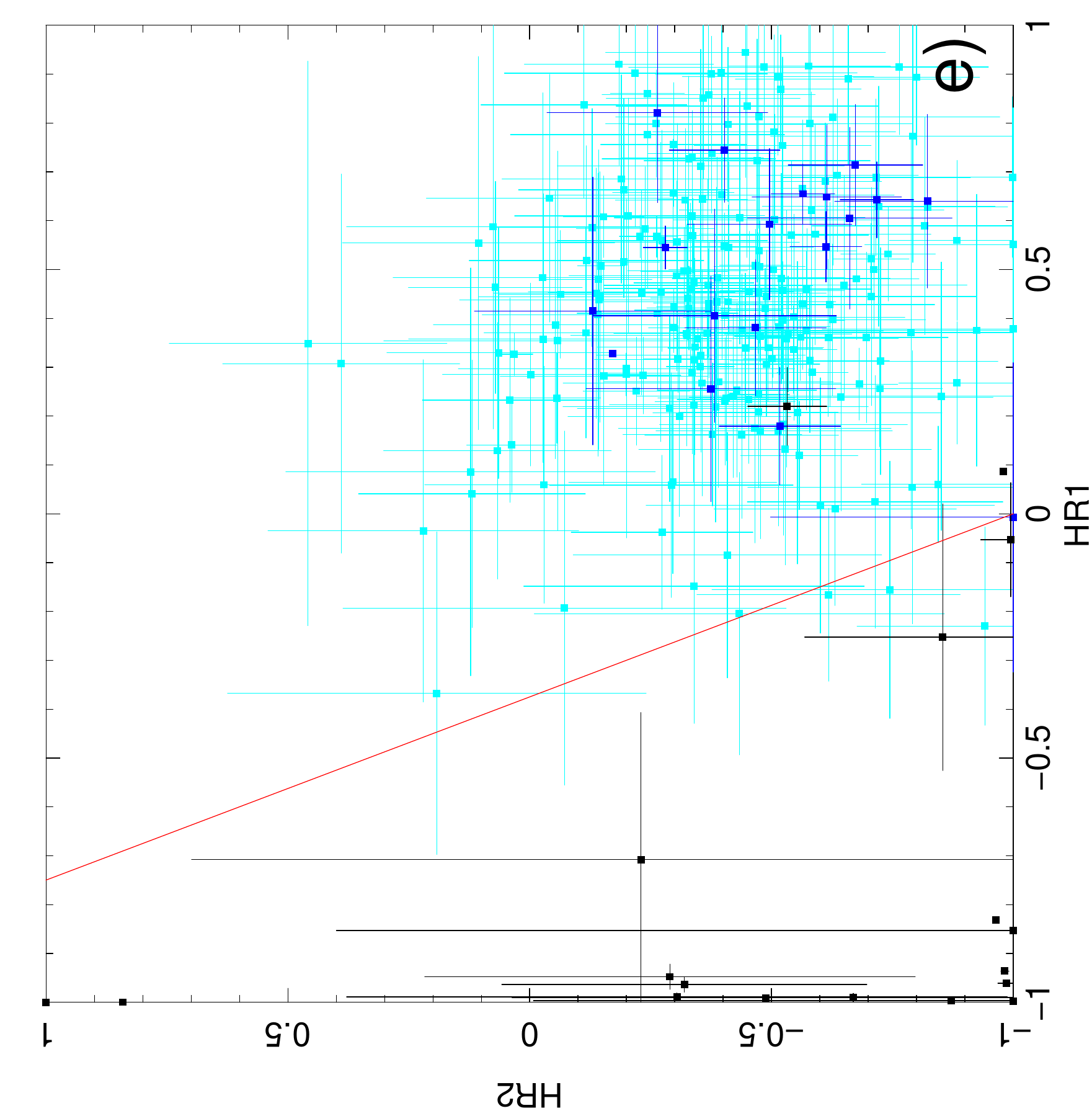}\includegraphics[angle=-90,clip=]{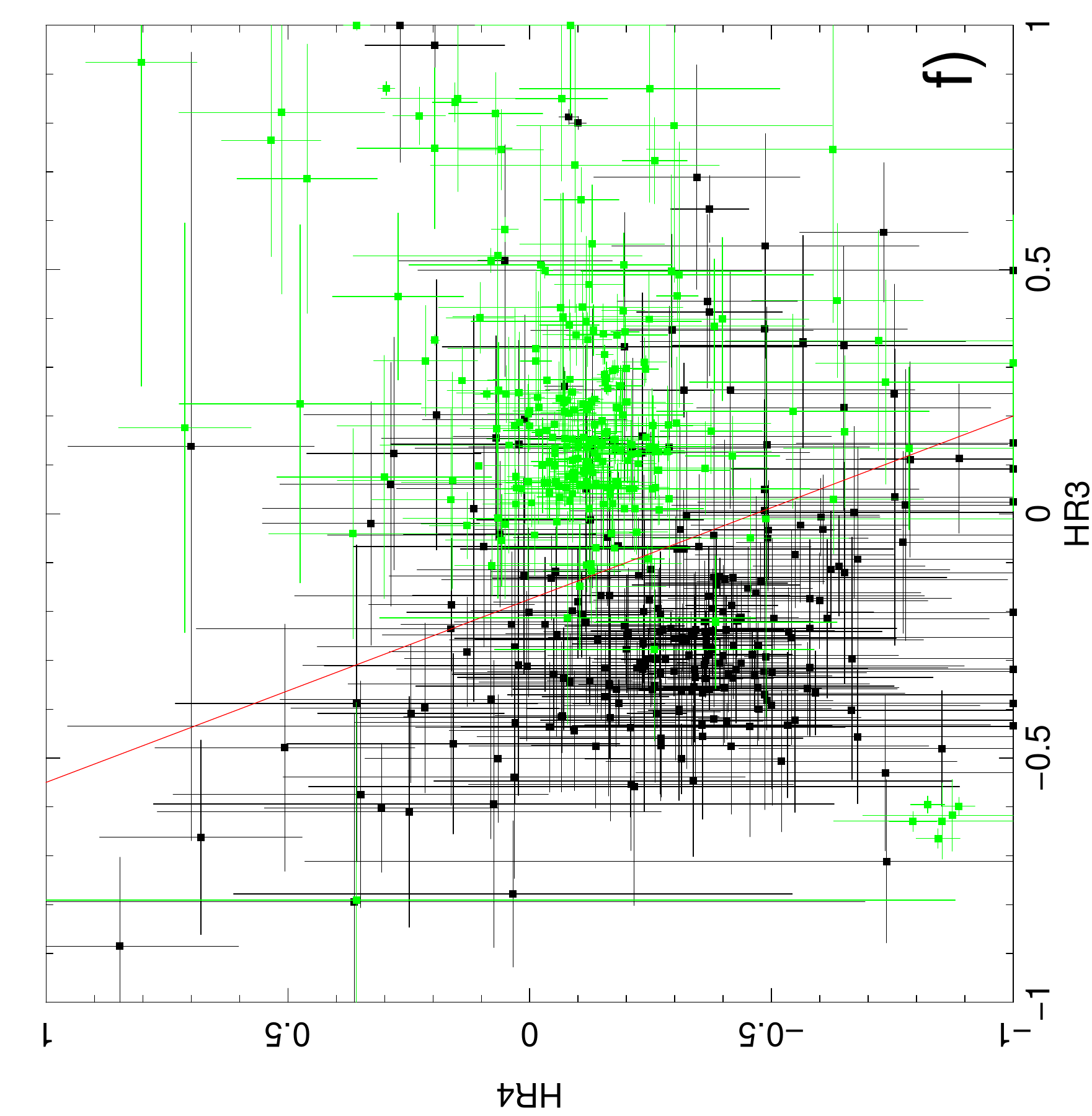}}
  \caption{
   Hardness-ratio diagrams for \xmm\ detections of various source classes. Red lines show HR-selection cuts as used for our source classification.
   $^{(a)}$~Detections with optical counterparts in the Tycho-2 catalogue. Galactic stars are shown in black and magenta, SMC-stars in green. Details are given in Sec. ~\ref{fgstars}.
   $^{(b)}$~Detections of known pulsars (black) and identified HMXBs (green). See Sec. ~\ref{classification:hmxb}.
   $^{(c)}$~Detections of spectroscopically confirmed AGN (black), radio background sources (magenta) and galaxies (green). See Sec.~\ref{sec:class:agn}.
   $^{(d)}$~Detections of sources, screened due to their extent. SNRs are shown in black, other sources (likely galaxy clusters) in green. See Sec.~\ref{sec:class:clg}.
   $^{(e)}$~Comparison of detected SSSs in the SMC and LMC (black) with identified (blue) and classified (cyan) stars in the $HR_1$-$HR_2$-plane. See Sec.~\ref{sec:SSS}.
   $^{(f)}$~Comparison of AGN and radio sources from c) (black) with pulsars and HMXBs from b) (green) in the $HR_3$-$HR_4$-plane. See Sec. ~\ref{classification:hmxb}.
   }
  \label{fig:hr2hr3-select}
\end{figure*}

\begin{table*}
\caption[]{Spectral classification of the X-ray source sample.}
\begin{center}
\begin{tabular}{lrc}
\hline\hline\noalign{\smallskip}
\multicolumn{1}{l}{Spectrum} &
\multicolumn{1}{c}{Classified} &   
\multicolumn{1}{c}{Selection criteria}  \\
\noalign{\smallskip}\hline\noalign{\smallskip}
super soft  & 18   &   ($8HR_1+3HR_2<-3$ or ($HR_1<-0.75$ \&\& $HR_2$ not def.))                                                                                                     \\
            &      &                                                            \&\& $F_1>3\Delta F_1$  \&\& $F_3<3\Delta F_3$ \&\& $F_4<3\Delta F_4$    \&\& $F_5<3\Delta F_5$       \\
soft        & 298  &  $8HR_2+3HR_3<-3$ or ($HR_2<-0.75$ \&\& $HR_3$ not def.)                                                                                                        \\
hard        & 2711 &  ($8HR_2+3HR_3>-3$ or ($HR_2>0 \&\& HR_3$ not def.))  \&\&  not super soft                                                                                      \\
ultra hard  & 945  &   $8HR_3+3HR_4>-1.4$ or ($HR_3>0.2 \&\& HR_4$ not def.)   \&\&  not soft   \&\&  not super soft                                                                 \\
unclass.    & 8    &  --                                                                                                                                                            \\
\noalign{\smallskip}\hline
\end{tabular}
\end{center}
\label{tab:specclass}
\end{table*}

\begin{table*}
\caption[]{Source classification criteria.}
\begin{center}
\begin{tabular}{lcrr}
\hline\hline\noalign{\smallskip}
\multicolumn{1}{l}{Class} &
\multicolumn{1}{c}{Classification criteria} &
\multicolumn{1}{c}{Identified} &
\multicolumn{1}{c}{Classified} \\
\noalign{\smallskip}\hline\noalign{\smallskip}
ClG          &    hard \&\& $HR_3$$<$0     \&\&     Ext$>$$\Delta$Ext  \&\&  $ML_{\rm Ext}$$>$10                                  	                &  12\tablefootmark{a}    & 13      \\
SSS          &    super soft \&\& no opt. loading \&\& $ML_{\rm det}>10$ \&\&  ($N_{\rm det}>1$ or ($ML_{\rm pn}>4$ \&\& $(ML_{\rm m1}>2$ or $ML_{\rm m2}>2)))$   &   4                    &  8        \\
fg-star      &    soft \&\& $\log{(f_X/f_o)} <-1$  \&\& ($B-V>1.2$ or $  (B-V>0.3 \&\& V>17))$                                              	        &  34                    & 128      \\
AGN          &    hard \&\& appropriate radio (r), infrared (i), X-ray (x) or optical (o) counterpart                                       	        &  72                    &  2106    \\
HMXB         &    ultra hard \&\&  $13.5 <V< 17.0$  \&\&   $-0.5<B-V<  0.5$   \&\&    $-1.5<U-B<-0.2$    \&\&   no AGN id        	                &  49                    &  45      \\
\noalign{\smallskip}\hline
\end{tabular}
\end{center}
\tablefoot{16 additional sources were identified with other source classes. 581 sources are unclassified.
\tablefoottext{a}{Not in this catalogue, see Table 3 of \citet[][]{2012A&A...545A.128H}.}
}
\label{tab:class}
\end{table*}

Besides X-ray sources within the SMC, the observed field contains Galactic X-ray sources and background objects behind the SMC.
To distinguish between these, we identified and classified individual sources.
For identification, we searched the literature as described below and selected secure cases only.

To classify the unidentified sources, we developed an empirical approach following \citet{2004A&A...426...11P}.
We derived classification criteria obtained from the parameters of
individual detections of identified sources as seen  with \xmm\ in our processing.
Individual source detections were used, instead of averaged source values, to increase the statistics and account for spectral variability, since $\sim$70\% of our sources were only detected once.
Classifications are marked by angle brackets ($<${\it class}$>$).
We note that classes give likely origins for the X-ray emission, but have to be regarded with care.

First, we distinguished between point sources and sources fitted with small, but significant, extent.
Most of these sources were classified as clusters of galaxies ($<$ClG$>$, see Sec.~\ref{sec:class:clg} and Table~\ref{tab:char:ext}).
Sources with extent too large to be modelled properly by {\tt emldetect} as one single source (e.g. SNRs with substructure),
were flagged beforehand and were not included in the final catalogue.
An overview of these sources can be found in \citet[][]{2012A&A...545A.128H}.

The remaining point sources were classified using X-ray hardness ratios and multi-wavelength properties.
Using the selection criteria given in Table~\ref{tab:specclass}, we divided our sample into super-soft, soft, hard, and ultra-hard sources.
We selected super-soft X-ray sources first, which are classified only if detector noise is an unlikely alternative explanation.
Soft X-ray sources are classified as foreground stars if they have an appropriately bright optical counterpart that is unlikely to be within the SMC on the basis of its brightness and colours.
Also, depending on the counterpart, hard X-ray sources were classified as either AGN or HMXB.
An overview of our classification criteria and results is presented in Table~\ref{tab:class}.
The hardness ratios of identified sources are compared in Fig.~\ref{fig:hr2hr3-select}.
By comparing our classification result with the source classification of \citet{2000A&AS..142...41H} and \citet{2008MNRAS.383..330M}, we found a good agreement.
Details for each source class are given in the following.

\subsection{X-rays from non-degenerate stars}
\label{fgstars}
Shocks in the wind of OB stars, coronal activity from F to M stars, accretion processes in T\,Tau stars
and interaction of close-binary stars can cause X-ray emission from non-degenerate stars \citep[for a review see ][]{2009A&ARv..17..309G}.
Because such stars are weak X-ray sources, most of them in the SMC are below the sensitivity limit of our survey.
Galactic stars are foreground sources, expected to be homogeneously distributed in the \xmm\ SMC field and 
due to their high galactic latitude ($b\sim-44.5$\degr), the sample is expected to be dominated by late-type stars.
Compared to distant Local-Group galaxies, the identification of Galactic stars as X-ray sources in front of the SMC is challenging,
because luminous SMC stars and faint Galactic stars can have a similar brightness, so are hard to differentiate.

\subsubsection{Identification of Galactic stars}
\label{fgstars-ident}

To identify the brightest ($V<11$ mag) foreground stars, we used our correlation with the Tycho-2 catalogue, where we expect one or two chance coincidences.
40 individual X-ray sources with a Tycho-2 counterpart
resulting in 84 XMM-Newton detections
with determined $HR_2$ and $HR_3$ are plotted in Fig. 8 a.
For three further detections of these sources and one additional X-ray source $HR_3$ is undefined and $HR_2\leq-0.9$.

There are 33 Tycho-2 sources with significant ($>$3$\sigma$) proper motions that are all $>$8 mas yr$^{-1}$.
These are obviously foreground stars.
Two more counterparts are stars with a late-type main-sequence classification \citep{2003AJ....125..359W}.
X-ray detections of these 35 confirmed foreground stars are plotted in black in Fig.~\ref{fig:hr2hr3-select}a.
Twenty-five detections of three Tycho-2 sources, correlating with SMC-stars (see Sec.~\ref{sec:class:smcstars}), are plotted in green.
The remaining three matches (\no{140}, 2008, and 2158) were classified as candidates for Galactic stars ($<$fg-star$>$, plotted in magenta).
Source \no{929} shows harder X-ray colours than the remaining foreground stars and is therefore not classified.
The optical and X-ray emission might correlate by chance, but also a foreground cataclysmic variable (CV) is possible.

\subsubsection{Classification of Galactic stars}

To classify an X-ray source as a foreground star candidate ($<$fg-star$>$), we require four criteria:

(i) Using the Tycho-2 set of 35 confirmed foreground stars, we defined a cut (red line in Fig.~\ref{fig:hr2hr3-select}a)
for the X-ray colour selection of fg-star candidates ($<$fg-star$>$),
which separates them from hard X-ray sources, such as AGN and HMXBs (see below and cf. Fig.~\ref{fig:hr2hr3-select}b and c).
For faint soft sources with a low $HR_2$ value, the count rate $R_4$ will not be well determined, leading to an unconstrained $HR_3$.
Our selection allows a less precise determined $HR_3$ for sources with lower $HR_2$.
From similar source samples, a correlation between X-ray plasma temperature and spectral type is not found \citep{2010ApJ...725..480W}.
Therefore, we do not expect a bias in our selection method, although the selection criteria on X-ray hardness ratios are defined using
the Tycho-2 catalogue that contains only the brightest stars in the $B$ and $V$ bands.
We find 258 unidentified soft X-ray sources in our catalogue.

\begin{figure}
 \resizebox{\hsize}{!}{\includegraphics[angle=-90,clip=]{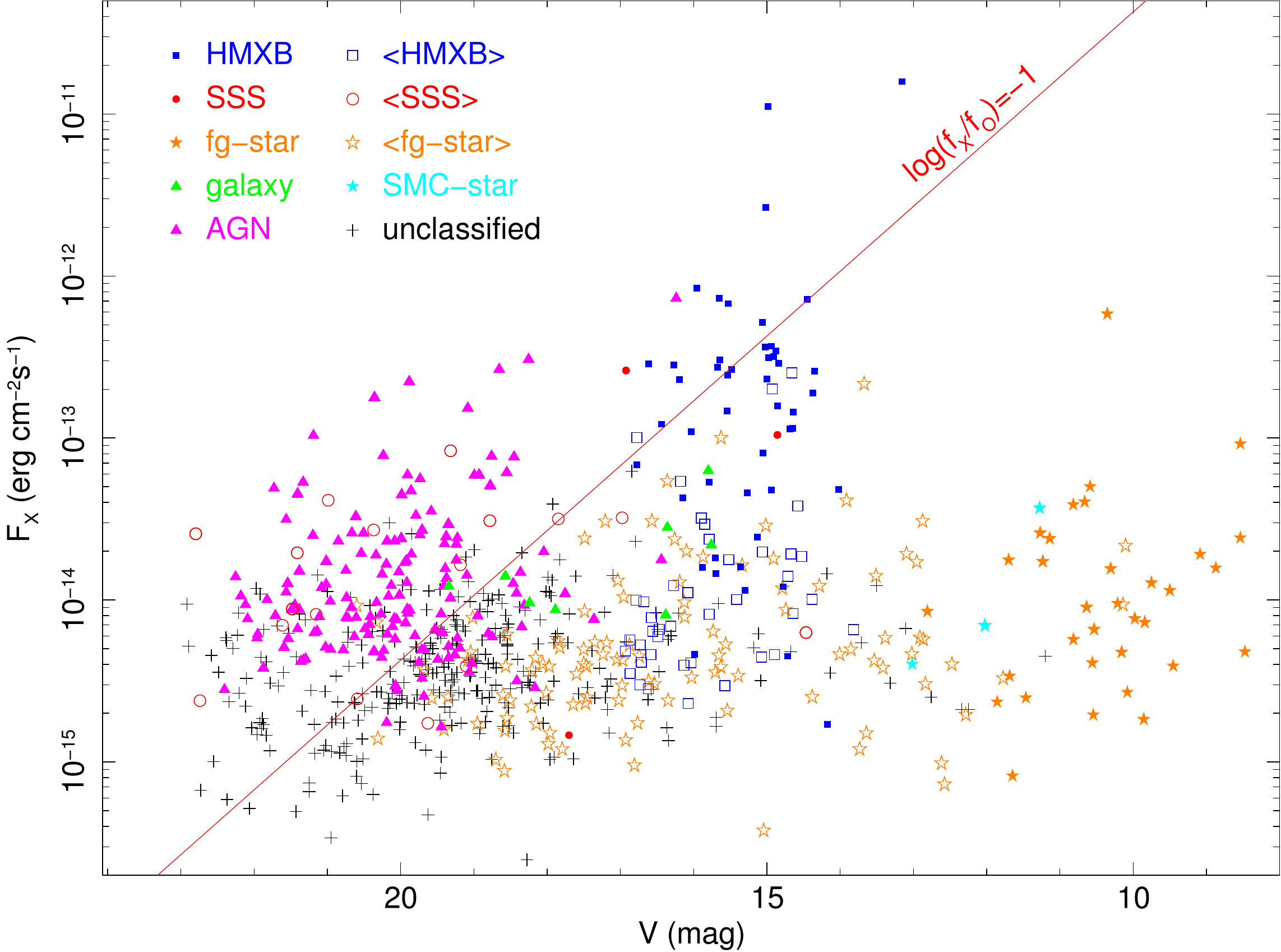}}
 \caption{
          Optical $V$-band magnitude vs. the detected X-ray flux in the (0.2--4.5) keV band for various source classes.
          For foreground stars, the brightest correlated source is plotted. For HMXB we selected the counterparts as in Sec.~\ref{classification:hmxb:search}.
          For all other sources, the nearest optical counterpart is plotted.
         }
 \label{fig:fgstar-fluxratio}
\end{figure}

(ii) For stars with fainter optical magnitudes, it becomes more complicated to discriminate between stars in the Galaxy and the SMC.
In addition to soft X-ray colours, the source must have a sufficiently bright optical counterpart.
Following \citet{1988ApJ...326..680M}, we calculated
$$ \log{(f_X/f_o)} = \log{(F_{(0.2-4.5) {\rm keV}})} + \frac{V}{2.5} + 5.37$$
for the MCPS correlations and
$$ \log{(f_X/f_o)} = \log{(F_{(0.2-4.5) {\rm keV}})} + \frac{R+B}{2 \times 2.5} + 5.37$$
for GSC correlations, where the X-ray flux is in units of erg cm$^{-2}$ s$^{-1}$.
We classified sources as foreground-star candidates ($<$fg-star$>$) only,
if they have an optical counterpart with $\log{(f_X/f_o)}<-1$.
Of the 258 sources, 197 have a sufficiently bright optical counterpart in the MCPS.
The dependence of X-ray flux on optical $V$ magnitude is plotted in Fig.~\ref{fig:fgstar-fluxratio}.
For foreground stars we expect to find an optical counterpart with the given sensitivity of the MCPS.

(iii) Since optical counterparts are still outnumbered by chance correlations with stars of the SMC, we used a colour selection to exclude most of them.
In Fig.~\ref{fig:Beselect}, we show the colour-magnitude and colour-colour diagram of all optical counterpart candidates of X-ray sources
with measured $U$, $V$, and $B$ magnitudes in the MCPS (black points).
To avoid main-sequence and horizontal-branch stars of the SMC, we only selected optical counterparts to the right of the blue dashed line,
which have $V<17$ mag and $B-V>0.3$ mag
or $B-V > 1.2$ mag without any magnitude selection.
This reduces the foreground star sample to 107 sources.

For source \no{548}, we found a Tycho-2 colour of $B-V=-0.38$.
However, this source is identified with the Galactic star Dachs\,SMC\,3-2
and other catalogues give $B-V=0.70$ \citep[e.g.][]{2002ApJS..141...81M}.
The Tycho-2 colour is regarded as an outlier and corrected with the magnitudes of \citet[][]{2002ApJS..141...81M} for Fig.~\ref{fig:Beselect}.

(iv) To avoid possibly erroneous correlations, we did not classify X-ray sources with a positional uncertainty of more than 3\arcsec\ (4 sources).

This allowed us to classify 103 candidates for foreground stars.
To estimate the number of chance coincidences, we shifted the coordinates of one catalogue.
For 26$\pm$5 of the 249 unidentified soft X-ray sources with $ePos<3\arcsec$, we find at least one counterpart candidate compatible with the selection criteria for stars by chance.
When we take into account that some true correlations cause chance correlation when their coordinates are shifted,
we estimate that $\sim$$(16.1\pm3.6)\%$ of the classified foreground stars are chance coincidences.
In addition, using the GSC in cases where the X-ray source did not have a counterpart in the Tycho-2 or MCPS catalogues, we classified 17 X-ray sources as foreground stars.

However, the X-ray emission of stars can become harder during flares \citep[e.g.][]{2004A&A...416..713G},
so that our hardness-ratio selection criteria can be violated.
Similarly, as the X-ray flux increases the $f_X/f_o$ criteria might also be violated.
This causes some overlap with AGN in hardness ratios and $f_X/f_o$.
To investigate this possibility, we searched for sources with short-term variability $Cst<0.5\%$ and $f_X/f_o>0$.
To exclude HMXBs and the bulk of AGNs we also required $HR_2<0.2$ and $HR_3<-0.3$.
We selected five additional sources (\no{146}, 1998, 2041, 2740, and 3059) as candidates ($<$fg-star$>$).

For 61 stars, which have a 2MASS counterpart in the MCPS catalogue,
we derived a spectral classification from the $J-K$ colour \citep[see ][]{2009A&A...507..705B}.
The resulting distribution of the spectral types is shown in Fig.\ref{fig:fgstarclass}
and follows the expected distribution, peaking around early M stars.

\subsubsection{Stars within the SMC}
\label{sec:class:smcstars}
In some extreme cases, X-ray emission from early-type stars within the SMC can be observed with \xmm,
e.g. from stellar-wind interaction of a Wolf-Rayet star in a binary system with an OB star.
\citet{2008ApJS..177..216G} found X-ray emission from SMC-WR5, SMC-WR6, and SMC-WR7 \citep{2003PASP..115.1265M},
which are the sources \no{150}, 237, and 1212 in our catalogue (cyan stars in Fig.~\ref{fig:fgstar-fluxratio} and Fig.~\ref{fig:Beselect}, left).
\no{1212} is visible in Fig.~\ref{fig:deepxmmcxo}.

For completeness, we note that source \no{145} is close to SMC-WR3, but due to a separation
of 3.29\arcsec\ (2.73$\sigma$), this correlation is doubtful.
Also, sources \no{294}, 1031, and 2963 formally correlate with the SMC stars
AzV\,369 (4.6\arcsec, 3.1$\sigma$),
AzV\,222 (4.1\arcsec, 2.8$\sigma$), and
2dFS\,3274 (1.8\arcsec, 1.0$\sigma$).

In the centre of the star cluster NGC\,346 we see an unresolved convolution
of X-ray bright stars \citep[see ][]{2002ApJ...580..225N,2004ApJ...608..208N},
which is source \no{535} in our catalogue.
Source \no{2706} has similar X-ray colours and correlates with the star cluster Lindsay\,66.
Also \no{294} can be associated with the star cluster Bruck\,125.
From our correlation  with the star cluster catalogue of \citet{2008MNRAS.389..678B},
we expect around 13$\pm$7 X-ray sources to be correlated with star clusters.
About half of these sources can be explained by HMXBs, which might have formed in these clusters \citep{2005MNRAS.358.1379C}.

\subsection{Super-soft X-ray sources}
\label{sec:SSS}

SSSs are a phenomenological class of X-ray sources, defined by a very soft thermal X-ray spectrum and with no emission above 1 keV.
Luminous SSSs are associated with CVs, planetary nebulae, symbiotic stars, and post-outburst optical novae.
The general scenario is steady thermonuclear burning on the surface of an accreting white dwarf \citep{2007ApJ...663.1269N}.
Less luminous SSSs can be observed in some CVs, cooling neutron stars and PG 1159 stars. For a review, see \citet{2006AdSpR..38.2836K}.

\begin{table*}
\caption[]{Faint SSS candidates in the SMC.}
\begin{center}
\begin{tabular}{lccccccccccc}
\hline\hline\noalign{\smallskip}
\multicolumn{1}{l}{No.} &
\multicolumn{1}{c}{RA\tablefootmark{a}} &
\multicolumn{1}{c}{Dec\tablefootmark{a}} &
\multicolumn{1}{c}{ePos\tablefootmark{b}} &
\multicolumn{1}{c}{$HR_1$} &
\multicolumn{1}{c}{$HR_2$}  &
\multicolumn{1}{c}{$F$\tablefootmark{c}}  &
\multicolumn{1}{c}{$ML_{\rm det}$\tablefootmark{d}}  &
\multicolumn{1}{c}{$ML_{\rm pn}$\tablefootmark{d}}  &
\multicolumn{1}{c}{$ML_{\rm m1}$\tablefootmark{d}}  &
\multicolumn{1}{c}{$ML_{\rm m2}$\tablefootmark{d}}  &
\multicolumn{1}{c}{Comment} \\
\noalign{\smallskip}\hline\noalign{\smallskip}
235  &   01 01 47.58   &  -71 55 50.7  &  0.85  &  -0.3$\pm$0.1  &  -0.8$\pm$0.1  &  6.0$\pm$0.6  &  146.5  &  54.7  &  33.2  &  4.2 & WD/Be?   \\      
1198  &  01 01 24.19   &  -72 00 37.9  &  1.82  &  -0.4$\pm$0.2  &  -1.0$\pm$0.5  &  2.5$\pm$0.7  &  16.2  &  9.4  &  6.2  &  2.8 &    \\
1531  &  00 39 45.58   &  -72 47 01.4  &  1.58  &  -1.0$\pm$0.1  &  0.8$\pm$0.4  &  1.8$\pm$0.4  &  28.6  &  10.7  &  7.8  &  12.2 &star?    \\       
1549  &  00 38 58.74   &  -72 55 10.4  &  1.68  &  -1.0$\pm$0.2  &  --            &  3.6$\pm$0.9  &  18.3  &  13.6  &  4.5  &  2.9 & star?   \\
2132  &  00 57 45.29   &  -71 45 59.7  &  0.93  &  -0.4$\pm$0.1  &  -0.5$\pm$0.2  &  4.1$\pm$0.5  &  121.1  &  85.4  &  20.4  &  18.8 &    \\
2178  &  00 55 37.71   &  -72 03 14.0  &  0.74  &  -0.9$\pm$0.0  &  0.2$\pm$0.4  &  6.7$\pm$0.5  &  391.5  &  282.6  &  54.5  &  58.7 &   \\
2218  &  00 55 08.45   &  -71 58 26.7  &  1.38  &  -0.2$\pm$0.2  &  -0.9$\pm$0.4  &  1.3$\pm$0.3  &  14.0  &  13.0  &  3.2  &  0.9 &  WD/Be?, star?  \\
3235  &  00 55 03.65   &  -73 38 04.1  &  0.64  &  -0.9$\pm$0.0  &  0.2$\pm$0.2  &  13.6$\pm$0.8  &  787.0  &  649.0  &  48.1  &  96.2 &   \\
\noalign{\smallskip}\hline
\end{tabular}
\end{center}
\tablefoot{
\tablefoottext{a}{Sexagesimal coordinates in J2000.}
\tablefoottext{b}{Positional uncertainty in arcsec.}
\tablefoottext{c}{Detected flux in the (0.2--1.0) keV band in $10^{-15}$ erg cm$^{-2}$ s$^{-1}$.}
\tablefoottext{d}{Source detection likelihood for combined and the individual instruments.}
}
\label{tab:SSS}
\end{table*}

\subsubsection{Identification of super-soft X-ray sources}
\label{sec:SSSindent}
Two bright SSSs in the SMC,
the planetary nebula SMP SMC 22 (\no{686}) and the symbiotic nova SMC3 (\no{616}),
were observed during our survey \citep{2010A&A...519A..42M,2011A&A...529A.152S}.
In addition, \citet{2010A&A...519A..42M} confirmed SMP SMC 25 as a faint SSS in the survey data (\no{1858}),
that was discovered with ROSAT by \citet{1999A&AS..136...81K}.
Other SSSs known from ROSAT (RX\,J0059.1-7505, RX\,J0059.4-7118, RX\,J0050.5-7455),
were previously observed with \xmm\ \citep{2006A&A...452..431K}.
The first source is the symbiotic star LIN 358 (\no{1263}),
the second was suggested to be a close binary or isolated neutron star (\no{324}),
for the third source \citet{2006A&A...452..431K} give an upper limit.
In our survey analysis, this latter source is detected (\no{1384}),
but is very probably associated with the Galactic star TYC 9141-7087-1 and affected by optical loading.
Other ROSAT sources from \citet{1996A&A...312..919K} are the transient SSS RX\,J0058.6-7146 and the candidate SSS RX\,J0103.8-7254.
For neither source can we find a detection in our catalogue.
The position of the variable SSS 1E0035.4-7230 is not covered by any \xmm\ observation yet.
Source \no{235} was found as a new faint SSS candidate (see Sec.~\ref{sec:SSScandidates}) and is proposed to be a binary system consisting of a white dwarf and a Be star \citep{2012A&A...537A..76S}.
The position of the super-soft transient MAXI\,J0158-744  \citep{2012ApJ...761...99L} was not covered with \xmm.
New luminous SSS transients were not found during the \xmm\ SMC survey.

\subsubsection{Search for faint SSS candidates}
\label{sec:SSScandidates}
The \xmm\ survey enables a search for faint SSSs.
Analogously to our division into soft and hard X-ray sources in Sec.~\ref{fgstars},
we separate super-soft from soft X-ray sources in the $HR_1$-$HR_2$-plane, as shown in Fig.~\ref{fig:hr2hr3-select}e.
Detections of identified SSSs from Sec.~\ref{sec:SSSindent}, are plotted in black.
To increase the reference sample, we also used detections of identified SSSs in the LMC \citep[see][and references therein]{2008A&A...482..237K},
from an identical data processing method as used for the SMC data.
In general, $HR_1$ is negative for SSSs and depends strongly on photo-electric absorption.
$HR_2$ is expected to be close to -1, but due to low count rates in the energy bands 2 and 3,
$HR_2$ is only poorly determined for weak SSSs.
We also demand no significant ($<$3$\sigma$) emission in the energy bands 3--5, but significant emission in the energy band 1,
to designate the spectrum as super soft.
The two LMC SSSs outside our selection area are CAL\,87 and RX\,J0507.1-6743, which are both affected by high absorption \citep[][]{2008A&A...482..237K}
causing a $HR_1$ of $0.087\pm0.003$ and $0.22\pm0.08$, respectively.
Identified Tycho-2 stars (Sec.~\ref{fgstars}), which are not affected by optical loading, are plotted in blue.
In cyan, we show all sources, which fulfil our selection criteria for candidate foreground stars, have a detection likelihood of $ML_{\rm det}>10$ and are not affected by optical loading.
Three of these sources fulfil the selection criteria of both SSS and stars.
Here a X-ray spectral analysis is necessary to discriminate between them.

Unfortunately, optical loading and detector noise cause spurious detections with characteristics similar to SSS.
EPIC-MOS is less sensitive below 500 eV by a factor of 6 compared to EPIC-pn.
Therefore, we demanded a conservative total detection likelihood of $ML_{\rm det}>$10
and rejected candidates affected by optical loading in EPIC-pn.
Further, we required that the source has at least a slight detection in another instrument or observation.
The selection procedure yielded a total of 8 candidate faint SSSs,
which are listed in Table~\ref{tab:SSS}.
Source \no{2218} has a optical counterpart candidate with a separation of 4.6\arcsec ($3.2\sigma$) and typical colours
for B stars in the SMC (see Sec.\ref{classification:hmxb:search}).

\subsection{High-mass X-ray binaries}
\label{classification:hmxb}

The SMC hosts a remarkably large population of HMXBs \citep[e.g. ][]{2010ASPC..422..224C},
which is probably caused by a high recent star-formation rate \citep{2010ApJ...716L.140A} and low metallicity \citep{2006MNRAS.370.2079D}.
With the exception of SMC X-1 (super-giant system, source \no{1}) and SXP\,8.02 \citep[anomalous X-ray pulsar, source \no{48},][]{2008ApJ...680L.133T},
all known X-ray pulsars in the SMC are presumably Be/X-ray binaries.
Here, matter is ejected in the equatorial plane of a fast rotating Be star, resulting in the build up of a decretion disc.
These systems can have a persistent or transient X-ray behaviour.
Outbursts occur when the neutron star in the system accretes matter during periastron passage (Type I) or due to decretion-disc instabilities (Type II).
For a recent review, see \citet{2011Ap&SS.332....1R}.

From the X-ray sources correlating with bright SMC stars (Sec.~\ref{sec:class:smcstars})
only \no{1031} might also be explainable by a supergiant HMXB from optical and X-ray colours.
A search for supergiant systems resulted in no further candidate.

\begin{figure*}
  \resizebox{\hsize}{!}{\includegraphics[angle=-90,clip=]{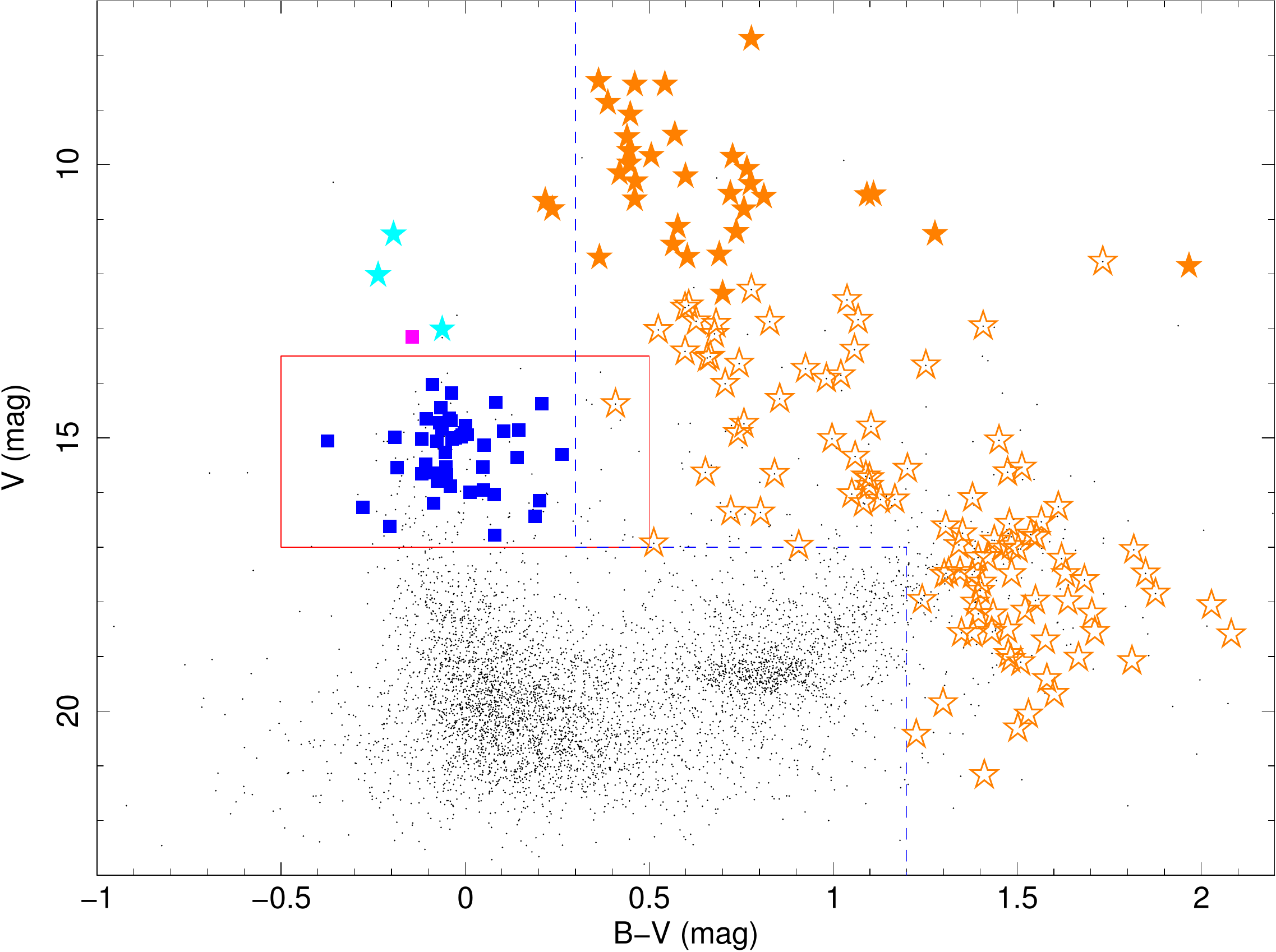}\includegraphics[angle=-90,clip=]{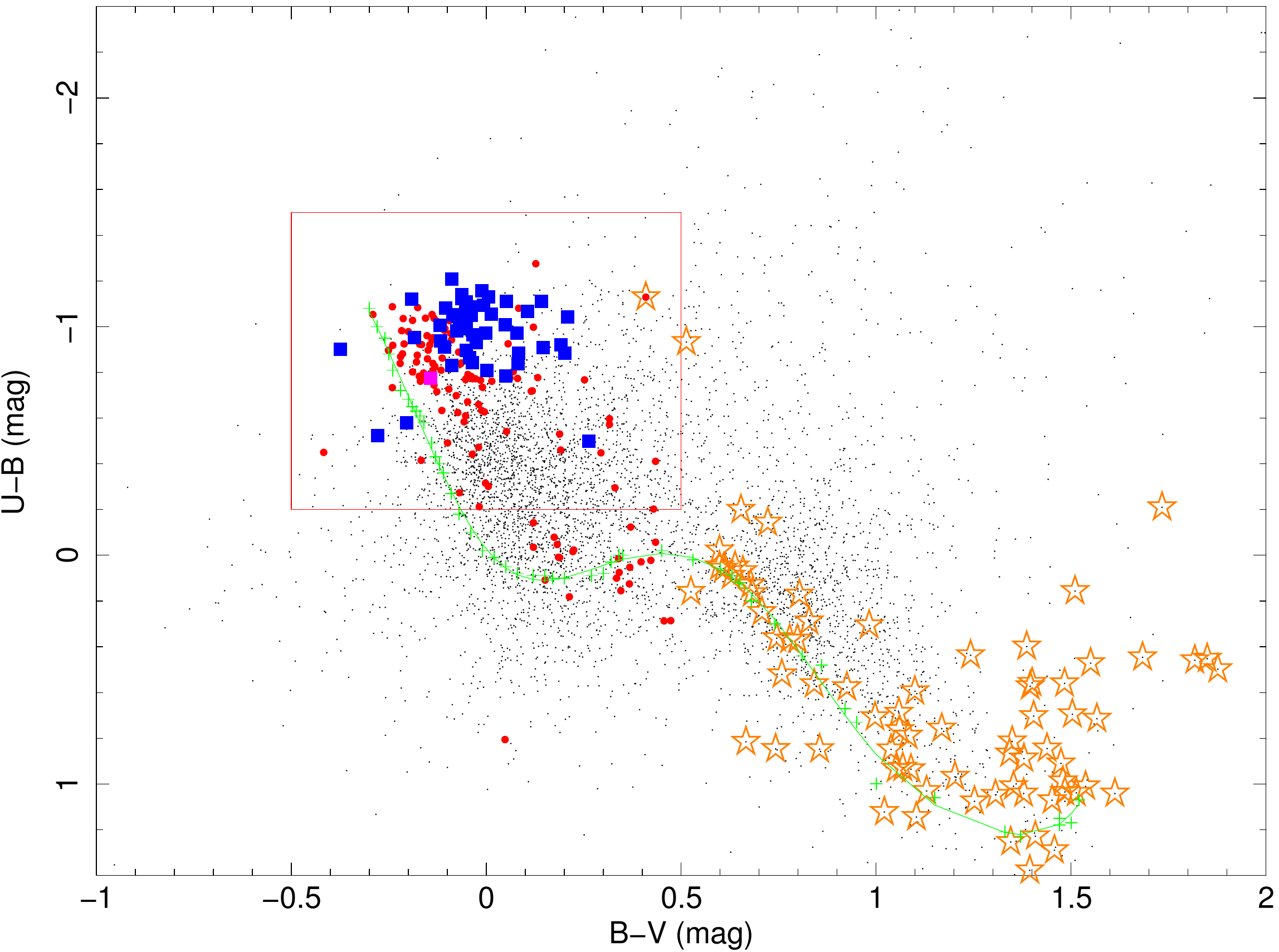}}
  \caption{Colour-magnitude ({\it left}) and colour-colour ({\it right}) diagram.
    Black points show all possible optical counterparts of X-ray sources with measured $U$, $V$, and $B$ magnitudes in the MCPS found inside the $3\sigma$ positional uncertainty.
    Counterparts for Galactic foreground star candidates were selected redwards of the blue dashed line only.
    The red boxes mark the selection region of counterparts for BeXRBs in both plots.
    Identified foreground stars (filled orange stars), classified foreground stars (open orange stars), identified BeXRBs (blue squares) are marked.
    WR stars in the SMC and SMC X-1 are shown by cyan stars and the magenta square.
    Red dots in the right diagram mark all sources (black dots) within the red box of the left diagram.
    The green line gives the colours for the unreddened main sequence according to \citet{1970A&A.....4..234F}.
  }
  \label{fig:Beselect}
\end{figure*}

 \begin{figure}
  \resizebox{\hsize}{!}{\includegraphics[angle=0,clip=]{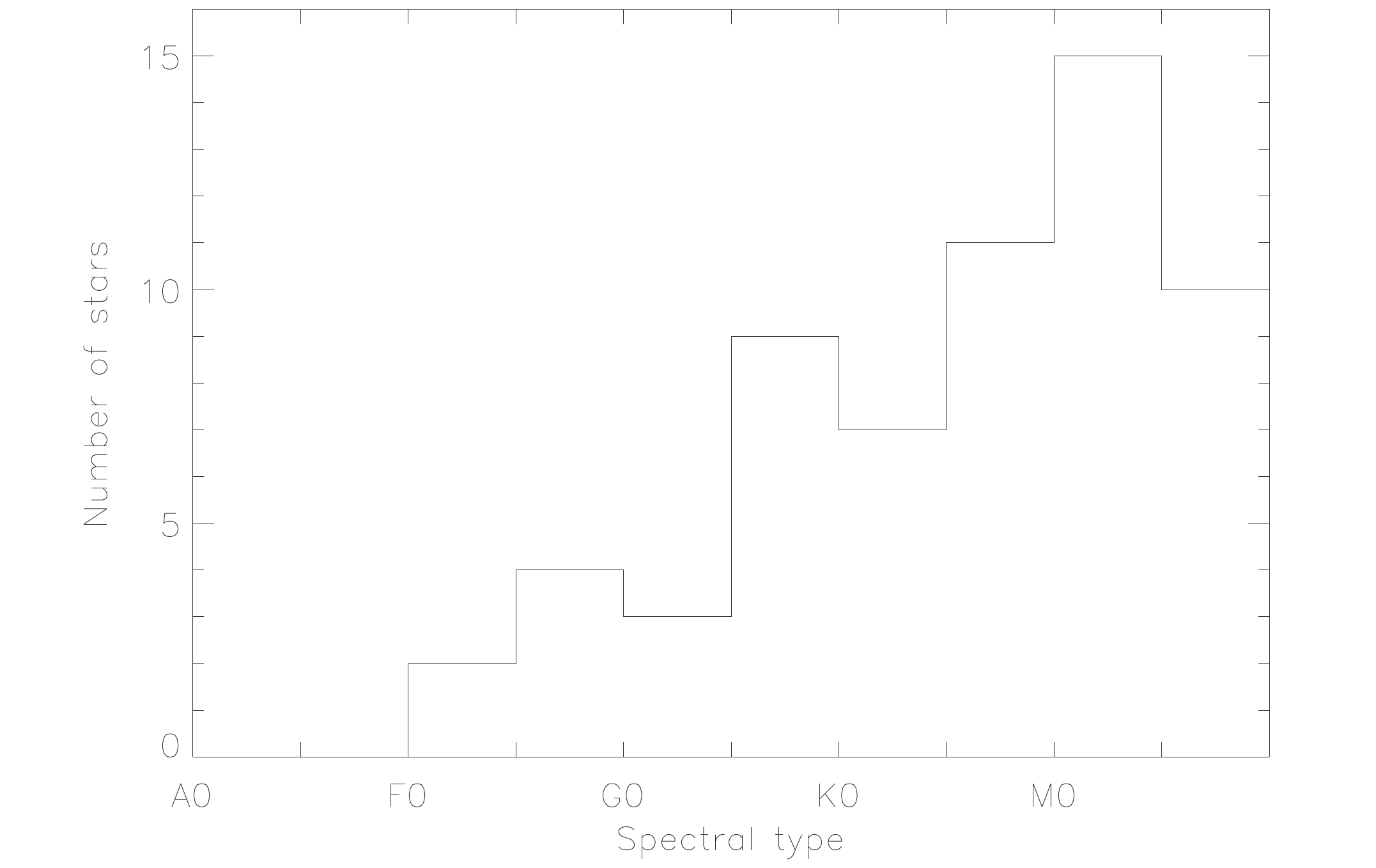}}
   \caption{
     Distribution of spectral classes of X-ray emitting stars.
   }
   \label{fig:fgstarclass}
 \end{figure}

\subsubsection{Identification of Be/X-ray binaries}
We identified 49 HMXBs listed in literature \citep[e.g.][]{2000A&A...359..573H,2008ApJS..177..189G}.
During our survey, two new X-ray pulsars were discovered \citep{2011MNRAS.414.3281C,2011A&A...527A.131S},
as well as two further bright BeXRB transients \citep[sources \no{2732} and 3115, ][]{2012MNRAS.424..282C}.
Although detected with \xmm, the catalogue does not contain the source SXP\,11.5,
since it was not observed in the nominal field of view and therefore was not accessible to our processing \citep{2011MNRAS.410.1813T}.
The same holds for SXP\,1062, which was recently discovered in the outer wing of the SMC \citep{2012MNRAS.420L..13H,2012A&A...537L...1H} after our data processing.
All other pulsars with known position are within our main field.
A detailed analysis of the observed BeXRB population will be discussed in a forthcoming study.
Our catalogue contains 200 detections of 42 X-ray pulsars.
X-ray pulsations confirm the neutron star nature of the accreting object.
Hardness ratios for all X-ray pulsars are shown in Fig.~\ref{fig:hr2hr3-select}b in black.
All other 17 detections of 8 HMXBs with unknown pulse period are plotted in green.

\subsubsection{Search for Be/X-ray binary candidates}
\label{classification:hmxb:search}
Sources are classified as HMXB candidates ($<$HMXB$>$),
if they fulfil the following criteria:

(i) Because of the power-law-like X-ray spectrum, with a typical photon index of $\Gamma \sim 1$ \citep{2004A&A...414..667H},
HMXBs can easily be discriminated from soft X-ray sources, by using the same dividing line as in Fig.~\ref{fig:hr2hr3-select}a.
In general, HMXB show a harder X-ray spectrum than AGN ($\Gamma \sim1.7$), thus providing a lower limit for $HR_3$ at $-0.3$ (see Fig.~\ref{fig:hr2hr3-select}b).
There is a notable exception, SXP\,8.02, where all detections of this pulsar lie outside the selection region of Fig~\ref{fig:hr2hr3-select}b.
This can be explained on the basis of the anomalous X-ray pulsar (AXP) nature of this object \citep{2008ApJ...680L.133T}.
The colours of SXP\,22.1 (\no{935}) have large uncertainties.
In total 1536 \xmm\ sources that are not identified as HMXB or AGN (see Sec.~\ref{sec:class:agn})
are hard (see Table~\ref{tab:specclass}) X-ray sources with $HR_3>-0.3$.

(ii) To avoid chance correlations with sources having a high positional uncertainty,
we only used X-ray sources with a positional uncertainty $<$2.5\arcsec. This excludes 33 X-ray sources.

(iii) In addition to the selection of X-ray colours, we searched for an early-type star as counterpart.
We used the loci of the confirmed BeXRBs (shown with blue squares)
on the colour-magnitude and colour-colour diagram of Fig.~\ref{fig:Beselect},
to define the selection area for candidate BeXRB systems.
The loci are indicated with red boxes and correspond to 13.5 mag $<V<$17 mag and colours of $-0.5$ mag $<B-V<$0.5 mag  and $-1.5$ mag$<U-B<$$-0.2$ mag.
The MCPS catalogue comprises 16\,605 entries, which fulfil these criteria and are in the \xmm\ field.

(iv) To further improve the discrimination between AGN and HMXB, we use a third dividing line in the $HR_3$-$HR_4$-plane (Fig.~\ref{fig:hr2hr3-select}f),
where the difference in average power-law photon index has most effect.
We note, that the separation between BeXRB and AGN population is not clear-cut.
Highly obscured AGN are shifted towards larger $HR_3$,
and there are also some detections of HMXB on the left side of the cut.
We find 34 sources fulfilling the criteria i$-$iv above.
By using subsamples of \xmm\ and MCPS sources fulfilling the criteria i$-$iv
and shifting the coordinates of one catalogue as described in Sec.\ref{sec:correlations:chance},
 we estimate $\sim$16.6$\pm$3.4 chance coincidences.

(v) Two weak candidates (\no{154} and 1408) correlate with emission-line objects, confirming the possible HMXB nature of these sources.

Five sources that fulfil the criteria i$-$iii, but violate criteria  iv and v, are considered as weaker candidates, and are marked with ``{\tt ?}''.
In addition, we found one source, \no{66}, in the young star cluster NGC\,330, but due
to the high stellar density, no optical counterpart could be identified in the MCPS at this position.
However, the source correlates with the Be star NGC\,330:KWBBe 224 \citep{1999A&AS..134..489K}.
The hardness ratios and short-term variability further support the HMXB nature of this source.
Source \no{1605} is in the star cluster NGC\,376 and was rejected because of a $B-V$=0.71 in the MCPS.
However, this colour might be influenced by confusion with other stars in the cluster.
We find $B-V = -0.15$ in the OGLE catalogue \citep{1998AcA....48..147U} and a classification of B2e by \citet{2010A&A...509A..11M}.
Source \no{823} was rejected because of a $B-V$ colour of 2.4 mag in the MCPS, although this source was classified as B1-5\,III\,e by \citet{2004MNRAS.353..601E} and has $B-V = -0.13$ in the OGLE catalogue.
Source \no{3003} is outside the MCPS field. Its X-ray properties and optical colours from \citet{2002ApJS..141...81M} are also consistent with a HMXB \citep{2013A&A...551A..96S}.
\citet{2004MNRAS.353..601E} classified the optical counterpart as B1-3\,III.
Therefore, we also add these four sources to the catalogue of HMXB candidates.

All 45 candidate HMXBs ($<$HMXB$>$) are listed in Table~\ref{tab:HMXB}.
This list includes also the weak candidates as they are useful to set an upper limit to the BeXRB luminosity function.

\begin{table*}
\caption[]{HMXB candidates in the SMC.}
\begin{center}
\begin{tabular}{rcccrrrcrrccc}
\hline\hline\noalign{\smallskip}
\multicolumn{6}{c|}{X-ray} &
\multicolumn{4}{c|}{MCPS} &
\multicolumn{2}{c|}{MA93}  &
\multicolumn{1}{c}{Comments}\\
\multicolumn{1}{l}{No} &
\multicolumn{1}{c}{RA} &
\multicolumn{1}{c}{Dec} &
\multicolumn{1}{c}{ePos} &
\multicolumn{1}{c}{$V$}  &
\multicolumn{1}{c|}{$Cst$}  &
\multicolumn{1}{c}{$d$}  &
\multicolumn{1}{c}{$V$}  &
\multicolumn{1}{c}{$B-V$}  &
\multicolumn{1}{c|}{$U-B$}  &
\multicolumn{1}{c}{$d$}  &
\multicolumn{1}{c|}{No}  &
\multicolumn{1}{c}{and} \\
\multicolumn{1}{l}{} &
\multicolumn{1}{c}{(J2000)} &
\multicolumn{1}{c}{(J2000)} &
\multicolumn{1}{c}{(\arcsec)} &
\multicolumn{1}{c}{} &
\multicolumn{1}{c|}{(\%)}  &
\multicolumn{1}{c}{(\arcsec)}  &
\multicolumn{1}{c}{(mag)}  &
\multicolumn{1}{c}{(mag)}  &
\multicolumn{1}{c|}{(mag)}  &
\multicolumn{1}{c}{(\arcsec)}  &
\multicolumn{1}{c|}{}  &
\multicolumn{1}{c}{references} \\
\noalign{\smallskip}\hline\noalign{\smallskip}

12    &  01 19 38.94   &  -73 30 11.4  &  0.7  & 1.8 &  34.4  &  0.3  &  15.8  &  -0.1  &  -0.8  &  0.9  &  1867 &  H00, SG05   \\
65    &  00 57 23.66   &  -72 23 55.8  &  0.8  & 6.4 &  13.0  &  1.7  &  14.7  &  -0.1  &  -1.0  &  --  &  --  &  ?, SG05, A09\\
66    &  00 56 18.85   &  -72 28 02.7  &  0.7  & 1.3 &  0.2  &  --  &  --  &  --  &  --  &  --  &  --  &  in NGC\,330, SG05 \\
94    &  00 55 07.72   &  -72 22 40.3  &  0.9  & 1.9 &  20.7  &  0.8  &  14.4  &  -0.1  &  -1.0  &  --  &  -- & L10 \\
117   &  00 48 18.73   &  -73 20 59.9  &  0.6  & 2.3 &  4.4  &  0.2  &  16.2  &  0.3  &  -0.8  &  --  &  -- &  SG05, A09, K09 \\
133   &  00 50 48.06   &  -73 18 17.6  &  0.9  & 3.4 &  8.3  &  0.3  &  15.1  &  0.1  &  -1.0  &  2.8  &  396  &  SG05, A09 \\
137   &  00 52 15.06   &  -73 19 16.3  &  0.6  & 2.2 &  0.0  &  2.2  &  15.9  &  -0.1  &  -1.0  &  5.7  &  552\tablefootmark{c} & L10  \\
154   &  01 00 30.26   &  -72 20 33.1  &  1.0  & 4.4 &  12.4  &  0.7  &  14.6  &  -0.1  &  -1.0  &  0.3  &  1208  & SPH03, SG05\\
160   &  01 00 37.31   &  -72 13 17.4  &  0.9  & 2.7 &  73.5  &  2.4  &  16.7  &  -0.2  &  -0.9  &  --  &  -- &   N03,SG05  \\
247   &  01 02 47.51   &  -72 04 50.9  &  0.8  & 5.1 &  14.5  &  0.5  &  16.0  &  -0.3  &  -1.1  &  --  &  --  &  SXP\,523 (W12,S13a)   \\
259   &  01 03 28.54   &  -72 06 51.4  &  0.7  & 6.3 &  4.4  &  1.9  &  16.5  &  -0.2  &  -0.9  &  --  &  --  &  SG05, eclipsing (W04)\\
287   &  01 01 55.89   &  -72 10 27.9  &  0.9  &12.1 &  0.1  &  0.9  &  15.1  &  -0.2  &  -0.9  &  --  &  --  \\
337   &  00 56 14.65   &  -72 37 55.8  &  0.8  & 1.8 &  3.9  &  0.7  &  14.6  &  0.1  &  -1.3  &  1.9  &  922  &  SG05\\
474   &  00 54 25.99   &  -71 58 24.1  &  0.8  &--   &  53.3  &  2.4  &  16.6  &  -0.1  &  -0.8  &  --  &  --  & ? \\
562   &  01 03 31.73   &  -73 01 44.4  &  1.0  & 3.3 &  0.3  &  1.5  &  15.4  &  -0.2  &  -1.1  &  --  &  --  \\
823   &  01 00 55.85   &  -72 23 20.3  &  1.0  & 6.6 &  71.9  &  1.1  &  15.6  &  2.4\tablefootmark{a}  &  --  &  --  &  -- & B1-5\,III\,e \\
1019  &  00 49 02.67   &  -73 27 07.4  &  1.6  & 3.3 &  41.9  &  3.5  &  15.8  &  -0.2  &  -0.9  &  --  &  --  \\
1189  &  01 03 33.62   &  -72 04 17.5  &  1.7  &--   &  --  &  4.9  &  16.1  &  -0.1  &  -1.0  &  --  &  --  \\
1400  &  00 53 41.76   &  -72 53 10.1  &  0.8  &12.4 &  2.8  &  2.2  &  14.7  &  0.1  &  -1.1  &  --  &  --  \\
1408  &  00 54 09.28   &  -72 41 43.2  &  1.4  & 1.7 &  53.4  &  1.3  &  13.8  &  -0.0  &  -0.7  &  1.0  &  739  \\
1481  &  00 42 07.77   &  -73 45 03.4  &  0.7  & --  &  0.0  &  1.5  &  16.8  &  -0.1  &  -0.5  &  --  &  --  &  B1-5\,III (E04)\\
1524  &  00 45 00.20   &  -73 42 46.7  &  1.7  & --  &  10.4  &  1.5  &  15.6  &  0.0  &  -0.3  &  --  &  --  \\
1605  &  01 03 55.08   &  -72 49 52.7  &  1.5  & --  &  89.2  &  3.7  &  16.2  &  0.7\tablefootmark{a}  &  -0.6\tablefootmark{a}  &  --  &  --  & B2e, in NGC\,376\\
1762  &  01 03 38.00   &  -72 02 15.5  &  1.6  & 3.8 &  13.7  &  4.5  &  16.3  &  -0.2  &  -0.8  &  --  &  --  \\
1817  &  00 54 08.68   &  -72 32 07.5  &  1.4  & --  &  12.6  &  1.1  &  16.9  &  -0.1  &  -0.3  &  --  &  --  \\
1820  &  00 53 18.52   &  -72 16 17.6  &  1.6  & --  &  98.9  &  2.3  &  16.6  &  -0.2  &  -0.8  &  --  &  --  \\
1823  &  00 53 14.81   &  -72 18 47.6  &  1.7  & --  &  12.8  &  4.9  &  16.6  &  -0.0  &  -0.8  &  --  &  --  & L10\\
1826  &  00 52 35.29   &  -72 25 20.8  &  1.6  & --  &  32.0  &  5.7  &  14.9  &  -0.2  &  -0.9  &  --  &  --  \\
1859  &  00 48 55.55   &  -73 49 46.4  &  0.6  & --  &  8.8  &  1.3  &  14.9  &  -0.2  &  -0.7  &  --  &  -- & SG05 \\
1955  &  00 55 35.02   &  -71 33 40.9  &  1.3  & --  &  17.8  &  4.7  &  16.1  &  -0.1  &  -0.8  &  --  &  --  \\
2100  &  01 04 48.54   &  -71 45 41.5  &  1.6  & --  &  84.0  &  4.3  &  16.9  &  0.4  &  -0.2  &  --  &  --  \\
2208  &  00 56 05.48   &  -72 00 11.1  &  2.0  & --  &  6.8  &  1.3  &  16.7  &  -0.1  &  -0.9  &  --  &  -- &  N11 \\
2211  &  00 55 07.25   &  -72 08 25.7  &  1.7  & --  &  18.8  &  3.9  &  16.9  &  -0.1  &  -0.7  &  --  &  --  \\
2300  &  00 56 13.87   &  -72 29 59.7  &  1.0  & 3.9 &  2.0  &  0.7  &  14.5  &  0.0  &  -1.0  &  --  &  --  & B0.5\,V\,e (E06)\\
2318  &  00 56 19.02   &  -72 15 06.1  &  1.8  & --  &  97.9  &  5.2  &  16.1  &  -0.1  &  -0.9  &  4.7  &  928  \\
2497  &  00 43 15.87   &  -73 24 39.2  &  1.5  & --  &  11.8  &  2.7  &  16.7  &  -0.1  &  -0.8  &  --  &  --  \\
2569  &  00 51 46.12   &  -73 07 04.3  &  1.1  & 1.4 &  7.0  &  2.9  &  16.7  &  -0.0  &  -0.7  &  --  &  -- & ?  \\
2587  &  00 52 59.47   &  -72 54 02.1  &  2.1  & 7.4 &  --  &  1.7  &  16.8  &  0.2  &  -0.5  &  --  &  --  \\
2675  &  00 55 49.77   &  -72 51 27.1  &  1.5  & 1.4 &  --  &  1.0  &  16.5  &  -0.0  &  -0.6  &  --  &  -- & eclipsing (W04) \\
2721  &  01 06 00.78   &  -72 33 03.7  &  1.9  & 4.5 &  11.8  &  2.0  &  16.3  &  -0.1  &  -0.9  &  --  &  --  \\
2737  &  01 08 20.18   &  -72 13 47.1  &  0.7  & --  &  72.0  &  2.2  &  14.7  &  -0.1  &  -0.7  &  --  &  -- & ?, B5\,II (E04)  \\
3003  &  01 23 27.46   &  -73 21 23.4  &  1.1  & --  &  20.9 &   1.3\tablefootmark{b}  &  15.5\tablefootmark{b}  &  -0.1\tablefootmark{b}  &  -0.9\tablefootmark{b}  &  --  &  -- & B1-5\,III (E04), S13b \\
3052  &  01 11 08.59   &  -73 16 46.1  &  0.7  & --  &  36.1  &  0.1  &  15.5  &  -0.1  &  -1.0  &  --  &  -- &  SXP\,31.0 ?,  B1-5\,II\,e (E04)  \\
3271  &  00 51 33.27   &  -73 30 12.2  &  1.5  & --  &  16.4  &  4.4  &  16.6  &  0.1  &  -0.8  &  --  &  --  \\
3285  &  01 04 29.42   &  -72 31 36.5  &  1.3  & 8.2 &  70.8  &  1.4  &  15.8  &  -0.2  &  -1.1  &  --  &  --  \\
\noalign{\smallskip}\hline
\end{tabular}
\end{center}
\tablefoot{
\tablefoottext{a}{Colour questionable.}
\tablefoottext{b}{Source is outside MCPS area. Values are from \citet{2002ApJS..141...81M}.}
\tablefoottext{c}{ Only a formal correlation. [MA93]\,522 is associated with the nearby BeXRB SXP\,15.3 (see L10).}
}
\tablebib{
(H00)~\citet{2000A&A...359..573H};
(SPH03)~\citet{2003A&A...403..901S};
(E04)~\citet{2004MNRAS.353..601E};
(E06)~\citet{2006A&A...456..623E};
(W04)~\citet{2004AcA....54....1W};
(SG05)~\citet{2005MNRAS.362..879S};
(K09)~\citet{2009ApJ...701..508K};
(A09)~\citet{2009ApJ...697.1695A};
(N03)~\citet{2003ApJ...586..983N};	
(L10)~\citet{2010ApJ...716.1217L};
(N11)~\citet{2011A&A...532A.153N};
(W12)~\citet{2012ATel.4628....1W};
(S13a)~\citet{2013ATel.4719....1S};
(S13b)~\citet{2013A&A...551A..96S}.
}
\label{tab:HMXB}
\end{table*}

\subsection{Active galactic nuclei}
\label{sec:class:agn}

Galaxies with an AGN are bright X-ray sources at cosmological distances,
and constitute the majority (71\% classified) of point sources in our catalogue.
X-rays are caused by accretion onto a super-massive black hole.
In the \xmm\ energy band, AGN show power-law-like spectra with a typical photon index of 1.7.
Different spectral properties of AGN strongly depend on the inclination of the AGN \citep{1995PASP..107..803U}. 
In addition to studying the AGN itself, identified AGN in the background of the SMC offer reference positions for proper motion studies,
and might be used to probe the absorption by the interstellar medium of the SMC.

\subsubsection{Identification of AGN}
Forty seven spectroscopically confirmed quasars could be identified in our catalogue, mainly from \citet[][]{2006A&A...455..773V} and \citet[][]{2011ApJS..194...22K}.
All \xmm\ detections of these sources are plotted in Fig.~\ref{fig:hr2hr3-select}c in black.
Point sources, emitting X-rays and radio, are also dominated by AGN.
We identified 25 X-ray sources, which correlate with radio background sources of \citet{2004MNRAS.355...44P}.
Also in this case the number of expected chance correlations is low.
These sources are appended with an additional {\tt r} to their classification.
In Fig.~\ref{fig:hr2hr3-select}c, detections of these sources are plotted in magenta.

\subsubsection{Classification of AGN}
AGN can be separated well from stars in the $HR_2$--$HR_3$-plane (Fig.~\ref{fig:hr2hr3-select}c).
We selected AGN candidates ($<$AGN$>$) among hard X-ray sources, where we use the same cut as for stars to discriminate between soft and hard X-ray sources (red line in Fig.~\ref{fig:hr2hr3-select}c).
We could classify 16 AGN, which have a SUMSS radio counterpart (noted with {\tt r}), but no correlation with a radio source of the SMC or foreground in \citet{2004MNRAS.355...44P}.
For 110 sources, we found a hard X-ray source correlating with an infra-red selected AGN candidate of  \citet[][noted with {\tt i}]{2009ApJ...701..508K}, in addition to the already identified AGN.
Using the \cxo\ Wing survey, the optical counterparts can be determined more precisely, and we found 126 hard X-ray sources, correlating with \cxo\ sources and classified as AGN by \citet[][noted with {\tt x}]{2008MNRAS.383..330M}.

In general, one expects for the optical and X-ray flux of an AGN a ratio of  $-1<\log(f_X/f_O)<1$ \citep{1988ApJ...326..680M}.
Another 1861 X-ray sources were classified with $<$AGN$>$, if the source has an optical counterpart candidate with  $\log(f_X/f_O)>-1$ in the MCPS (noted with {\tt o}).
We stress, that this last classification is very general, because of the high source density in the MCPS.
Chance correlations with stars in the SMC can result in fulfiling the same $\log(f_X/f_O)$ criterion.
Also, for weak X-ray sources, the optical luminosity of the AGN can be below the completeness limit of the MCPS.
Since the bulk of hard X-ray sources are expected to be of the AGN class, this classification will be correct in most cases (cf. Sec.~\ref{sec:LF}), but some sources may be of a different nature.
Therefore we mark AGN classifications, based only on the optical criterion with a ``{\tt ?}''.

\subsection{Galaxies}
Galaxies behind the SMC can be seen in X-rays, comprising an unresolved combination of different X-ray sources, e.g. X-ray binaries, SNRs, diffuse emission, and a contribution of a central AGN.
In the 6dFGS \citep{2009MNRAS.399..683J}, we found 6 entries, correlating with X-ray sources (\no{365}, 376, 645, 1726, 2905, and 3208).
These sources were classified as galaxies, with the exception of \no{365} (6dFGS\,gJ005356.2-703804), which was identified as AGN in the previous section.
Source \no{1711} was fitted as an extended source in X-rays and also has a counterpart in the 2MASS extended source catalogue \citep[2MASX, ][]{2006AJ....131.1163S}, similar to the nearby source \no{1726}.
There is an indication of diffuse emission in the mosaic image connecting both sources.
Also sources \no{708} and \no{709} are inside a cluster of galaxies (ClG) and have 2MASX counterparts.
Therefore, we also classified these sources as galaxies.
Sources classified as galaxies are plotted in green in Fig.~\ref{fig:hr2hr3-select}c.
We did not find any redshift-confirmed galaxies in the SMC bar.

\subsection{Clusters of galaxies}
\label{sec:class:clg}

Clusters of Galaxies and galaxy groups contribute to the background sources.
For a review see \citet{2002ARA&A..40..539R}.
The hot intra-cluster medium with temperatures of $kT\sim (2-10)$ keV causes thermal X-ray emission.
Just like SNRs in the SMC, ClGs have an extent detectable with \xmm.
Since the temperature of SNRs is significantly lower, these two source classes can be separated by hardness ratios.
The hardness ratios of all detections, which were flagged as significantly extended ({\tt QFLAG=E}) in the X-ray images, are plotted in Fig.~\ref{fig:hr2hr3-select}d.
Only SNRs, super bubbles, and ClGs are expected as X-ray sources with such a large extent in the SMC field.
Diffuse emission of the hot interstellar medium in the SMC is modelled by spline maps and treated as background.
Identified SNRs and new candidates of \citet[][]{2012A&A...545A.128H} are plotted in black. They have similar soft X-ray colours as stars.
All other sources are plotted in green and show X-ray colours typical of ClGs in the mosaic image \citep[cf.][]{2012A&A...545A.128H}.
The red line marks our selection cut for the ClG classification. The only SNRs within this cut are IKT\,2, IKT\,4 and IKT\,25.
In addition to X-ray colours, we require a significant extent of the X-ray source of $Ext>\Delta Ext$ and a maximum likelihood for the extent of $ML_{\rm ext}>10$ for a CIG classification.
Using these criteria, we classified 13 of 19 sources with significant extent as ClG candidates ($<$ClG$>$), in addition to the 11 ClGs not included in the point-source catalogue (because of their very large extent).
All sources with significant extent are listed in Table \ref{tab:char:ext}.

\subsection{Other source classes}

The search for additional source classes is more extensive and will be discussed in other studies.
This includes fainter low-mass X-ray binaries or cataclysmic variables in the SMC that are  at the detection limit of the \xmm\ survey.
Extended sources, such as SNRs and ClGs, which are not included in our catalogue, are presented in \citet[][]{2012A&A...545A.128H}.
A search for highly absorbed X-ray binaries in the survey data was presented by \citet{2011A&A...532A.153N}.
Candidates for highly absorbed white dwarf/Be systems are listed in \citet{2012A&A...537A..76S}.
We assigned a specific source class to some individual sources:
Source \no{48} as an anomalous X-ray pulsar \citep[AXP, ][]{2008ApJ...680L.133T}, 
source \no{54} as a pulsar wind nebula or micro quasar \citep[PWN?/MQ?, ][]{2011A&A...530A.132O},
source \no{324} as an isolated neutron star candidate \citep[INS?, ][]{2006A&A...452..431K},
source \no{551} as PWN candidate \citep[PWN?, ][]{2008A&A...485...63F},
and source \no{535} as a star cluster (Cl*, Sec.~\ref{sec:class:smcstars}).

\section{General characteristics of the dataset}
\label{sec:discussion}

With the \xmm\ catalogue of the SMC, the central field is covered completely down to a luminosity of 5\ergs{33} in the (0.2$-$4.5) keV band, 
deeper than with previous imaging X-ray telescopes.
The comparison with previous ROSAT and \cxo\ surveys, as well as with the XMM-Newton Serendipitous Source Catalogue, shows
that $\sim$1200 sources have been detected for the first time during the large-programme SMC survey.
Some basic properties of the dataset will be discussed in the following sections.

\subsection{Spatial distribution}

\begin{figure*}
\resizebox{0.95\hsize}{!}{
  \resizebox{\hsize}{!}{\includegraphics[width=0.5\textwidth,angle=0,clip=]{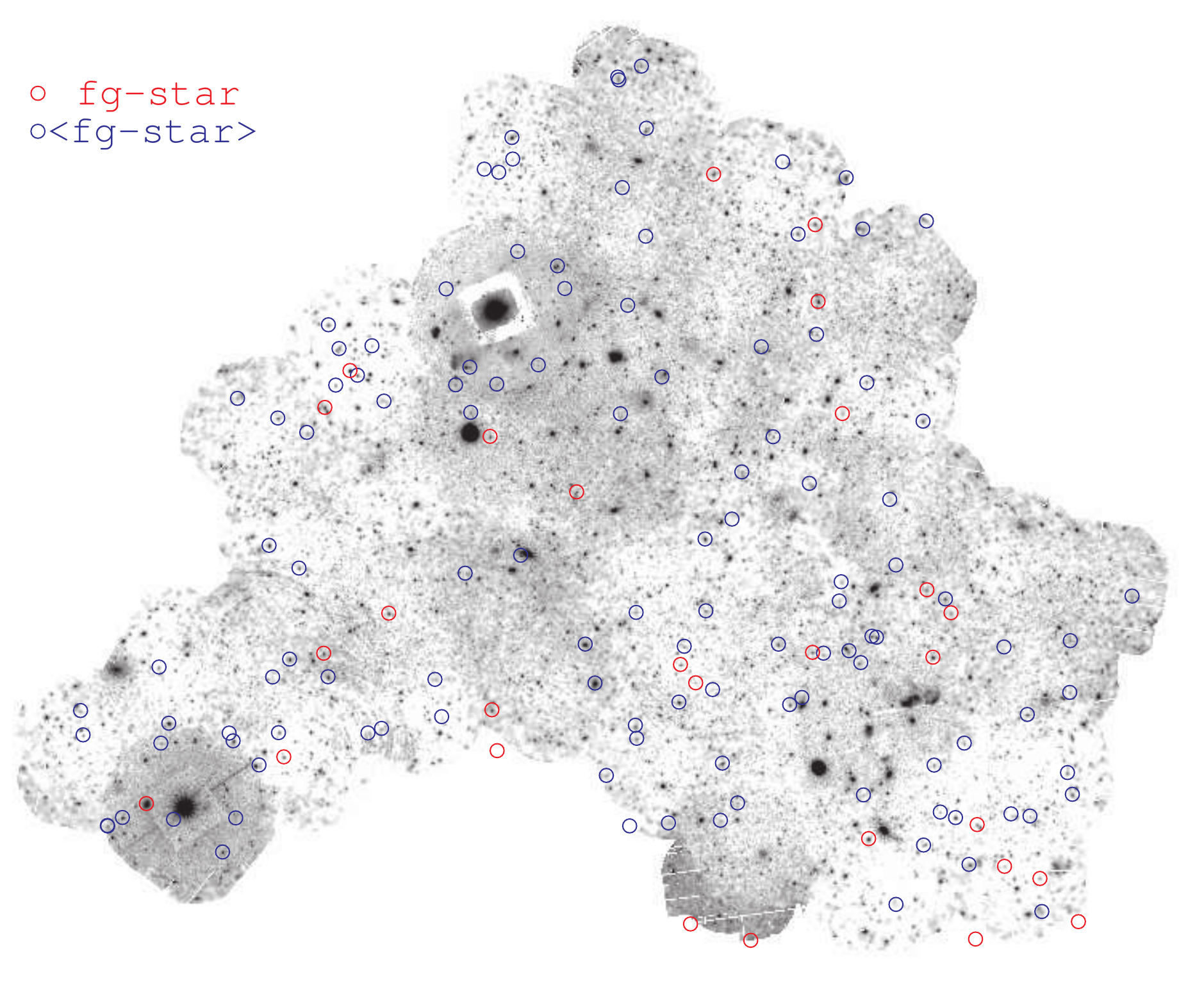}}
  \resizebox{\hsize}{!}{\includegraphics[width=0.5\textwidth,angle=0,clip=]{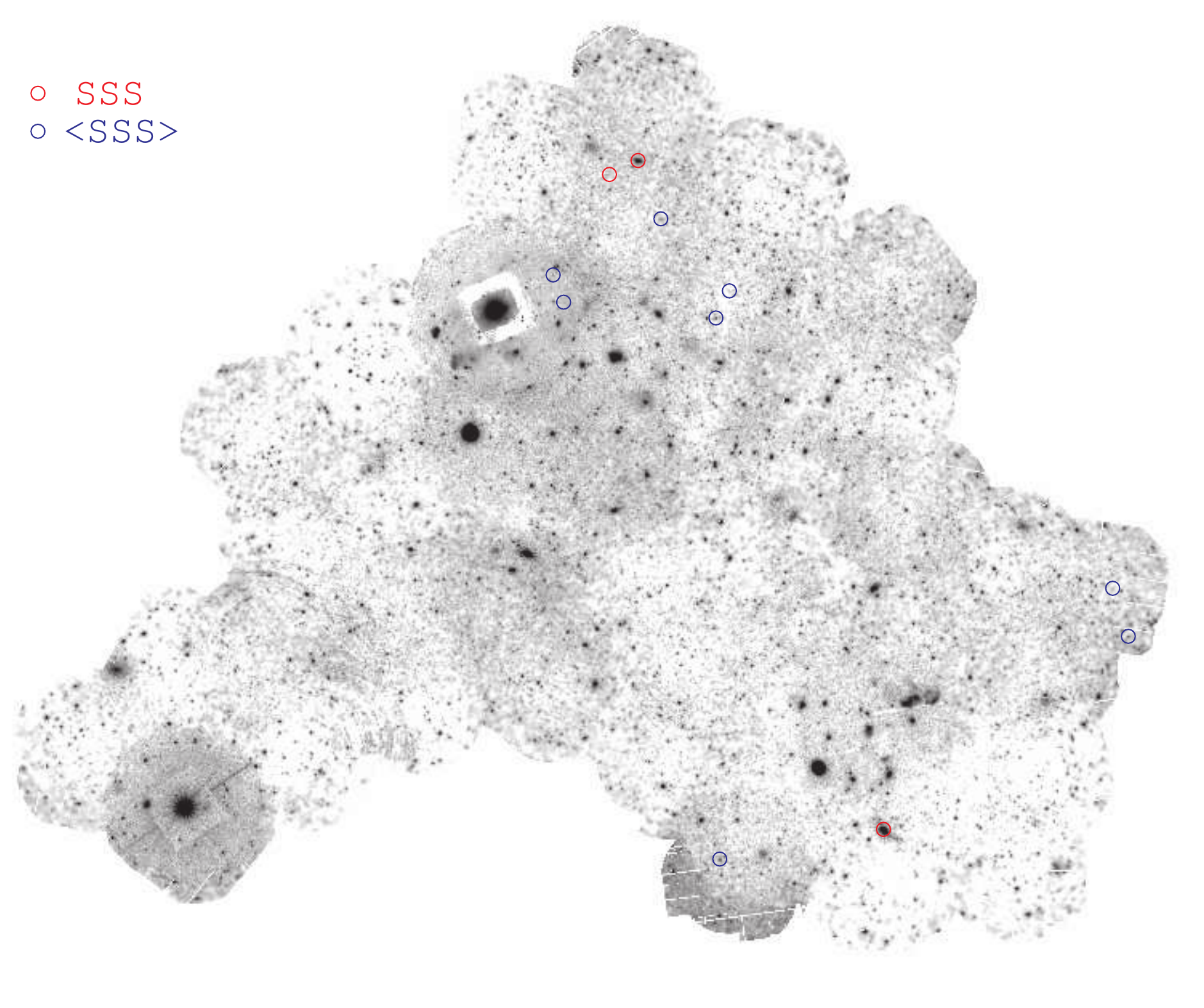}}
}
\newline
\resizebox{0.95\hsize}{!}{
  \resizebox{\hsize}{!}{\includegraphics[angle=0,clip=]{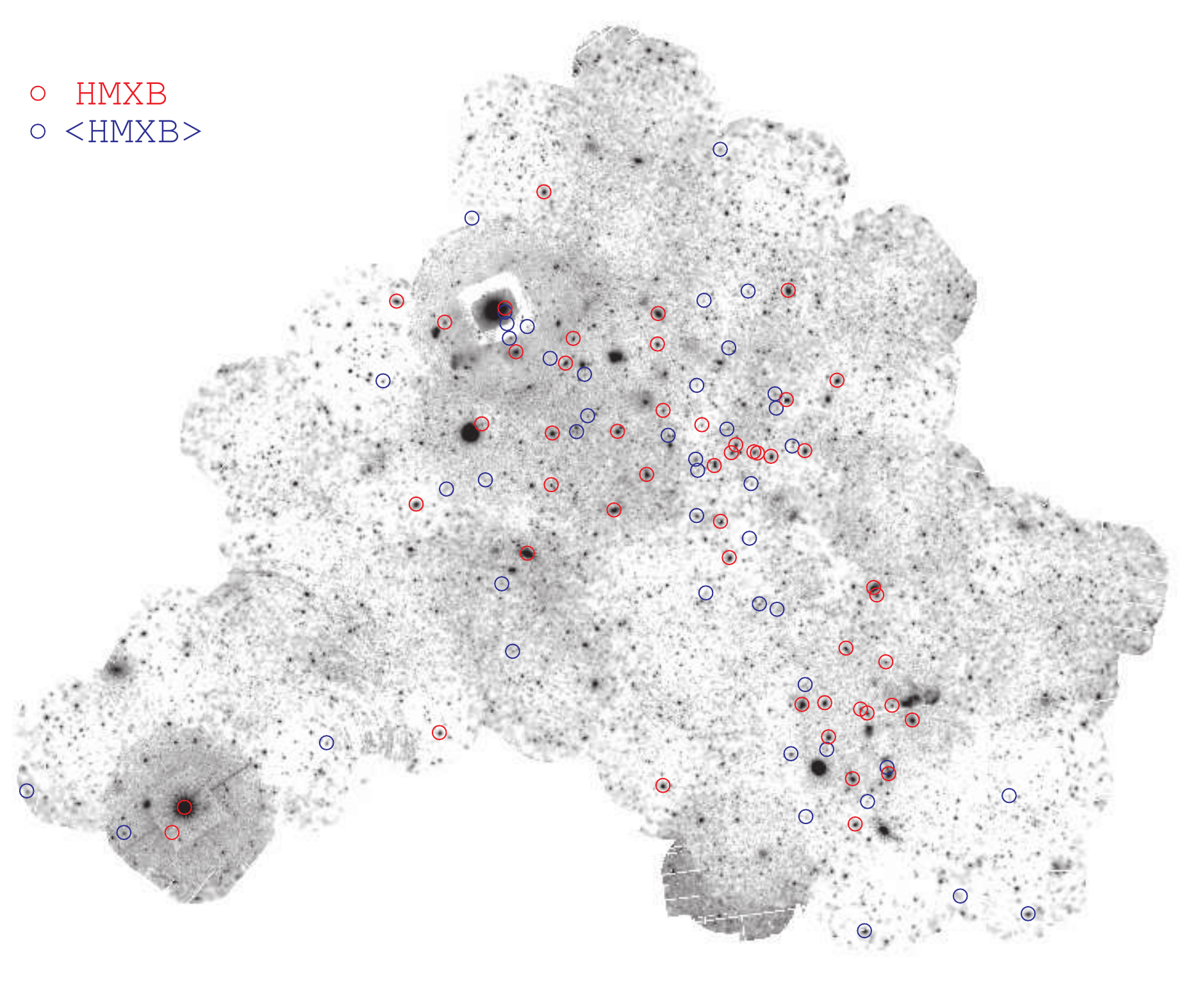}}
  \resizebox{\hsize}{!}{\includegraphics[angle=0,clip=]{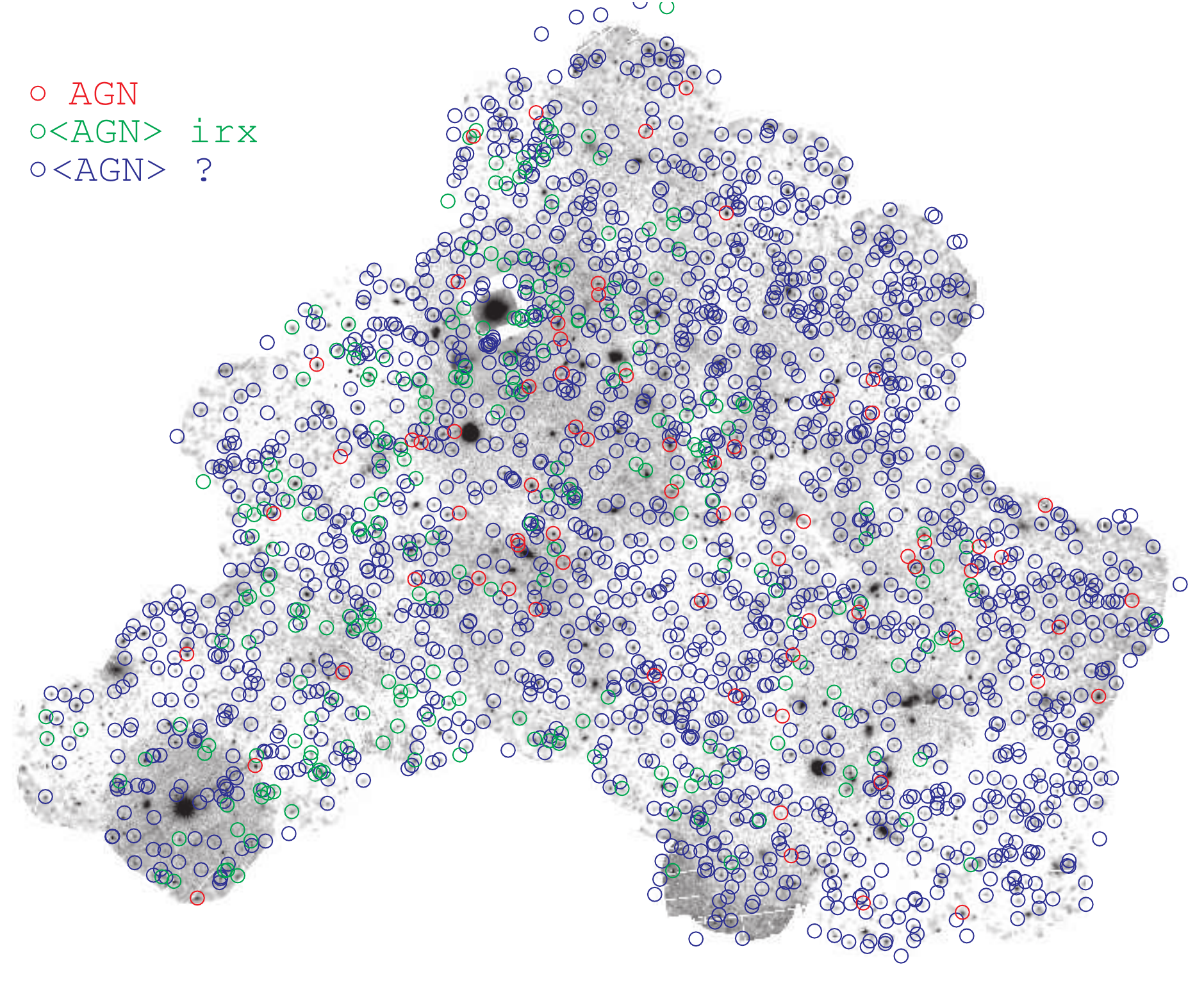}}
}
\newline
\resizebox{0.95\hsize}{!}{
  \resizebox{\hsize}{!}{\includegraphics[angle=0,clip=]{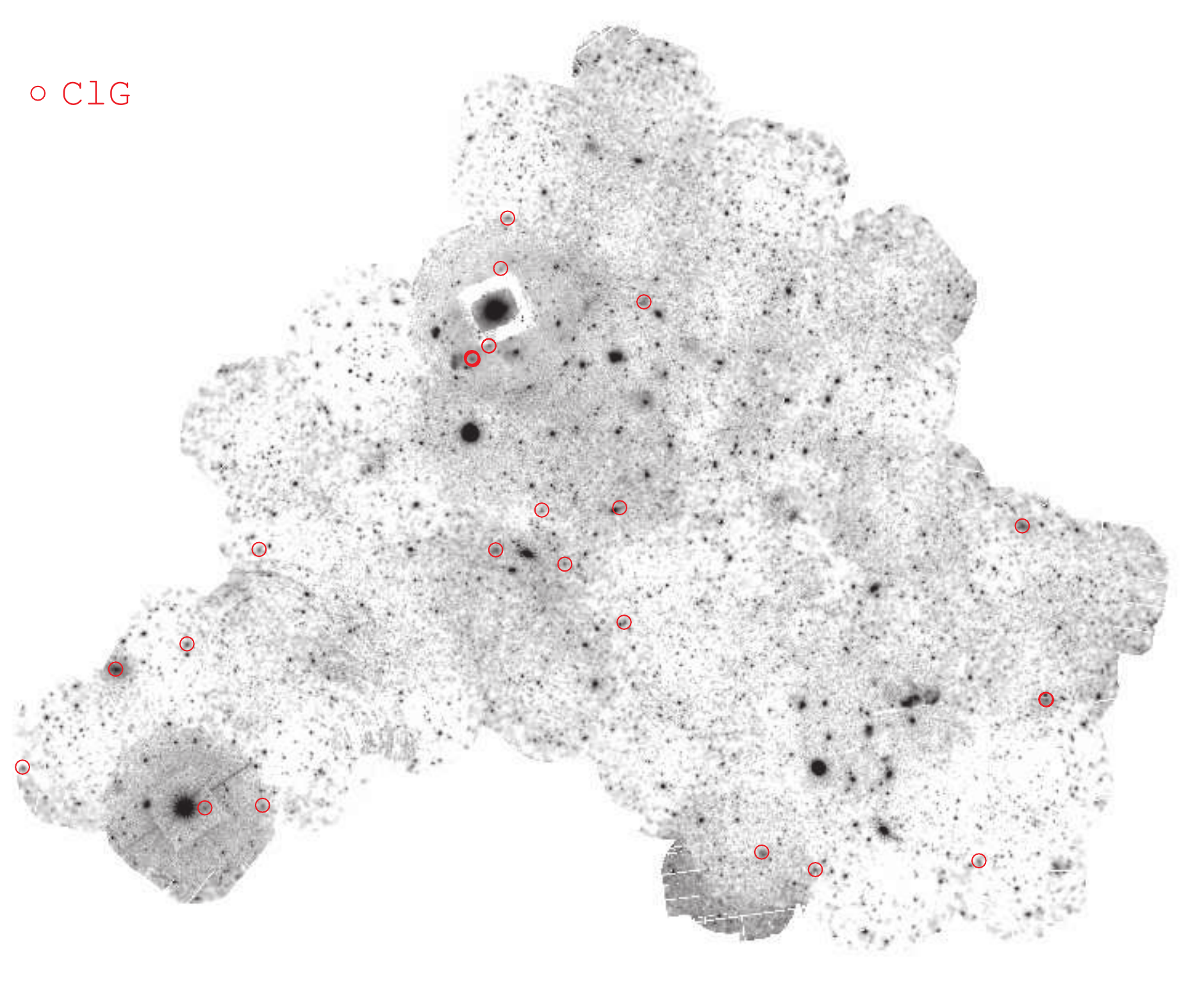}}
  \resizebox{\hsize}{!}{\includegraphics[angle=0,clip=]{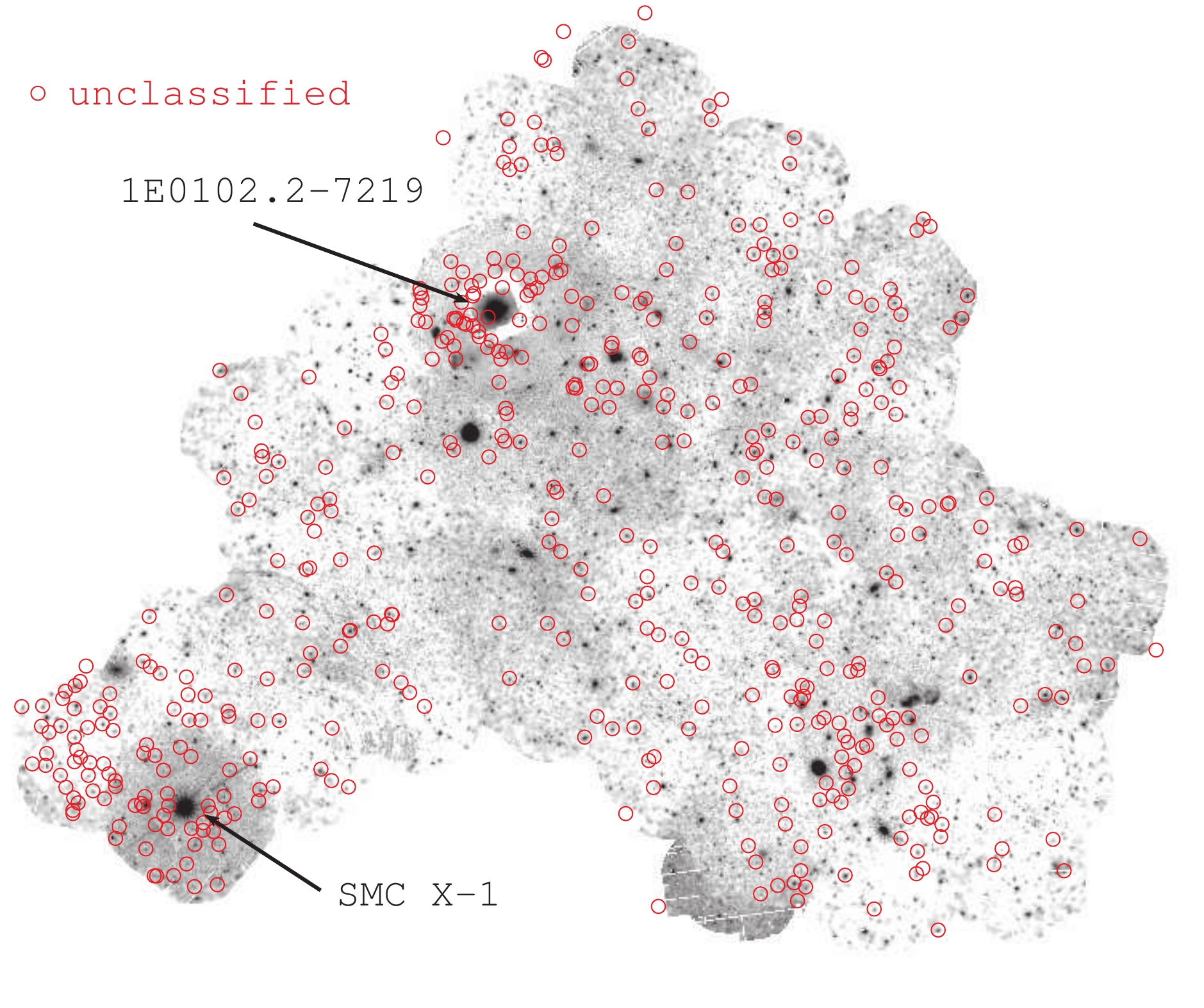}}
}
  \caption{Spatial distribution of identified and classified X-ray sources in the SMC main field.
           The underlying mosaic image shows logarithmically scaled intensities in the (0.2--4.5) keV band.
           North is up, east is left.
          }
  \label{fig:space-dist}
\end{figure*}

The spatial distribution of individual source classes in the main field is shown in Fig.~\ref{fig:space-dist}.
In the upper left, sources identified as Galactic stars (red) or classified as candidates for Galactic stars (blue) are marked.
The distribution is homogeneous over the entire field.
SSSs as well as SSS candidates (upper right) are found in the outer regions of the bar, especially in the northern part. We did not find any in the SMC wing. 
As expected, HMXBs and their candidates follow the SMC bar (middle left).
Since the bar harbours most of the blue main-sequence stars, we find here also most of the chance correlations with background AGN that contribute to the HMXB candidates.
AGN show a homogeneous distribution over the observed field (middle right).
Infrared selected AGN candidates are restricted to the smaller {\it Spitzer} S$^3$MC field, AGN candidates from \cxo\ are only in the \cxo\ Wing fields.
Clusters of galaxies that could be identified or classified, are shown in the lower left.
Unclassified sources are marked in the lower right.
Here we see some enhancement at the eastern rim, where the MCPS does not cover the field, and around SMC~X-1, which may cause some spurious detections due to its brightness.
Also around \calsrc, an enhancement of unclassified sources is observed, as expected, due to the high number of observations, which lead to a higher number of spurious detections.

\subsection{Luminosity functions}
\label{sec:LF}

\begin{figure*}
  \resizebox{\hsize}{!}{\includegraphics[height=3cm,angle=0,clip=]{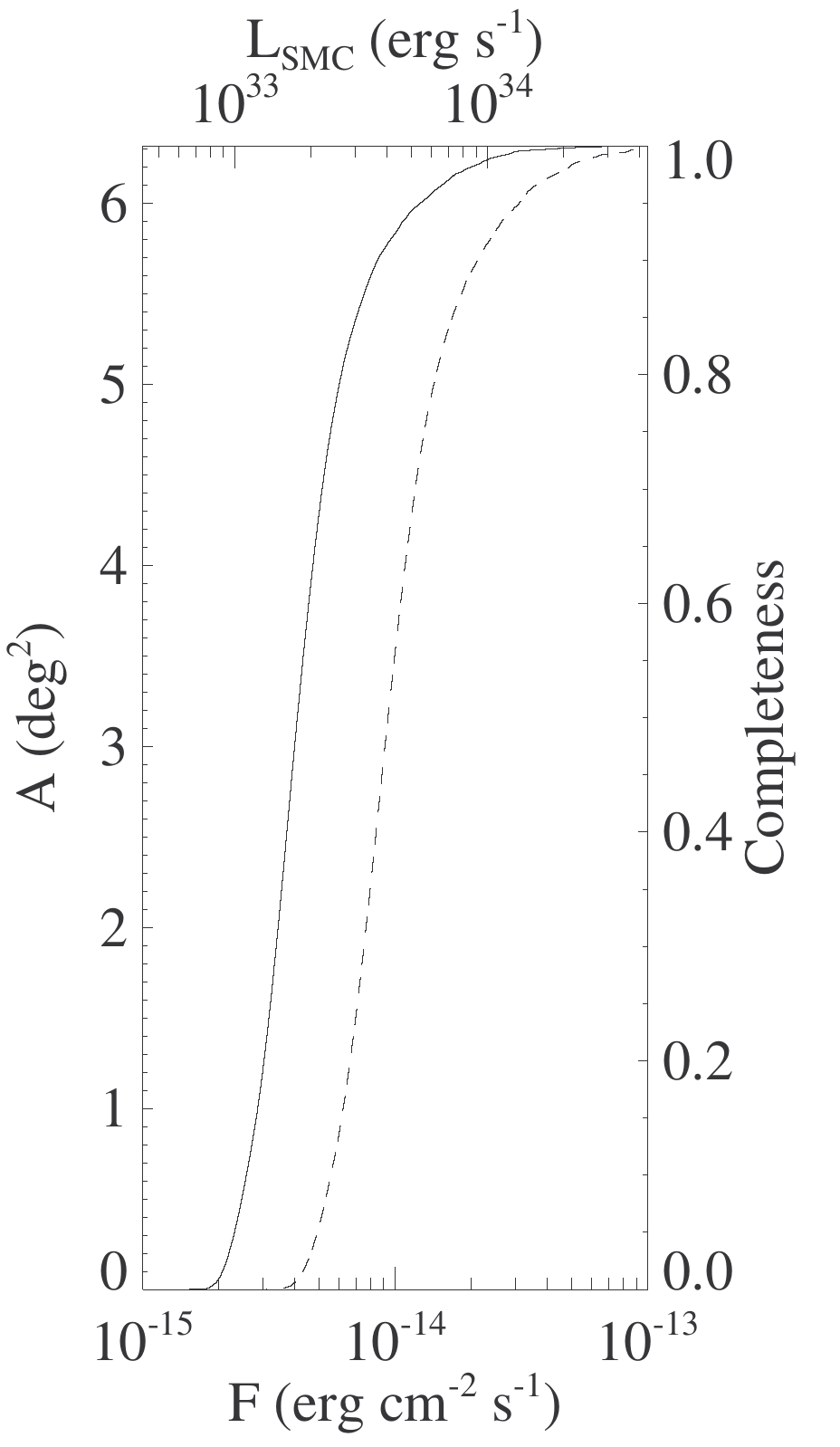}
                        \includegraphics[height=3cm,angle=0,clip=]{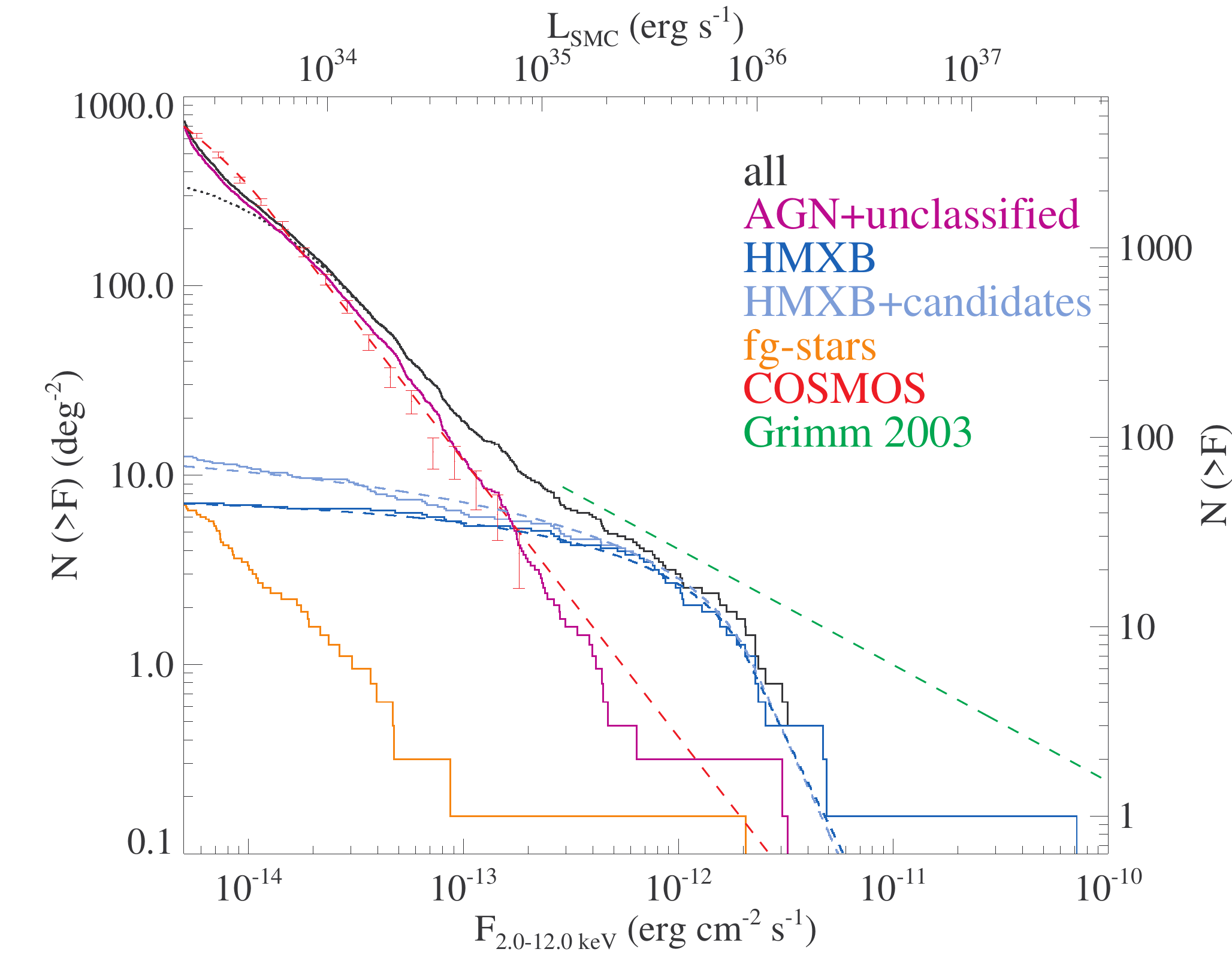}}
  \caption{
   {\it Left:} Sky coverage as function of flux is shown for the (0.2--12.0) keV and (2.0--12.0) keV band by the solid and dashed line, respectively.
   {\it Right:} Cumulative luminosity function for sources and their classification in our catalogue (solid lines) and according models (dashed lines) as described in Sec.~\ref{sec:LF}.
    }
  \label{fig:LF}
\end{figure*}

We constructed the luminosity function of the various classes of objects detected in the SMC fields.
For sources with high long-term time variability, taking the average or maximal flux would not represent the source luminosity distribution of the galaxy at one time.
For each source, we selected the flux from the observation with the highest sensitivity at this position, i.e. with minimal detection-limit flux.
If the source was not detected in this observation, the source was not taken into account for the luminosity function.
None of the selected detections is from an observation that was triggered by an outburst of the corresponding source.
Therefore, this method selects one of several measured fluxes of transient sources in a quasi-random manner and thus represents the flux distribution
as measured in one single observation of the whole galaxy.
Also, this method minimises the effect of spurious detections at higher fluxes,
since the source has to be detected in the most sensitive observation.

Depending on exposure time and observation background, the sensitivity varies between individual observations.
Diffuse emission and vignetting also cause a spatial dependence of the sensitivity within each observation.
The calculation of sensitivity maps is described in Sec.~\ref{sec:sensmaps}.
To estimate the sky coverage, we merged all sensitivity maps,
by selecting the observation with highest sensitivity at each position.
The corresponding completeness function is presented in Fig.~\ref{fig:LF}, left.

Especially for background sources, the completeness for the full energy band is clearly overestimated,
since the ECFs adopted from the universal spectrum (Sec.~\ref{sec:cat:compilation}) only account for galactic absorption,
but not for absorption in the SMC, reaching line-of-sight column densities of up to $1.4\times10^{22}$ cm$^{-2}$.
To minimise this effect, we use the (2.0$-$12.0) keV band in the following.
The flux reduction by Galactic absorption ($\sim$$6\times10^{20}$ cm$^{-2}$) is 0.5\% for the assumed universal spectrum, so we use the observed fluxes here.
The completeness-corrected cumulative distribution of all sources is shown by the solid black line in Fig.~\ref{fig:LF}, right.
The correction mainly affects the number of sources with fluxes below $\sim3\times10^{-14}$ erg cm$^{-2}$ s$^{-1}$,
as can be seen by the uncorrected distribution (dotted line).

For HMXBs, we see a break around $10^{-12}$ erg cm$^{-2}$ s$^{-1}$, similar to that inferred by \citet{2005MNRAS.362..879S}.
As suggested by these authors, this can be caused by the propeller effect, which can inhibit accretion at low accretion rates.
Using C statistics, we parameterise the flux distribution of the total (i.e. not normalised by area) HMXB populations by fitting a broken power law to the unbinned source counts:

$$
n(F)=\frac{dN}{dF}=
                    \left\{\begin{array}{lll}
                                              N_1\,F^{-\alpha_{1}} &{\rm if} & F \leq  F_{b} \\
                                              N_2\,F^{-\alpha_{2}} &{\rm if} & F >     F_{b} \\
                             \end{array}\right.
$$
with the faint and bright end slopes $\alpha_1$ and $\alpha_2$,
the normalisation $N_2=N_1 F_{b}^{\alpha_{2}-\alpha_{1}}$, and
the break flux $F_b$ and flux $F$ in 10$^{-12}$ erg cm$^{-2}$ s$^{-1}$.
For HMXBs, we obtain
$\alpha_{1} =  0.64_{-0.17}^{+0.13}$,
$\alpha_{2} =  3.30_{-1.38}^{+2.17}$,
$ F_b =  2.09_{-1.21}^{+0.74}\times 10^{-12}$ erg cm$^{-2}$ s$^{-1}$, and
$N_1 =   11.9_{-3.5}^{+7.5}$.
Including the HMXB candidates as well, we obtain
$\alpha_{1} =  0.87_{-0.10}^{+0.08}$,
$\alpha_{2} =  3.52_{-1.64}^{+1.91}$,
$ F_b =  2.25_{-1.49}^{+0.62} \times 10^{-12}$ erg cm$^{-2}$ s$^{-1}$, and
$N_1 =   13.9_{-3.4}^{+7.6}$.

Uncertainties are for 90\% confidence.
These models are shown by the blue dashed lines in Fig.~\ref{fig:LF} and give an upper and lower limit for the luminosity function.
The bright-end slope is significantly steeper than found for HMXB populations of nearby galaxies above a luminosity of $10^{38}$ erg s$^{-1}$ \citep[$\alpha = 1.61 \pm 0.12$,][]{2003MNRAS.339..793G}.
The extrapolation of this model to lower luminosities is shown by a dashed green line in Fig.~\ref{fig:LF},
where we used a star-formation rate of SFR$_{\rm SMC} = 0.15$ M$_{\sun}$ yr$^{-1}$ \citep[as in ][]{2003MNRAS.339..793G}
and a correction factor of 1.24 (as expected for a photon index of $\Gamma=1$) to obtain fluxes in the (2.0--12.0) keV band.
\citet{2012MNRAS.419.2095M} suggest that this model is valid down to $L_X \sim 10^{35}$ erg s$^{-1}$.
The deviation is probably caused by different source types.
Our sample is dominated by BeXRBs, which show outbursts above luminosities of $10^{36}$ erg s$^{-1}$,
whereas for more distant galaxies, due to higher flux limits only the brightest HMXBs can be detected.
These contain a higher fraction of supergiant HMXBs which are, compared to BeXRBs, rather persistent and can contain a black hole instead of a neutron star.

Indeed, the presence of one supergiant system in the SMC, SMC X-1, is consistent with the Grimm model.
In this context, the turn over might be interpreted as the transition from transient to persistent BeXRBs.
We do not discuss the luminosity function below $\sim 2 \times 10^{-14}$ erg cm$^{-2}$ s$^{-1}$, as we have only 3 sources here and run into incompleteness issues.

For background sources, we compare the SMC field with the  \xmm\ COSMOS field \citep{2007ApJS..172...29H}.
The COSMOS-field source counts and broken power-law model of \citet{2009A&A...497..635C} is shown in red in Fig.~\ref{fig:LF}.
As these values are given in the (2.0--10.0) keV band, we applied a factor of 1.14 to estimate fluxes for the (2.0--12.0) keV band,
as expected for a power law with photon index of $\Gamma=1.7$.
For such a power law, we expect a flux decrease in the (2.0--12.0) keV band by 1.2\% and 2.4\% when crossing a column density of 5 and 10 $\times 10^{21}$ cm$^{-2}$ of ISM of the SMC, respectively.
Analogously, we would expect a flux decrease by 35\% and 51\% (0.5--2.0) keV, for an AGN with negligible intrinsic absorption.
We find a general agreement with the distribution of unidentified and AGN sources (magenta line in Fig.~\ref{fig:LF}). Small deviations can be explained by the slightly different data processing.
The contribution of Galactic stars (orange line in Fig.~\ref{fig:LF}) is negligible above 2 keV.

\subsection{Spectral properties}

\begin{figure}
\centering
  \resizebox{8.2cm}{!}{\includegraphics[angle=-90,clip=]{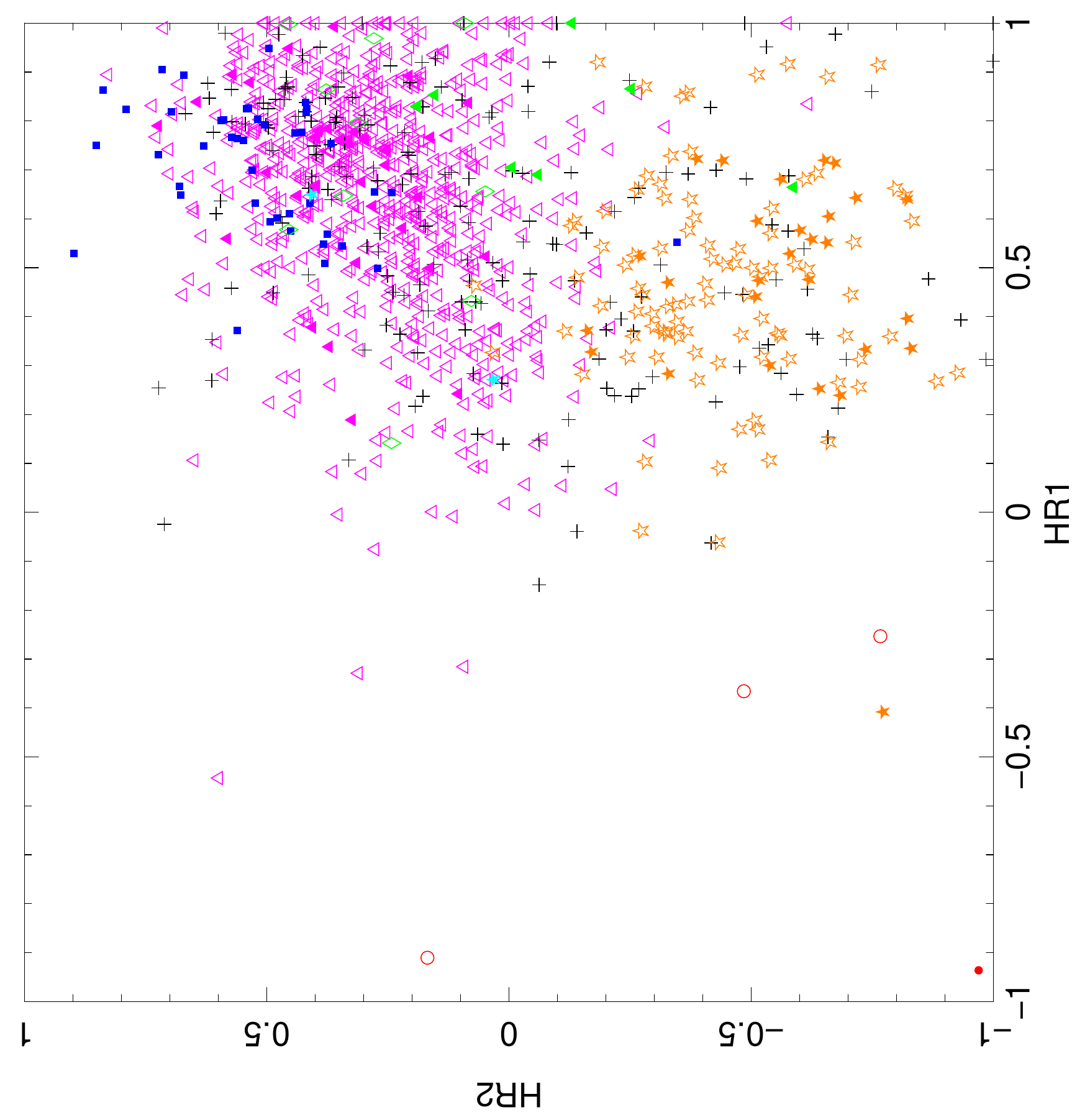}}
  \resizebox{8.2cm}{!}{\includegraphics[angle=-90,clip=]{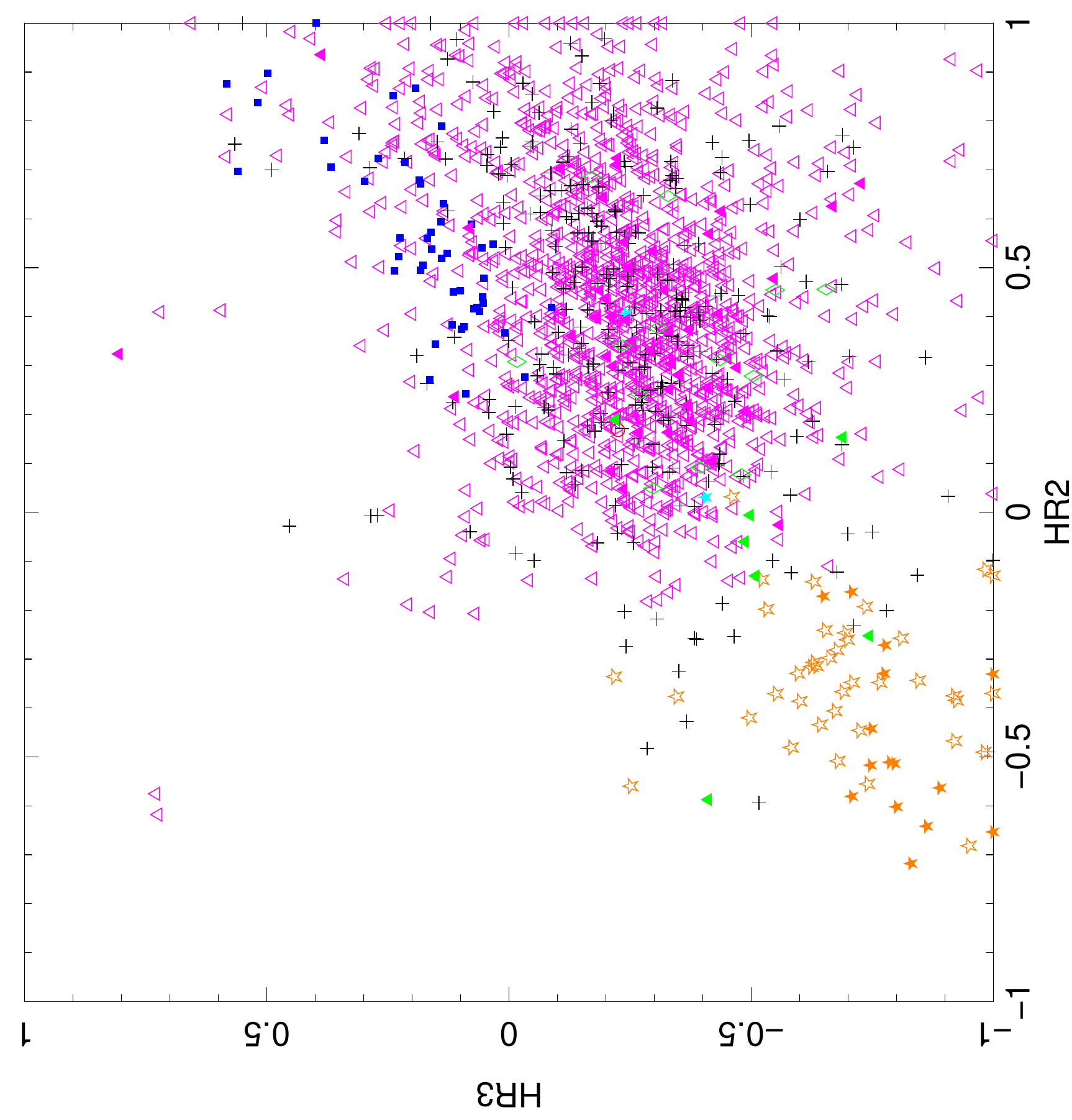}}
  \resizebox{8.2cm}{!}{\includegraphics[angle=-90,clip=]{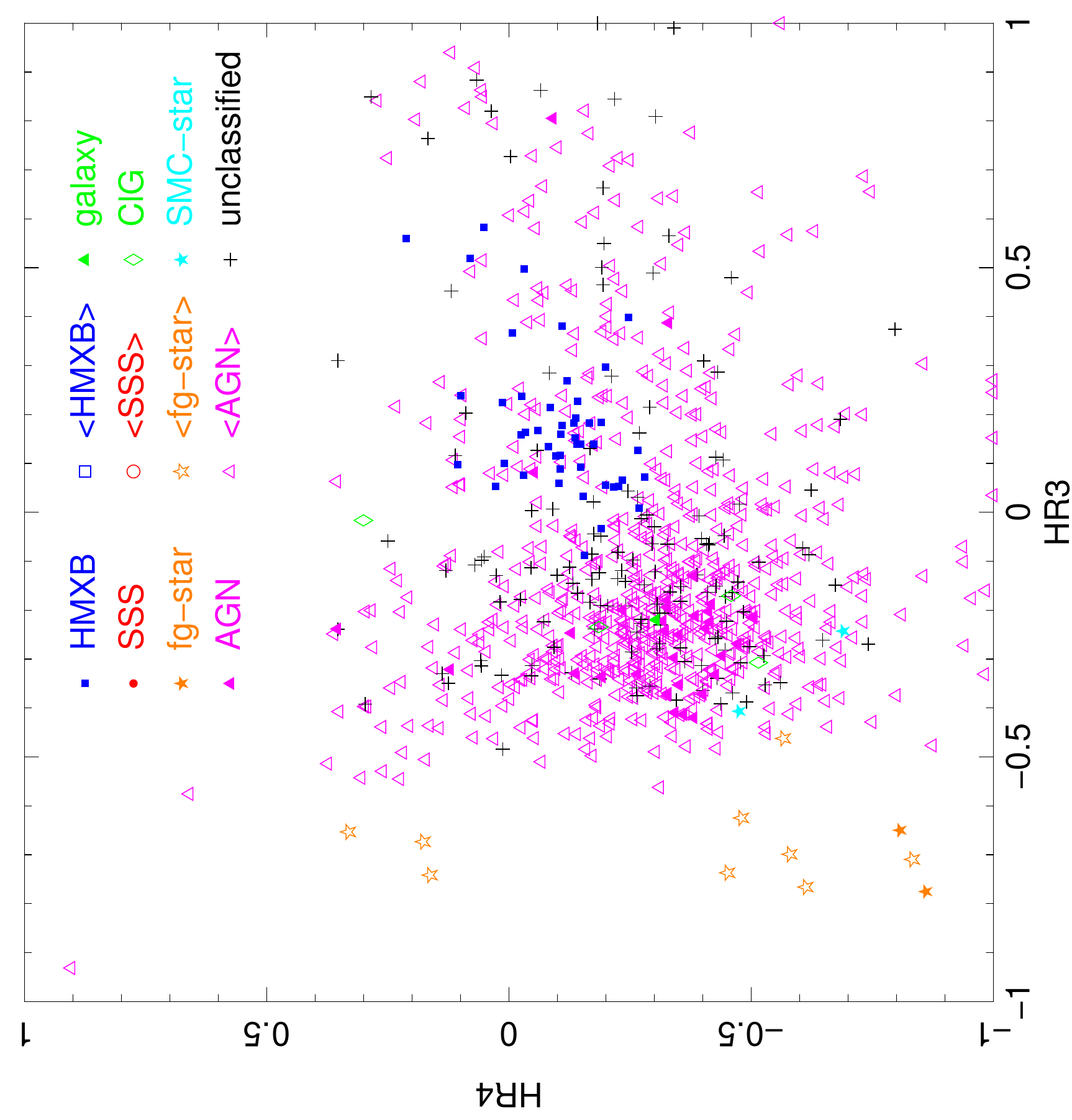}}
  \caption{
   Hardness-ratio diagrams  of all catalogue sources with respective uncertainties of $\Delta HR_i<0.25$, illustrating the spectral distribution of the catalogue sources.
   }
  \label{fig:hrs}
\end{figure}

To characterise the spectral properties of our source sample, we calculated hardness ratios as described in Sec.~\ref{sec:cat:compilation}.
Hardness-ratio diagrams of sources with well measured hardness ratios ($\Delta HR_i<0.25$) are presented in Fig.~\ref{fig:hrs}.
Of course, the classified sources follow our selection criteria (cf. Fig.~\ref{fig:hr2hr3-select}).
In total we find 436 hard unclassified sources which are presumably mainly background AGN, where we expect a faint optical counterpart.
If we assume a completeness of the MCPS catalogue of $V\approx21$ mag,
we can miss optical counterparts of AGN with $\log{(f_x/f_o)}=1$ if they are fainter than $F_x=10^{-12.7}$ erg cm$^{-2}$ s$^{-1}$.
Nearly all of the unclassified hard sources have such a low X-ray luminosity.
A few hard sources might be caused by foreground flare stars, where the X-ray photon statistics is insufficient to detect variability.
Also, the comparison with the COSMOS field (Fig. 12) seems to be consistent with most of these sources being background.
Sources in the upper left of the $HR_3$-$HR_4$-plane are good candidates for highly absorbed AGN \citep{2012MNRAS.422.1166B},
where the absorption can exceed \nh$>10^{24}$ cm$^{-2}$, which is significantly higher than expected from the interstellar medium of the SMC along the line of sight.
From 130 unclassified soft sources, 67 have a detection likelihood of $ML_{\rm det}<10$.
Here we expect most to be spurious detections.
Other soft X-ray sources might originate from distant K or G stars
that cannot be distinguished well from bright SMC stars and violate criterion iii of our foreground star selection.

\subsection{Source extent}

\begin{table*}
\caption[]{Sources with significant extent.}
\begin{center}
\begin{tabular}{lccccccccccc}
\hline\hline\noalign{\smallskip}
\multicolumn{1}{l}{SRC} &
\multicolumn{1}{c}{RA} &
\multicolumn{1}{c}{Dec}&
\multicolumn{1}{c}{Ext (\arcsec)}&
\multicolumn{1}{c}{$ML_{\rm ext}$}&
\multicolumn{1}{c}{$HR_2$}&
\multicolumn{1}{c}{$HR_3$}&
\multicolumn{1}{c}{Class}&
\multicolumn{1}{c}{Comment}\\
\noalign{\smallskip}\hline\noalign{\smallskip}

1  &  01 17 05.2   &  -73 26 36  &  6.34  $\pm$  0.02  &  69614.5  &  0.27  $\pm$  0.00  &  0.16  $\pm$  0.00  &  HMXB  &  pile-up    \\
307  &  01 03 28.6   &  -71 45 48  &  8.74  $\pm$  0.95  &  11.5507  &  0.34  $\pm$  0.08  &  -0.23  $\pm$  0.08  &  $<$ClG$>$  &     \\
362  &  00 53 08.7   &  -72 34 45  &  10.90  $\pm$  1.63  &  12.8268  &  -0.32  $\pm$  0.12  &  -0.35  $\pm$  0.20  &     &  in B0050-72.8    \\
566  &  01 01 25.0   &  -72 46 34  &  14.20  $\pm$  1.84  &  13.1577  &  0.31  $\pm$  0.15  &  -0.02  $\pm$  0.14  &  $<$ClG$>$  &     \\
571  &  01 04 08.1   &  -72 43 54  &  20.24  $\pm$  1.46  &  24.7586  &  0.65  $\pm$  0.09  &  -0.33  $\pm$  0.09  &  $<$ClG$>$  &     \\
604  &  00 52 53.4   &  -73 00 09  &  3.74  $\pm$  0.61  &  12.7043  &  -0.28  $\pm$  0.06  &  -0.68  $\pm$  0.09  &  $<$fg-star$>$  &  multiple stars?   \\  
638  &  01 09 53.3   &  -72 21 47  &  31.89  $\pm$  2.87  &  20.3845  &  -0.26  $\pm$  0.13  &  -0.38  $\pm$  0.22  &     &    extent spurious? \\
799  &  00 59 04.1   &  -72 56 45  &  8.51  $\pm$  1.05  &  29.0269  &  0.28  $\pm$  0.07  &  -0.50  $\pm$  0.08  &  $<$ClG$>$  &     \\
916  &  01 16 15.3   &  -73 26 57  &  11.94  $\pm$  1.11  &  45.9855  &  0.45  $\pm$  0.08  &  -0.55  $\pm$  0.09  &  $<$ClG$>$  &     \\
937  &  01 13 53.2   &  -73 27 08  &  8.51  $\pm$  1.00  &  11.7371  &  0.31  $\pm$  0.09  &  -0.44  $\pm$  0.10  &  $<$ClG$>$  &     \\
1174  &  01 03 45.9   &  -71 54 36  &  14.79  $\pm$  2.04  &  12.6019  &  0.69  $\pm$  0.17  &  -0.17  $\pm$  0.11  &  $<$ClG$>$  &     \\
1305  &  01 01 26.3   &  -75 05 06  &  9.12  $\pm$  1.40  &  18.3986  &  0.08  $\pm$  0.12  &  -0.48  $\pm$  0.15  &  $<$ClG$>$  &     \\
1436  &  00 53 03.5   &  -70 47 34  &  12.48  $\pm$  2.12  &  10.241  &  0.24  $\pm$  0.17  &  -0.27  $\pm$  0.17  &  $<$ClG$>$  &     \\
1505  &  00 44 19.4   &  -73 36 24  &  13.15  $\pm$  1.90  &  12.332  &  0.05  $\pm$  0.14  &  -0.30  $\pm$  0.17  &  $<$ClG$>$  &     \\
1562  &  00 58 22.2   &  -72 17 59  &  19.94  $\pm$  2.38  &  15.5316  &  0.08  $\pm$  0.15  &  -0.50  $\pm$  0.17  &  $<$AGN$>$ o?  &  in IKT 16   \\
1711  &  01 09 00.6   &  -72 29 03  &  8.56  $\pm$  1.03  &  23.5625  &  -0.25  $\pm$  0.08  &  -0.74  $\pm$  0.12  &  galaxy  &    in ClG?  \\
2695  &  01 02 18.3   &  -72 37 03  &  12.57  $\pm$  1.55  &  21.9732  &  0.38  $\pm$  0.11  &  -0.31  $\pm$  0.11  &  $<$ClG$>$  &     \\
2889  &  01 13 26.3   &  -72 42 19  &  13.25  $\pm$  1.64  &  26.7555  &  0.09  $\pm$  0.13  &  -0.40  $\pm$  0.17  &  $<$ClG$>$  &     \\
3030  &  01 23 32.1   &  -73 17 10  &  8.35  $\pm$  1.35  &  17.5979  &  0.46  $\pm$  0.11  &  -0.65  $\pm$  0.14  &  $<$ClG$>$  &     \\

\noalign{\smallskip}\hline
\end{tabular}
\end{center}
\tablefoot{This table does not contain sources with extent too large for the source detection. See \citet[][]{2012A&A...545A.128H}.}
\label{tab:char:ext}
\end{table*}

Table \ref{tab:char:ext} lists all sources, fitted with a significant extent, where we demand
$Ext>\Delta Ext$ and a likelihood for the extent of $ML_{\rm ext}>10$.
Most of them are consistent with a ClG classification.
Three other sources are inside or behind an extended X-ray source.
In the case of SMC X-1 (\no{1}), the source extent is caused by pile-up.
For \no{604}, the hardness ratios point to a star. In the optical, several bright counterpart candidates are found.
The extent might be due to a superposition of two or more stars.
\no{638} was detected as an extended source only in one of three detections. Therefore the extent might be spurious.
We note that 13 additional sources were fitted with extent at lower likelihood or with high uncertainty.

\subsection{Source variability}

\subsubsection{Intra-observational variability}
\label{sec:discussion:variabilityshort}

\begin{figure}
  \resizebox{\hsize}{!}{\includegraphics[angle=0,clip=]{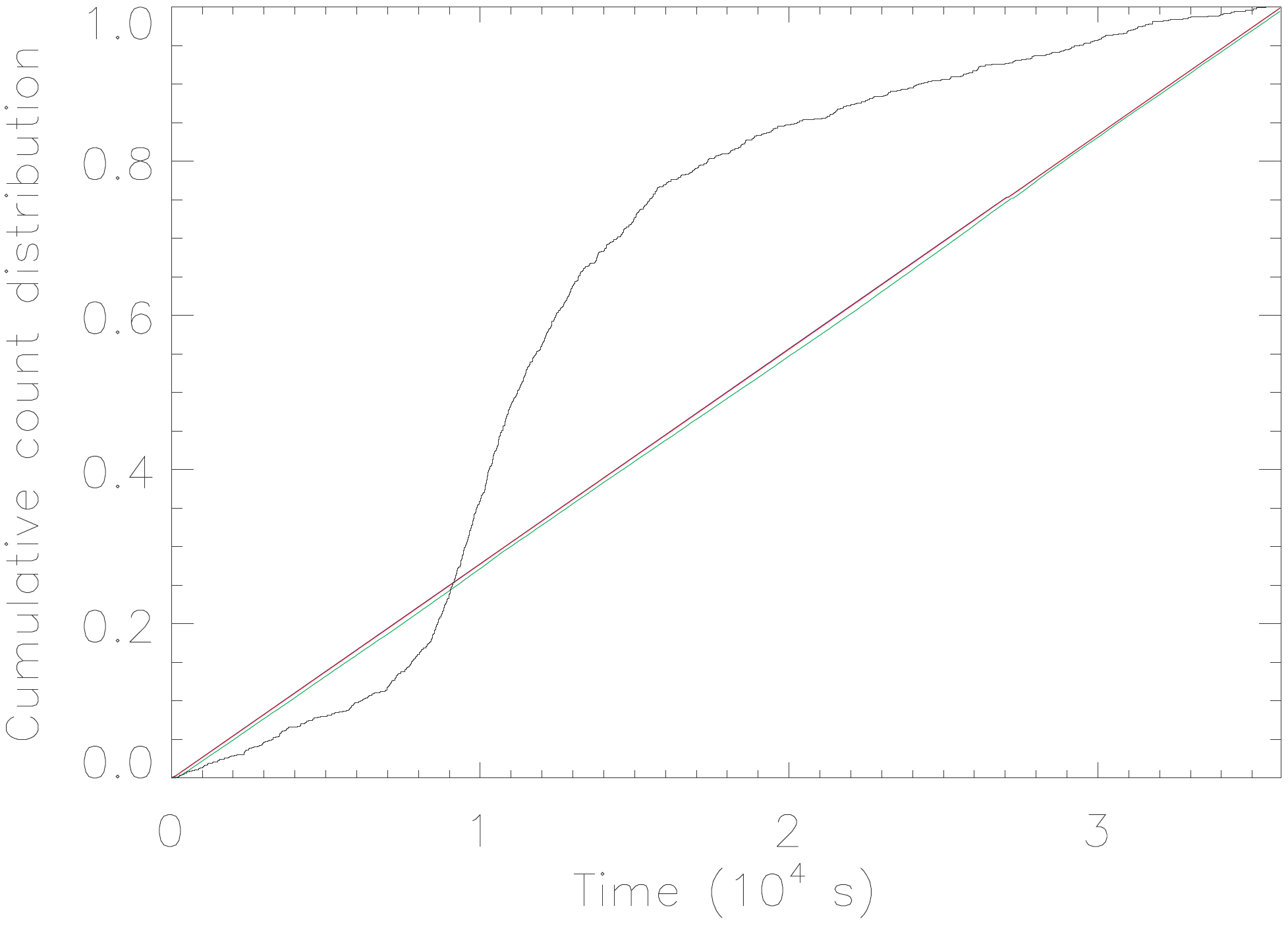}\includegraphics[angle=0,clip=]{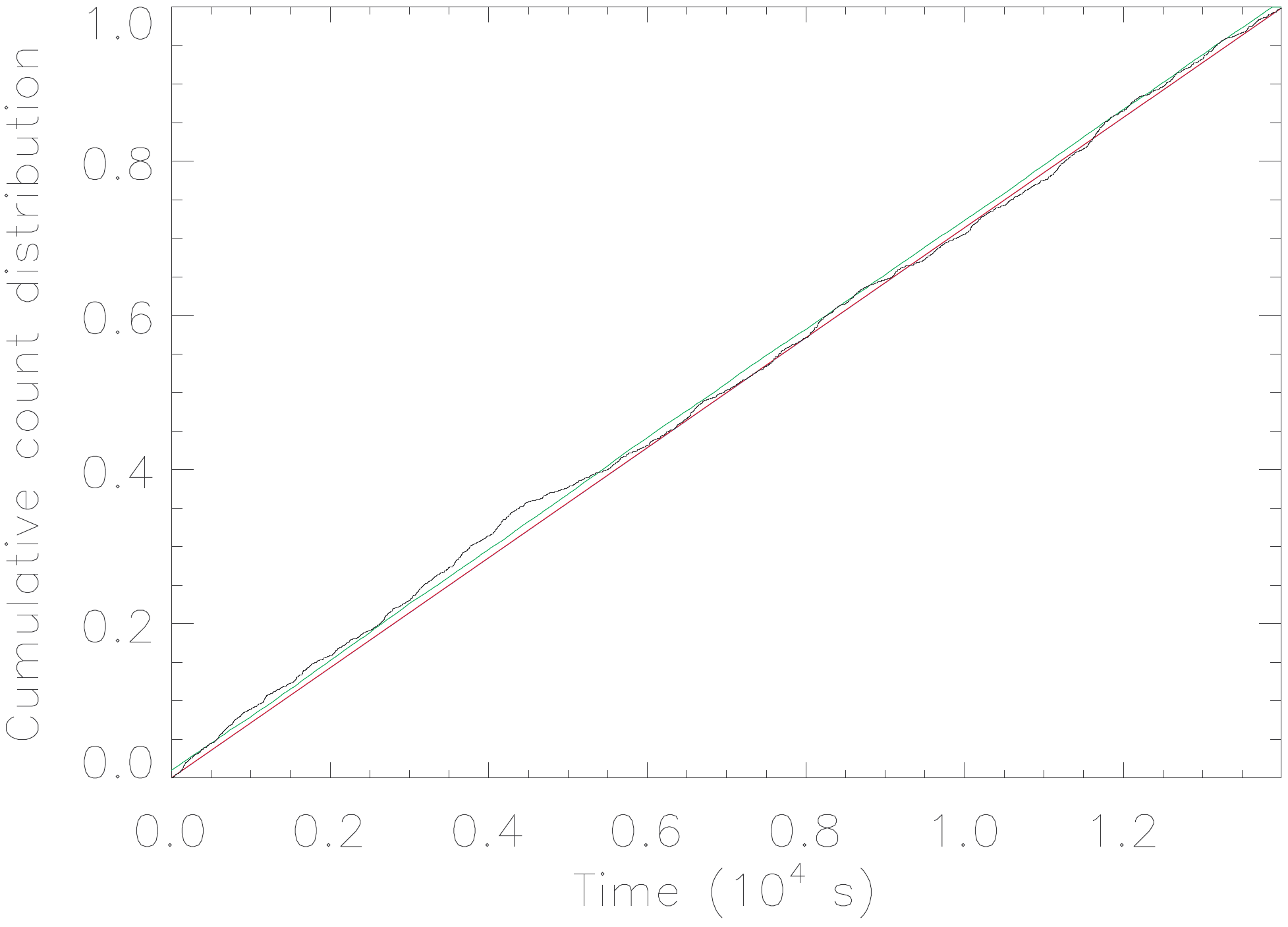}}
  \resizebox{\hsize}{!}{\includegraphics[angle=0,clip=]{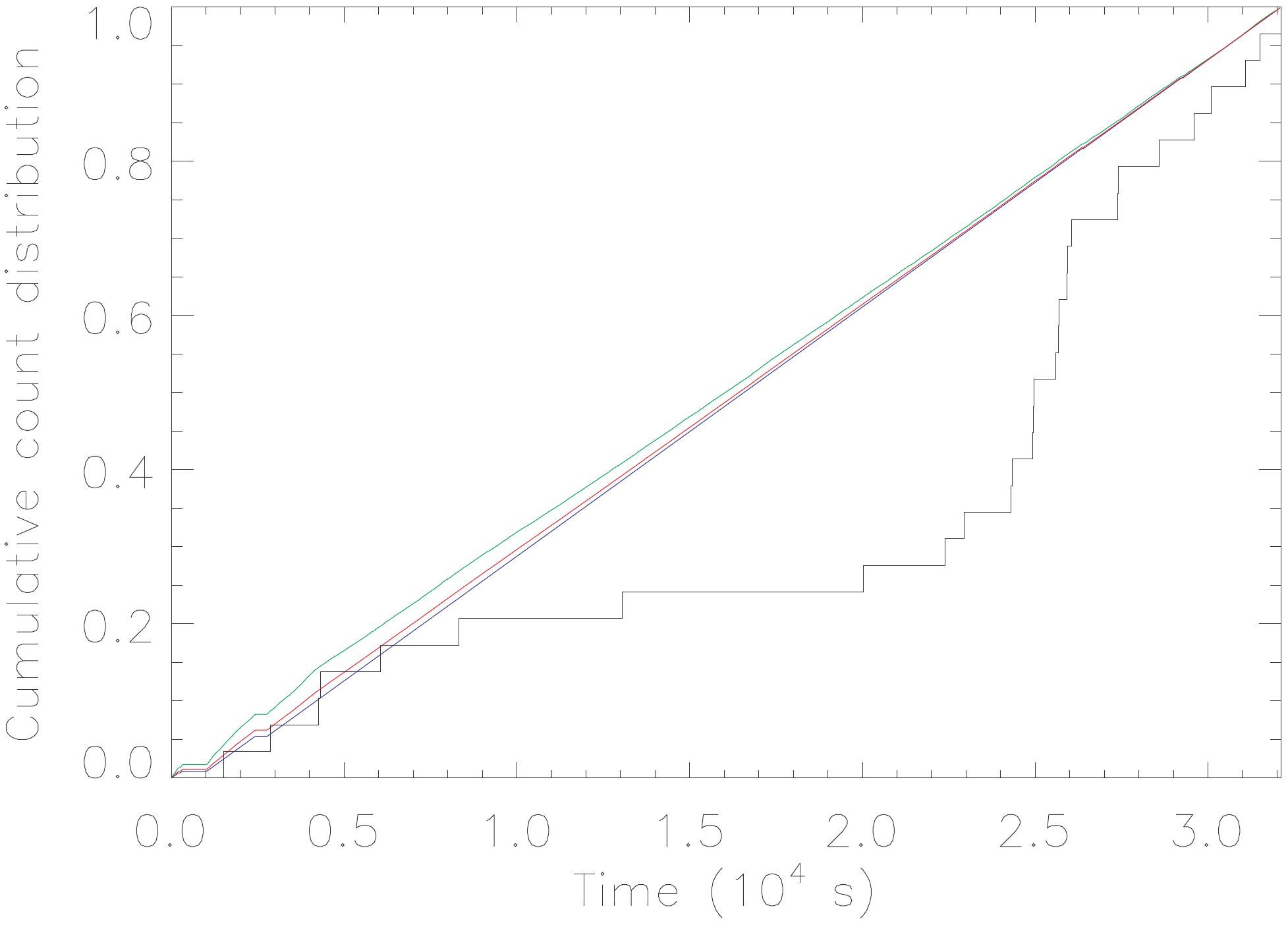}\includegraphics[angle=0,clip=]{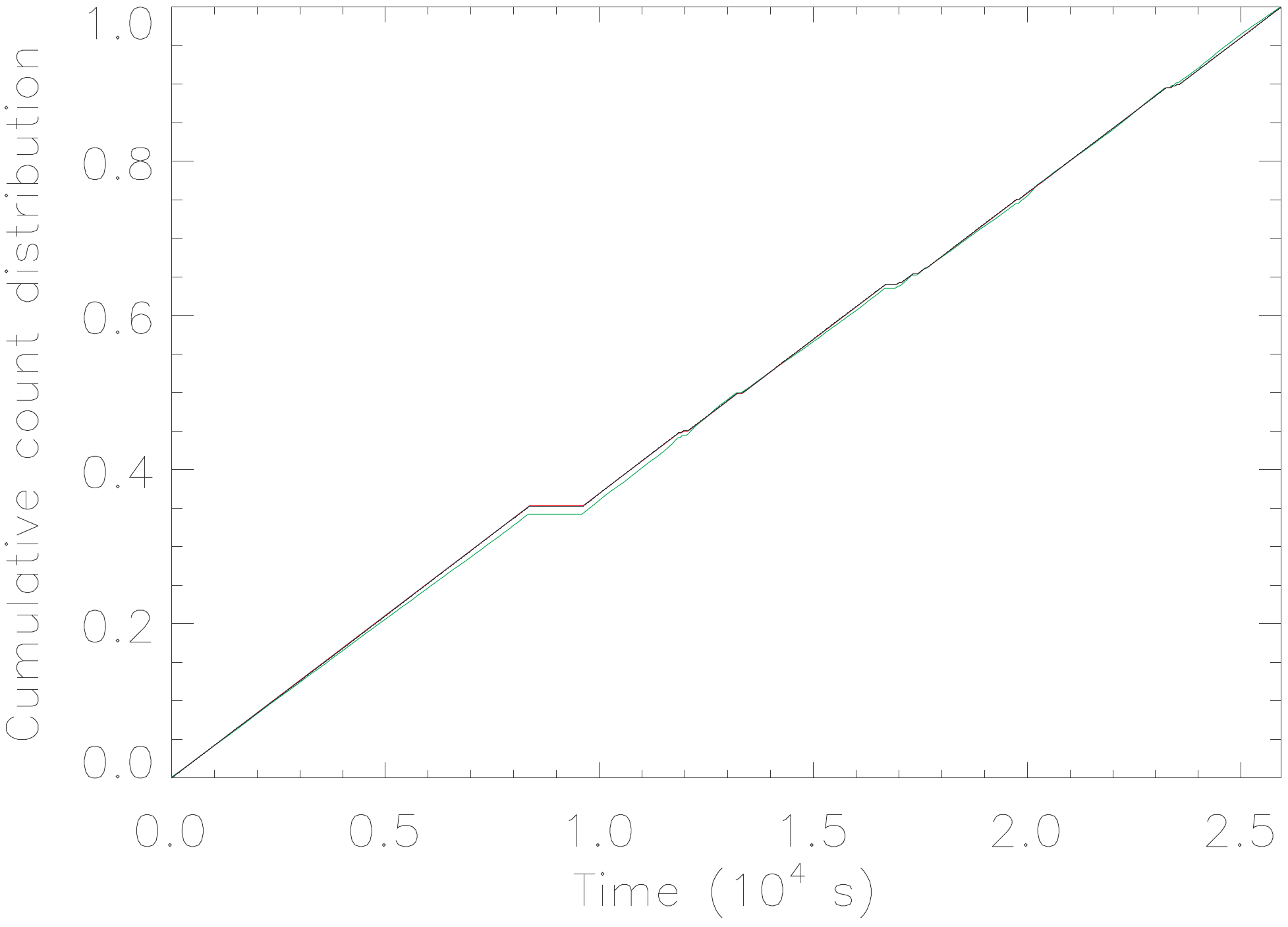}}

  \caption{Cumulative photon arrival time distributions as used for the KS tests. Source counts are plotted in black, the background light curve is shown in green.
           The distribution, expected from a constant source is shown in blue and the reference distribution in red. In case of a constant source and background light curve, all lines are blended.
           The source-count distribution of variable sources is significantly different from the expected distribution.
    {\it Upper left}: The bright foreground star \no{2041}.
    {\it Lower left}: The foreground star candidate \no{255}.
    {\it Upper right}: The HMXB RX\,J0054.9-7245 (\no{335}).
    {\it Lower right}: The SNR \calsrc.
  }
  \label{fig:Kstest}
\end{figure}

To estimate the variability of sources during the individual observations,
we used KS-tests as described in Sec.~\ref{sec:sttimevar}.
This allows us to estimate the source variability for sources with poor statistics.
Some examples of cumulative count distributions, as used for the KS-test, are presented in Fig.~\ref{fig:Kstest}.
Here, we give the example of a bright star (\no{2041}, upper left) showing a flare
and a variable HMXB (\no{335} = RX\,J0054.9-7245, upper right).
A foreground star candidate (\no{255}, lower left), detected with only 29 counts, also exhibits a flare.
As an example for a constant source, the SNR \calsrc\ is given in the lower right.
In the case of high variability, the photon time distribution of the source (black line)
shows a difference to the reference distribution for a constant source (red line), that is unlikely to be caused by statistical fluctuations.

\begin{figure}
  \resizebox{\hsize}{!}{\includegraphics[angle=0,clip=]{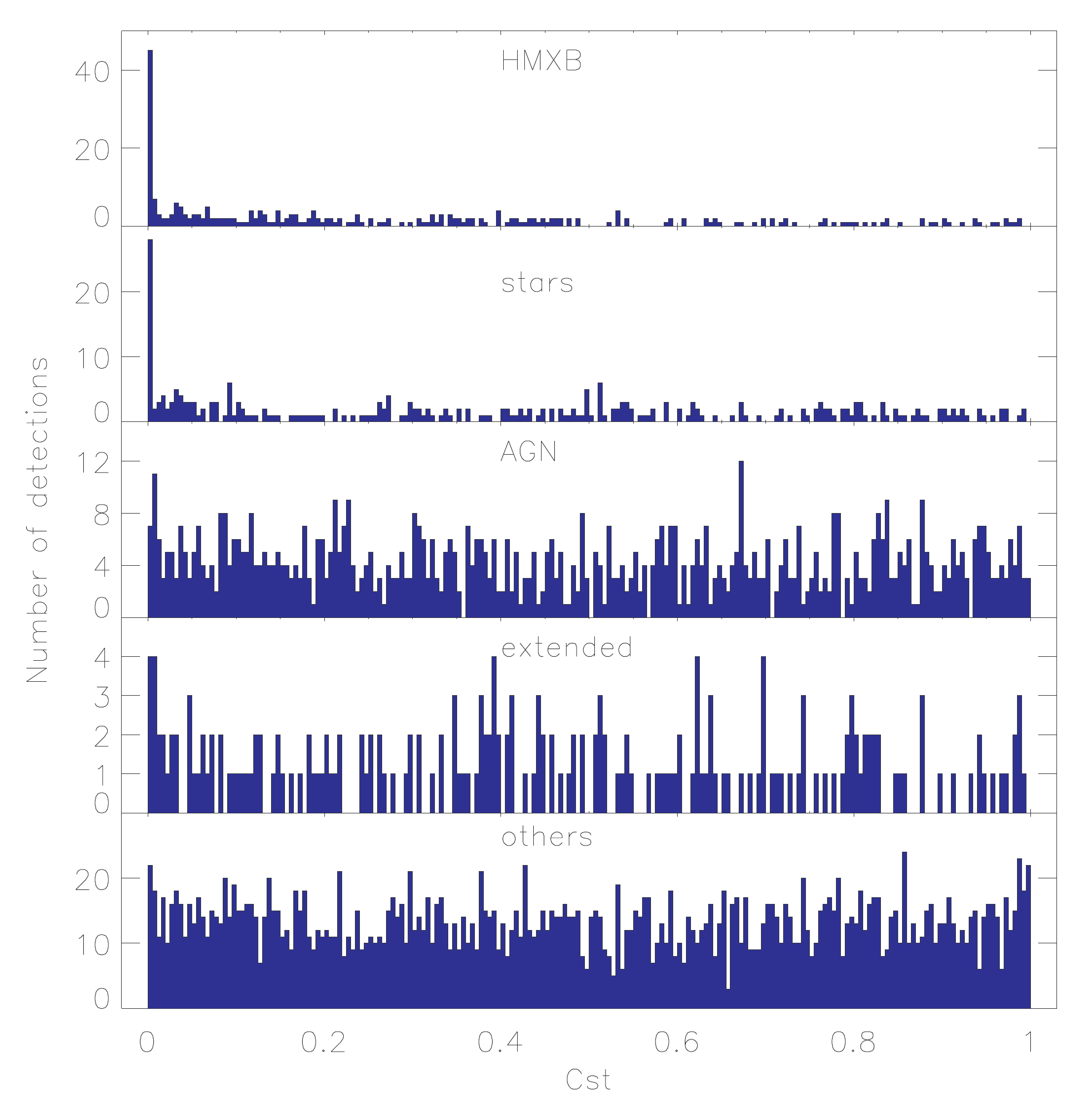}}
  \caption{
    Histogram of probability that the source flux was constant within the observation.
    The bin size is 0.005.
    Note the number of stars and HMXB showing short-term variability with $Cst<0.5\%$  caused by the flaring behaviour of these source classes.
  }
  \label{fig:Cst}
\end{figure}

The distribution of probabilities $Cst$ for constancy during individual observations is presented in Fig.~\ref{fig:Cst} for various source classes.
Here we see a uniform distribution, with the exception of stars and HMXBs.
These are expected to show  variability, whereas extended sources can be assumed to have a constant X-ray luminosity on short time scales.
All 89 sources with $Cst<0.5\%$ are listed in Table~\ref{tab:var_short}.
Assuming a uniform distribution we would expect $\sim$15 catalogue sources to be found with $Cst<0.5\%$ by chance.
Whereas 15.4\% of all HMXB detections and 9.3\% of all foreground-star detections have $Cst<0.5\%$,
this occurs only for 0.8\% of the remaining detections.

\begin{table*}
\caption[]{X-ray sources in the SMC field with significant short-term variability.}
\begin{center}
\begin{tabular}{rrrcrr|rrrcrr}
\hline\hline\noalign{\smallskip}
\multicolumn{1}{l}{SRC} &
\multicolumn{1}{c}{RA} &
\multicolumn{1}{c}{Dec} &
\multicolumn{1}{c}{Class} &
\multicolumn{1}{c}{$\log{(Cst)}$\tablefootmark{a}} &
\multicolumn{1}{c|}{$N$\tablefootmark{b}} &
\multicolumn{1}{l}{SRC} &
\multicolumn{1}{c}{RA} &
\multicolumn{1}{c}{Dec} &
\multicolumn{1}{c}{Class} &
\multicolumn{1}{c}{$\log{(Cst)}$\tablefootmark{a}} &
\multicolumn{1}{c}{$N$\tablefootmark{b}}\\
\noalign{\smallskip}\hline\noalign{\smallskip}

1  &  01 17 05.2   &  -73 26 36 & HMXB & $<$-40 & 316817  &        851  &  01 10 57.6   &  -73 05 14 & $<$fg-star$>$ & -13.06 & 290  \\
2  &  01 18 38.0   &  -73 25 27 & fg-star & -39.36 & 13873  &      852  &  01 16 33.5   &  -72 59 49 & AGN & -5.51 & 1008  \\
7  &  01 19 39.4   &  -73 27 33 & $<$fg-star$>$ & -3.63 & 171  &       884  &  01 00 13.9   &  -73 07 25 & $<$fg-star$>$ & -17.02 & 2022  \\
9  &  01 16 27.9   &  -73 14 47 & $<$AGN$>$ o? & -3.08 & 153  &        885  &  01 00 37.2   &  -73 00 36 & $<$fg-star$>$ & $<$-40 & 1067  \\
37  &  01 14 55.6   &  -73 15 33 & $<$fg-star$>$ & -2.58 & 280  &      888  &  01 00 39.2   &  -73 16 55 &   & -9.01 & 211  \\
51  &  01 01 20.7   &  -72 11 19 & HMXB & -3.73 & 1081  &          905  &  00 55 02.1   &  -73 21 15 & $<$fg-star$>$ & -15.30 & 290  \\
61  &  00 55 35.4   &  -72 29 07 & HMXB & -4.67 & 505  &           923  &  01 17 50.1   &  -73 30 09 &   & -4.18 & 55  \\
63  &  00 54 56.3   &  -72 26 47 & HMXB & -3.37 & 463  &           934  &  01 16 35.0   &  -73 37 43 & $<$AGN$>$ o? & -2.97 & 10  \\
66  &  00 56 18.9   &  -72 28 03 & $<$HMXB$>$ & -2.70 & 214  &         937  &  01 13 53.2   &  -73 27 08 & $<$ClG$>$ & -2.38 & 119  \\
92  &  00 54 46.3   &  -72 25 23 & HMXB & -8.26 & 1160  &          948  &  01 16 21.7   &  -73 29 32 &   & -13.49 & 33  \\
112  &  00 47 23.3   &  -73 12 28 & HMXB & -31.48 & 7094  &        1032  &  00 59 14.3   &  -72 20 12 & $<$fg-star$>$ & -4.87 & 21  \\
113  &  00 51 52.2   &  -73 10 34 & HMXB & -6.92 & 10189  &        1041  &  00 53 07.8   &  -74 39 06 & fg-star & $<$-40 & 241087  \\
137  &  00 52 15.1   &  -73 19 16 & $<$HMXB$>$ & -4.97 & 387  &        1194  &  01 05 46.8   &  -72 02 34 & $<$AGN$>$ o? & -8.93 & 63  \\
146  &  00 48 59.2   &  -72 58 17 & $<$fg-star$>$ & -13.08 & 214  &    1199  &  01 02 23.2   &  -72 11 35 & $<$fg-star$>$ & -2.59 & 11  \\
149  &  00 57 49.4   &  -72 02 36 & HMXB & $<$-40 & 6024  &        1399  &  00 55 18.3   &  -72 38 52 & HMXB & -5.35 & 1551  \\
165  &  00 59 40.8   &  -72 19 05 &   & -2.47 & 22  &              1462  &  00 52 03.0   &  -72 05 05 & $<$AGN$>$ o? & -2.40 & 219  \\
183  &  00 58 58.7   &  -72 01 12 & $<$fg-star$>$ & -3.36 & 20  &      1481  &  00 42 07.8   &  -73 45 03 & $<$HMXB$>$ & -6.64 & 315  \\
184  &  01 01 52.3   &  -72 23 33 & HMXB & -6.16 & 999  &          1483  &  00 41 55.3   &  -73 34 16 & $<$AGN$>$ o? & -2.35 & 131  \\
187  &  01 03 14.0   &  -72 09 14 & HMXB & -2.63 & 1322  &         1500  &  00 42 21.2   &  -73 27 58 & $<$fg-star$>$ & -8.80 & 29  \\
197  &  01 05 09.9   &  -72 11 47 & $<$AGN$>$ ox & -3.20 & 189  &      1525  &  00 44 18.5   &  -73 37 00 & $<$AGN$>$ o? & -2.50 & 28  \\
216  &  01 02 06.7   &  -71 41 16 & HMXB & -7.85 & 1948  &         1560  &  00 43 02.2   &  -72 52 34 & $<$AGN$>$ o? & -4.48 & 428  \\
227  &  01 03 37.6   &  -72 01 33 & HMXB & -9.15 & 7426  &         1617  &  01 02 08.0   &  -72 47 32 & $<$AGN$>$ o? & -2.51 & 19  \\
228  &  01 05 55.4   &  -72 03 49 & HMXB & -3.57 & 235  &          1700  &  00 50 05.6   &  -73 32 01 & $<$AGN$>$ o? & -2.31 & 30  \\
231  &  01 01 37.7   &  -71 54 15 & $<$fg-star$>$ & -8.88 & 199  &     1702  &  00 49 13.5   &  -73 25 60 & $<$fg-star$>$ & -2.51 & 21  \\
239  &  01 05 50.3   &  -71 57 60 & $<$fg-star$>$ & -2.74 & 65  &      1755  &  01 03 42.6   &  -71 57 58 &   & -17.86 & 36  \\
255  &  01 05 00.3   &  -72 11 48 & $<$fg-star$>$ & -4.26 & 29  &      1802  &  00 52 05.7   &  -72 26 05 & HMXB & -2.32 & 2592  \\
256  &  01 05 37.5   &  -71 57 19 &   & -2.34 & 22  &              1998  &  00 50 12.1   &  -71 46 54 & $<$fg-star$>$ & -4.35 & 28  \\
287  &  01 01 55.9   &  -72 10 28 & $<$HMXB$>$ & -2.95 & 28  &         2002  &  00 47 17.6   &  -71 55 20 & $<$AGN$>$ o? & -2.54 & 42  \\
335  &  00 54 55.9   &  -72 45 11 & HMXB & -3.58 & 3038  &         2041  &  01 03 16.5   &  -71 31 42 & $<$fg-star$>$ & $<$-40 & 1014  \\
352  &  00 53 24.2   &  -72 39 41 & $<$AGN$>$ o? & -2.32 & 10  &       2381  &  00 42 38.8   &  -72 33 27 & AGN r & -17.31 & 606  \\
392  &  00 52 58.5   &  -70 50 22 & $<$AGN$>$ o? & -2.55 & 19  &       2454  &  00 43 47.7   &  -73 02 08 & $<$AGN$>$ o? & -2.42 & 27  \\
402  &  01 09 30.0   &  -72 52 49 & $<$AGN$>$ o? & -2.94 & 580  &      2601  &  00 56 45.4   &  -72 59 32 &   & -5.50 & 154  \\
407  &  01 08 25.9   &  -72 54 31 & fg-star & -2.44 & 352  &       2651  &  00 54 50.8   &  -72 51 26 & $<$AGN$>$ o? & -4.05 & 14  \\
487  &  00 45 24.1   &  -73 29 07 & $<$fg-star$>$ & -4.39 & 451  &     2735  &  01 09 35.2   &  -72 11 45 & fg-star & $<$-40 & 708  \\
542  &  00 59 29.5   &  -71 58 09 & $<$AGN$>$ oi & -2.66 & 31  &       2738  &  01 10 50.8   &  -72 10 25 & AGN r & -2.66 & 332  \\
556  &  01 02 14.7   &  -72 49 17 & $<$AGN$>$ oi & -2.90 & 23  &       2740  &  01 09 18.6   &  -72 12 38 & $<$fg-star$>$ & -3.49 & 204  \\
562  &  01 03 31.7   &  -73 01 44 & $<$HMXB$>$ & -2.49 & 19  &         2845  &  01 13 02.3   &  -72 41 42 & $<$fg-star$>$ & -9.22 & 524  \\
615  &  00 49 30.6   &  -73 31 09 & HMXB & -2.83 & 223  &          2846  &  01 11 54.6   &  -72 45 57 & $<$fg-star$>$ & -11.28 & 304  \\
636  &  00 48 46.6   &  -73 30 00 & $<$AGN$>$ o? & -2.62 & 16  &       3059  &  01 13 04.2   &  -73 14 35 & $<$fg-star$>$ & -9.03 & 103  \\
654  &  00 53 23.9   &  -72 27 15 & HMXB & -3.16 & 2149  &         3115  &  01 06 33.0   &  -73 15 43 & HMXB & -11.24 & 811  \\
668  &  00 54 38.5   &  -72 22 09 & $<$AGN$>$ o? & -2.52 & 51  &       3167  &  01 02 28.0   &  -73 16 57 & $<$AGN$>$ o? & -3.35 & 27  \\
674  &  00 59 28.9   &  -72 37 04 & HMXB & -14.16 & 6464  &        3186  &  00 59 37.0   &  -73 25 41 & $<$AGN$>$ o? & -2.31 & 127  \\
730  &  00 49 22.8   &  -72 10 55 &   & -2.64 & 81  &              3190  &  00 58 35.7   &  -73 14 48 & $<$fg-star$>$ & -2.64 & 79  \\
760  &  00 42 45.8   &  -73 10 14 & $<$fg-star$>$ & -2.79 & 173  &     3267  &  00 50 34.6   &  -73 30 30 & $<$AGN$>$ o? & -6.26 & 214  \\
812  &  01 00 09.6   &  -72 57 49 & $<$AGN$>$ o? & -15.07 & 1087  \\

\noalign{\smallskip}\hline
\end{tabular}
\tablefoot{
\tablefoottext{a}{Probability $Cst$ that the source is constant during the observation. Minimum of all detections of the source is given.}
\tablefoottext{b}{Number of source counts of the detections with the given value of $Cst$.}
}
\end{center}
\label{tab:var_short}
\end{table*}

\subsubsection{Inter-observational variability}

\begin{figure}
  \resizebox{\hsize}{!}{\includegraphics[angle=-90,clip=]{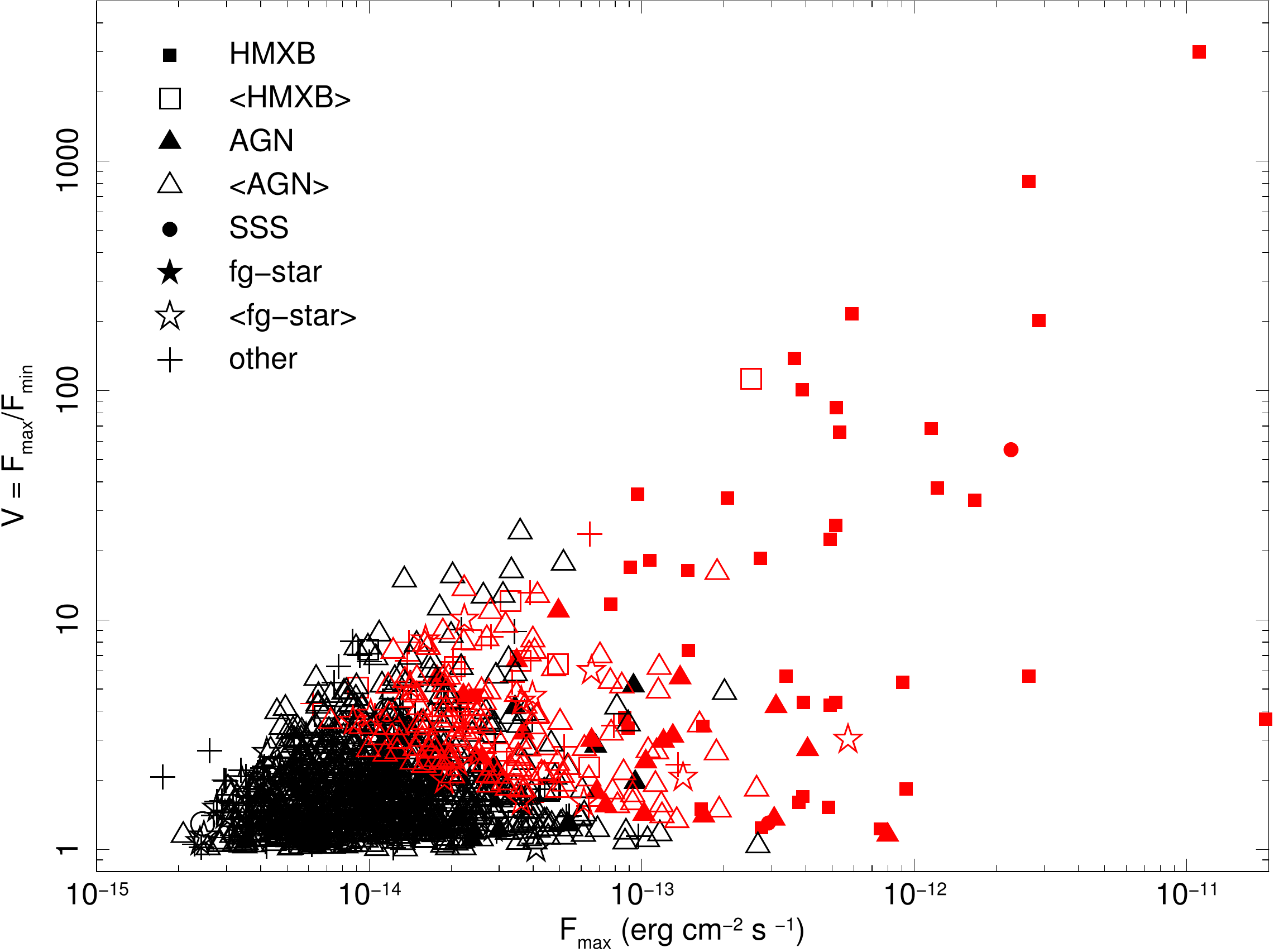}}
  \caption{
           Variability of SMC sources as observed with \xmm\ in the (0.2--4.5) keV band.
           Sources with a significance of variability greater (less) than 3 are plotted in red (black).
  }
  \label{fig:lttimevar}
\end{figure}

\begin{figure}
  \resizebox{\hsize}{!}{\includegraphics[angle=0,clip=]{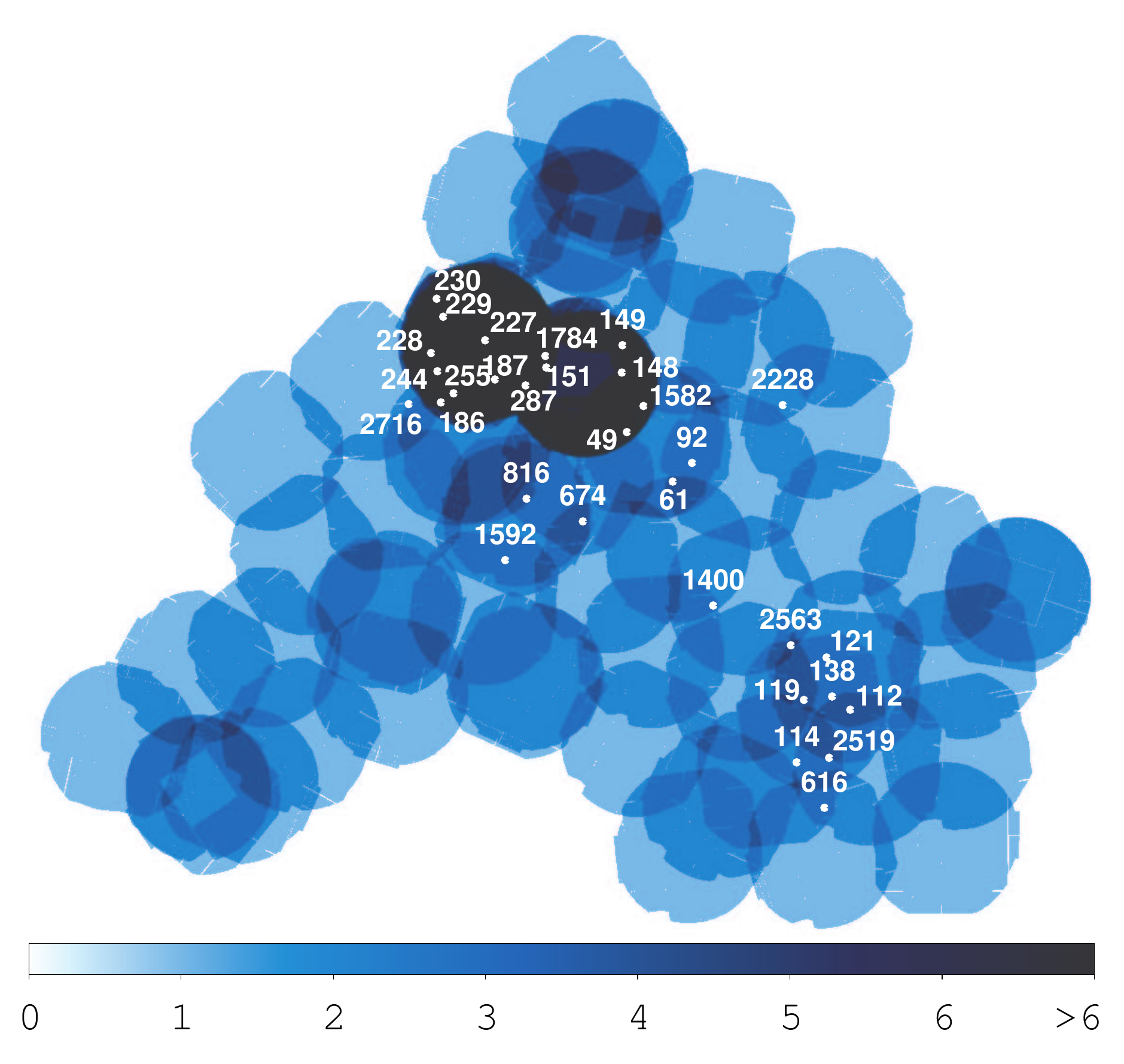}}
  \caption{
    The number of observations per field is compared with the distribution of significant long-term variable sources ($V\geq10$ and $S\geq3$), shown by white dots.
    Labels give the source numbers.  Variable sources are only found in the SMC bar in regions that have been observed several times.
  }
  \label{fig:summask}
\end{figure}

For sources which were observed several times, the long-term variability was calculated as described in Sec.~\ref{sec:cat:compilation}.
The dependence of variability on the maximal detected flux is plotted in Fig.~\ref{fig:lttimevar}.
For sources with $F_{\rm max} <10^{-14}$ erg cm$^{-2}$ s$^{-1}$, the variability is uncertain.
The calculation results in a significant ($S\geq3$) variability measurement for most sources with $F_{\rm max}$ above $10^{-13}$ erg cm$^{-2}$ s$^{-1}$.

Sources with high variability of $V\geq10$ are listed in Table~\ref{tab:var_long}.
As expected, most of these sources are HMXBs.
For two HMXB candidates, the high variability supports their classification.
In addition, we find the symbiotic nova SMC3, which has a known 1600 day variability,
one Galactic star, possibly observed during a flare in one observation, and seven more sources, unclassified or with AGN classification.
These sources show clear variability, which is rather high, but possible, for AGN
(e.g. \no{229} is identified as AGN).
Another explanation of such high variabilities might be given by an X-ray binary nature,
although we note that these sources do not have bright optical counterparts, needed for a HMXB classification.
In the case of low-mass X-ray binaries (LMXBs), we would not expect to find an optical counterpart.
Only very few LMXBs are expected in the SMC, as this population scales with the mass of the galaxy, and
so far none are known, although the X-ray variable sources might be considered as candidates.

The spatial distribution of the highly variable sources is presented in Fig.~\ref{fig:summask}.
Obviously it is more likely to find variable sources in fields which were observed more frequently.
Specifically, long-term variability cannot be measured in fields observed only once.
Since most variable sources are HMXBs, the distribution follows the bar of the SMC.

\begin{table*}
\caption[]{X-ray sources in the SMC field with high variability between individual observations.}
\begin{center}
\begin{tabular}{rccrrr}
\hline\hline\noalign{\smallskip}
\multicolumn{1}{l}{SRC} &
\multicolumn{1}{c}{Name} &
\multicolumn{1}{c}{Class} &
\multicolumn{1}{c}{$V$\tablefootmark{a}} &
\multicolumn{1}{c}{$S$\tablefootmark{a}} &
\multicolumn{1}{c}{$F_{\rm max}$} \\
\multicolumn{1}{l}{} &
\multicolumn{1}{c}{} &
\multicolumn{1}{c}{} &
\multicolumn{1}{c}{} &
\multicolumn{1}{c}{} &
\multicolumn{1}{c}{($10^{-14}\times$ erg cm$^{-2}$ s$^{-1}$)}\\
\noalign{\smallskip}\hline\noalign{\smallskip}
49   & SXP\,565                   & HMXB          &  216.4 &  6.5   & 59.22 $\pm$ 9.06  \\                 
61   & XMMU\,J005535.2-722906     & HMXB          &  18.6  &  25.0  & 27.33 $\pm$ 0.91  \\    
92   & CXOU\,J005446.2-722523     & HMXB          &  33.9  &  27.8  & 20.65 $\pm$ 0.70  \\   
112  & SXP\,264                   & HMXB          &  37.6  &  66.7  & 121.59 $\pm$ 1.74 \\               
114  & SXP\,756                   & HMXB          &  22.4  &  38.5  & 49.25 $\pm$ 1.09  \\                
119  & SXP\,892                   & HMXB          &  16.4  &  10.8  & 14.76 $\pm$ 1.25  \\                
121  & RX\,J0048.5-7302           & HMXB          &  16.9  &  12.0  & 9.08 $\pm$ 0.69   \\         
138  & SXP\,25.5                  & HMXB          &  18.2  &  14.3  & 10.70 $\pm$ 0.69  \\               
148  & SXP\,152                   & HMXB          &  65.9  &  37.9  & 53.32 $\pm$ 1.38  \\                
149  & SXP\,280                   & HMXB          &  33.3  &  51.4  & 167.22 $\pm$ 2.68 \\               
151  & SXP\,304                   & HMXB          &  25.9  &  20.2  & 51.52 $\pm$ 2.25  \\                
186  &                            & $<$AGN$>$ oxr &  16.1  &  14.6  & 18.90 $\pm$ 1.16  \\                 
187  & SXP\,348                   & HMXB          &  100.8 &  31.2  & 38.84 $\pm$ 1.22  \\               
227  & SXP\,1323                  & HMXB          &  68.2  &  63.1  & 115.24 $\pm$ 1.79 \\              
228  & RX\,J0105.9-7203           & HMXB          &  35.3  &  13.7  & 9.67 $\pm$ 0.68   \\         
229  & [VV2006]\,J010522.5-715650 & AGN           &  11.0  &  9.7   & 4.95 $\pm$ 0.42   \\ 
230  &                            &               &  23.7  &  9.3   & 6.46 $\pm$ 0.66   \\                           
244  &                            & $<$AGN$>$ o?  &  12.8  &  7.5   & 4.15 $\pm$ 0.50   \\                    
255  &                            & $<$fg-star$>$ &  10.0  &  6.7   & 2.23 $\pm$ 0.29   \\                   
287  &                            & $<$HMXB$>$    &  12.1  &  6.6   & 3.30 $\pm$ 0.45   \\                      
616  & SMC3                       & SSS           &  55.1  &  198.9 & 226.43 $\pm$ 1.08 \\                  
674  & XMMU\,J005929.0-723703     & HMXB          &  201.5 &  67.0  & 286.67 $\pm$ 4.23 \\ 
816  & SXP\,7.92                  & HMXB          &  11.7  &  6.9   & 7.71 $\pm$ 1.01   \\                 
1400 &                            & $<$HMXB$>$    &  112.4 &  10.3  & 25.21 $\pm$ 2.43  \\                  
1582 &                            & $<$AGN$>$ o?  &  10.9  &  3.8   & 2.79 $\pm$ 0.66   \\                   
1592 & SXP\,6.85                  & HMXB          &  2990.9&  167.0 & 1111.89 $\pm$ 6.65\\         
1784 &                            &               &  13.2  &  4.6   & 3.91 $\pm$ 0.78   \\                          
2228 & SXP\,91.1                  & HMXB          &  84.1  &  51.0  & 51.72 $\pm$ 0.98  \\              
2519 & SXP\,11.87                 & HMXB          &  815.5 &  99.5  & 263.96 $\pm$ 2.65 \\           
2563 & SXP\,214                   & HMXB          &  137.9 &  32.9  & 36.31 $\pm$ 1.09  \\              
2716 &                            & $<$AGN$>$ o?  &  13.7  &   3.6  & 2.23 $\pm$ 0.57   \\                   
\noalign{\smallskip}\hline
\end{tabular}
\tablefoot{\tablefoottext{a}{Variability $V$ and significance $S$ as calculated in Sec.~\ref{sec:cat:compilation}.}}
\end{center}
\label{tab:var_long}
\end{table*}

\section{Summary}
\label{sec:conclusions}

For the first time, the central field of the SMC is fully covered to a limiting flux of $\sim$$2\times10^{-14}$ erg cm$^{-2}$ s$^{-1}$ in the (0.2--12.0) keV band with an imaging X-ray telescope.
91 \xmm\ pointings between April 2000 and April 2010 cover the bar and eastern wing of the SMC with an area of 5.6 deg$^2$ and a total net exposure of $\sim$2 Ms.
We created a catalogue of 3053 unique X-ray point sources based on 5236 detections in the SMC field, including archival observations, providing spectral and temporal characteristics for all sources.
The typical positional uncertainty of the sources in the catalogue is 1.3\arcsec.
Cross correlations with other catalogues give distinct counterparts in X-ray and radio bands.
In the optical, the selection of counterparts is challenging, due to the high stellar density in the SMC field.
Most X-ray sources are background sources behind the SMC, namely AGN.
We were able to positively identify 49 HMXB and 4 SSSs in the SMC together with 34 foreground stars and 72 background AGN.
In addition we propose classifications for other sources based on their X-ray hardness ratios and positional cross-correlations with other catalogues, which often provided additional optical, infrared or radio characteristics. 
This resulted in likely identifications with foreground stars (128), HMXBs (45), faint SSSs (8) as well as AGN (2105) and galaxy clusters (13).
This has allowed us to further investigate the X-ray source population of the SMC, in particular to derive
the luminosity function of Be/X-ray binaries. This shows a significantly different bright end slope than expected from the universal HMXB X-ray luminosity function.

\begin{acknowledgements}
This publication is based on observations with XMM-Newton, an ESA Science Mission with instruments and contributions
directly funded by ESA Member states and the USA (NASA).
The XMM-Newton project is supported by the Bundesministerium f\"ur Wirtschaft und
Technologie/Deutsches Zentrum f\"ur Luft- und Raumfahrt (BMWI/DLR, FKZ 50 OX 0001)
and the Max-Planck Society.
N. La Palombara and A. Tiengo acknowledge financial
contributions by the Italian Space Agency through ASI/INAF agreements I/009/10/0 and I/032/10/0 for the data
analysis and the XMM-Newton operations, respectively.
R. Sturm acknowledges support from the BMWI/DLR grant FKZ 50 OR 0907.

\end{acknowledgements}

\bibliographystyle{aa}
\bibliography{../auto,../general}

\begin{appendix}
\section{Details of the catalogue creation}
\label{sec:cat:details}

\subsection{Catalogue screening}
\label{sec:cat:screening}

Each observation was screened individually and a quality flag {\tt QFLAG} was set manually
to indicate if the detection was most likely not caused by a point source.
Spurious detections can be caused by single reflections of SMC X-1 ({\tt QFLAG=S}),
or by out-of-time events of bright sources ({\tt QFLAG=O}).
The substructure of extended sources, 
or residuals of the PSF of the brightest sources, 
can lead to multiple detections ({\tt QFLAG=M}).
Also, sources with significant extent,  like SNRs and the largest clusters of galaxies,  were marked ({\tt QFLAG=E}) and not used for the catalogue.

Using a mosaic colour image of the SMC \citep[][]{2012A&A...545A.128H},
we looked, in a second screening-step, for wrongly correlated detections and uncorrelated detections obviously originating from the same source.
This occurred, for example, in the case of bright sources with low statistical uncertainty, if the astrometric solution was not yet sufficiently accurate.
The EPIC-MOS CCDs show a significantly increased low-energy noise in some observations, if they are in an  {\it anomalous state} \citep{2008A&A...478..575K}.
Affected CCDs were identified with {\tt emtaglenoise} and the data were used for source detection, but not for the mosaic image.
To ensure that the detections in these noisy CCDs are real,
we required that the detection likelihood for the noisy CCD in the (1.0--12.0) keV band is sufficient for an independent detection,
or that the source is found by an instrument not in {\it anomalous state} within the same, or another, observation.
If, in addition, no source is visible in the mosaic image, the detection was flagged with {\tt QFLAG=N}.
Detections found in regions with enhanced diffuse emission, where the reliability is doubtful, were flagged ({\tt QFLAG=D})
if no clear point source was visible in the deep mosaic image.
The task {\tt epreject} abolishes EPIC-pn offset map corrections for optical loading
caused by bright optical sources with $V<12$ mag or $V<6$ mag for observations with thin or medium filter,
respectively.\footnote{M. J. S. Smith 2008, PN optical loading, XMM-SOC-CAL-TN-0051, http://xmm2.esac.esa.int/docs/documents/CAL-TN-0051-1-2.ps.gz}
We find 57 detections of 36 sources, which fulfil this criterion.
If the source was not detected with a high level of significance by another instrument, in another observation, or in the EPIC-pn energy bands 2$-$5,
we rejected this source ({\tt QFLAG=L}). This caused a rejection of only 4 sources. All other 32 potentially affected sources show evidence for real X-ray emission.
Sources, detected only in one observation and by only one instrument with detection likelihood $ML_{\rm det}>10$ in energy band 1, but not detected above 1 keV,
were checked for hot pixels, which might have been missed by {\tt badpixfind} and flagged ({\tt QFLAG=P})
if a bright pixel was found in the detector image or
if the detection was close to a hot pixel or column and had a peculiar, non-PSF-like, shape.
All other detections have {\tt QFLAG=G}, by default.
Only the G-flagged sources  were used for the creation of the point-source catalogue.
The flags and the number of occurrences are listed in Table~\ref{tab:screening}.

\begin{table}
\caption[]{Screening of detections.}
\begin{center}
\begin{tabular}{lcrrrrr}
\hline\hline\noalign{\smallskip}
\multicolumn{1}{l}{QFLAG} &
\multicolumn{1}{c}{Description} &
\multicolumn{1}{c}{Number} \\
\noalign{\smallskip}\hline\noalign{\smallskip}
G        & good (default)                 & 5236  \\
M        & multiple detection             & 503   \\
S        & single reflection              & 232   \\
E        & extended source                & 207   \\
P        & hot pixel                      & 129   \\
N        & MOS CCD in anomalous state     & 110   \\
O        & out-of-time events             & 22    \\
D        & diffuse emission               & 20    \\
L        & optical loading                & 4     \\
\noalign{\smallskip}\hline
\end{tabular}
\end{center}
\label{tab:screening}
\end{table}

\subsection{Astrometric corrections}
\label{sec:astcor}

The accuracy of the astrometric frame of \xmm\ can be improved
since the positions of bright sources have a higher statistical precision than the initial \xmm\ attitude solution.
However, the standard boresight correction, using a simple comparison with a complete optical reference catalogue, cannot be applied,
because the number of chance correlations overwhelms the number of real counterparts in the SMC field.
A first correction was applied to the event files, before the image creation \citep[][]{2012A&A...545A.128H}.
Using the autocorrelated detection list, we further improved the positions.
We selected optical counterparts, mainly from \citet{2002AJ....123..855Z},
of identified HMXB and spectroscopically confirmed AGN,
as well as identified foreground stars from the Tycho-2 catalogue \citep{2000A&A...355L..27H}, as described in Sec.~\ref{sec:class}.
In the latter case, the proper motion was taken into account.
We used this information to correct the spacecraft attitude for a linear boresight shift in right ascension and declination.
The overall applied boresight corrections for all observations are listed in Table~\ref{tab:observations}.
As we only used identified sources instead of a general correlation with a reference catalogue, we also accepted coordinate corrections if only one identified source was available.
This allowed a coordinate improvement for all observations.

\subsection{Calculation of energy conversion factors}
\label{sec:ecf}

For the calculation of energy conversion factors (ECFs) $f_i=R_i / F_i$, we assumed a universal spectrum for all sources, described by a power-law model with a photon index of $\Gamma=1.7$
and a photo-electric foreground absorption by the Galaxy of $N_{\rm H, Gal} = 6\times 10^{20}$ cm$^{-2}$ \citep[average for SMC main field in \Hone\ map of ][]{1990ARA&A..28..215D}.
This spectrum was simulated  with standard EPIC response matrix files
(RMFs)\footnote{available at\\ http://xmm2.esac.esa.int/external/xmm\_sw\_cal/calib/epic\_files.shtml}.
We note that this universal spectral shape is only a rough assumption.
Since the correction is from counts to detected flux, i.e. we do not report unabsorbed fluxes, and
since the fluxes are calculated independently for each of the relatively narrow energy bands, deviations from the source spectrum over the total energy band are reduced.
Therefore the calculated fluxes give a good approximation of the true detected flux in most cases \citep[see also ][]{2004A&A...426...11P}.

For EPIC-MOS, an increased redistribution of measured photon energies is known to occur around the EPIC-pn and RGS prime pointing position, where most targets are placed \citep{2006ESASP.604..925R}.
This has some effect on the ECFs for the energy bands 1 and 2.
We decided to use the {\it off-patch} single to quadruple event RMF, since in the case of the SMC survey, the majority of detections lies outside this patch area.
The differences between {\it on-} and {\it off-patch} ECFs is $<$10\% for energy band 1 and lower for the other bands.
For EPIC-pn, the dependence of the spectral resolution on the detector position has an even smaller influence on the ECFs ($<$2\%)
and we used the RMF for CCD rows 81-100 as average.
The dependence of the ECFs on data mode is $<$2\% \citep{2009A&A...493..339W} and we used the {\it full-frame} RMFs for all instruments.
The ancillary response files (ARFs) were calculated with {\tt arfgen} for each filter and instrument and do not contain corrections,
which have already been applied by {\tt emldetect}. For EPIC-pn, $f_5$ is corrected for the screened (7.2--9.2) keV sub-band, thus translating to (4.5--12.0) keV fluxes.
The derived ECFs are listed in Table~\ref{tab:ecf}. Since energy band 5 does not contribute substantially to the total flux,
but rather increases its uncertainty in most cases, it is not used for the so-called XID flux (sum from band 1 to 4).

\begin{table}
\caption[]{Energy conversion factors.}
\begin{center}
\begin{tabular}{lcrrrrr}
\hline\hline\noalign{\smallskip}
\multicolumn{1}{l}{Detector} &
\multicolumn{1}{c}{Filter} &
\multicolumn{1}{c}{$f_1$} &
\multicolumn{1}{c}{$f_2$} &
\multicolumn{1}{c}{$f_3$} &
\multicolumn{1}{c}{$f_4$} &
\multicolumn{1}{c}{$f_5$} \\
\noalign{\smallskip}\hline\noalign{\smallskip}
pn      & thin    & 11.150  & 8.132 & 5.839 & 1.943 & 0.463  \\
        & medium  &  9.976  & 7.897 & 5.758 & 1.926 & 0.465  \\
        & thick   &  6.329  & 6.071 & 4.964 & 1.824 & 0.459  \\
MOS1    & thin    &  2.118  & 1.937 & 2.064 & 0.744 & 0.144  \\
        & medium  &  1.891  & 1.883 & 2.029 & 0.735 & 0.144  \\
        & thick   &  1.273  & 1.530 & 1.807 & 0.705 & 0.142  \\
MOS2    & thin    &  2.117  & 1.938 & 2.070 & 0.748 & 0.152  \\
        & medium  &  1.886  & 1.884 & 2.034 & 0.740 & 0.152  \\
        & thick   &  1.264  & 1.530 & 1.812 & 0.710 & 0.150  \\
\noalign{\smallskip}\hline
\end{tabular}
\tablefoot{
Energy conversion factors $f_i$ in units of 10$^{11}$ cts cm$^{2}$ erg$^{-1}$ for the standard bands 1$-$5, respectively:
(0.2$-$0.5) keV,
(0.5$-$1.0) keV,
(1.0$-$2.0) keV,
(2.0$-$4.5) keV, and
(4.5$-$12.0) keV.
A power law with photon index of $\Gamma=1.7$ and foreground absorption of $N_{\rm H, Gal} = 6\times 10^{20}$ cm$^{-2}$ was used.
}
\end{center}
\label{tab:ecf}\end{table}

\subsection{Short-term time variability}
\label{sec:sttimevar}

To investigate the short-term time variability of all sources with KS-tests,
we extracted time series of events of the EPIC-pn and both EPIC-MOS in the (0.2--4.5) keV band.
Source counts were selected within an ellipse that approximates the PSF of the source at a surface brightness 
(in cts pix$^{-1}$) equal to the background surface brightness, as defined by the SAS task {\tt region}.
We merged the event lists in common GTIs of the individual EPIC instruments to obtain higher statistics.
To estimate the background variability caused, for example, by residual soft-proton flares,
we assumed a spatial independence of the background time variability.
Background time series were extracted in a similar manner,
but excluding regions around each point source where the PSF brightness is larger than 10\% of the background value.

In most cases, the KS-test works well by using
a linear increasing function during GTIs and a constant otherwise.
In some cases, the background time series exhibits significant variability.
To ensure that this has no influence on the variability estimation of faint sources,
we created a reference function, combining the background and constant source distribution.
The expected relative background contribution was estimated from the background maps (Sec.~\ref{sec:cat:detection}).
Since the statistics of the background time series is high compared to that of faint sources, it can be used as a quasi-continuous function.
We added this cumulative light curve to the constant-source function and performed a one-sample KS-test.
The resulting probability, $Cst$, that the temporal photon distribution can be explained by a constant source was calculated for each detection.
For sources with several detections, we give the minimum of $Cst$ over all detections.
Sources that can be considered variable show values of $Cst<0.5\%$ (see Sec.\ref{sec:discussion:variabilityshort}).

\end{appendix}

\begin{appendix}
\section{List of used observations}

\longtabL{1}{
\begin{landscape}
\begin{longtable}{rclr@{\,}c@{\,}lr@{}r@{}rr@{}r@{}lrrcr@{:}r@{:}lccrr}
\caption{\label{tab:observations} \xmm\ observations of the SMC.}\\
\hline\hline
\multicolumn{1}{c}{ID}&
\multicolumn{1}{c}{ObsID}&
\multicolumn{1}{c}{Target}&
\multicolumn{3}{c}{Date}&
\multicolumn{3}{c}{RA}&
\multicolumn{3}{c}{Dec}&
\multicolumn{1}{c}{$\Delta$RA}&
\multicolumn{1}{c}{$\Delta$Dec}&
\multicolumn{1}{c}{ExpID}&
\multicolumn{3}{c}{Time}&
\multicolumn{1}{c}{Filter}&
\multicolumn{1}{c}{Mode}&
\multicolumn{1}{c}{Exp}&
\multicolumn{1}{c}{Net Exp}\\
\multicolumn{1}{c}{}&
\multicolumn{1}{c}{}&
\multicolumn{1}{c}{}&
\multicolumn{3}{c}{}&
\multicolumn{3}{c}{(J2000)}&
\multicolumn{3}{c}{(J2000)}&
\multicolumn{1}{c}{(\arcsec)}&
\multicolumn{1}{c}{(\arcsec)}&
\multicolumn{1}{c}{}&
\multicolumn{3}{c}{(UT)}&
\multicolumn{1}{c}{}&
\multicolumn{1}{c}{}&
\multicolumn{1}{c}{(s)}&
\multicolumn{1}{c}{(s)}\\
\hline
\endfirsthead
\caption{Continued.} \\
\hline\noalign{\smallskip}
\multicolumn{1}{c}{ID}&
\multicolumn{1}{c}{ObsID}&
\multicolumn{1}{c}{Target}&
\multicolumn{3}{c}{Date}&
\multicolumn{3}{c}{RA}&
\multicolumn{3}{c}{Dec}&
\multicolumn{1}{c}{$\Delta$RA}&
\multicolumn{1}{c}{$\Delta$Dec}&
\multicolumn{1}{c}{ExpID}&
\multicolumn{3}{c}{Time}&
\multicolumn{1}{c}{Filter}&
\multicolumn{1}{c}{Mode}&
\multicolumn{1}{c}{Exp}&
\multicolumn{1}{c}{Net Exp}\\
\hline
\endhead
\hline
\endfoot
   S1 & 0601210101 & SMC survey 1 & 2009 & May & 14 & 00\rahour & 58\ramin & 16\fs4 & -71\degr & 28\arcmin & 45\arcsec & 1.42 & 0.85 & M1S001 & 09 & 14 & 24 & medium & ff & 29419 & 4280\\
   &  &  &  &  &  &  &  &  &  &  &  &  &  & M2S002 & 09 & 14 & 26 & medium & ff & 29425 & 4284\\
   &  &  &  &  &  &  &  &  &  &  &  &  &  & PNS003 & 09 & 36 & 46 & thin & ff & 27512 & 3551\\
   S2 & 0601210201 & SMC survey 2 & 2009 & Sep. & 25 & 00\rahour & 53\ramin & 58\fs9 & -71\degr & 43\arcmin & 48\arcsec & -0.91 & 0.1 & M1S001 & 00 & 15 & 42 & medium & ff & 37420 & 37362\\
   &  &  &  &  &  &  &  &  &  &  &  &  &  & M2S002 & 00 & 15 & 42 & medium & ff & 37425 & 37368\\
   &  &  &  &  &  &  &  &  &  &  &  &  &  & PNS003 & 00 & 38 & 05 & thin & ff & 35834 & 35782\\
   S3 & 0601210301 & SMC survey 3 & 2009 & May & 18 & 00\rahour & 49\ramin & 01\fs8 & -71\degr & 56\arcmin & 42\arcsec & 0.53 & -0.25 & M1U002 & 10 & 29 & 15 & medium & ff & 31614 & 27589\\
   &  &  &  &  &  &  &  &  &  &  &  &  &  & M2U002 & 10 & 29 & 34 & medium & ff & 31603 & 27580\\
   &  &  &  &  &  &  &  &  &  &  &  &  &  & PNU002 & 10 & 47 & 01 & thin & ff & 30304 & 26332\\
   S4 & 0601210401 & SMC survey 4 & 2009 & Sep. & 25 & 01\rahour & 02\ramin & 59\fs3 & -71\degr & 36\arcmin & 41\arcsec & -1.27 & 1.63 & M1S001 & 11 & 22 & 24 & medium & ff & 37521 & 37464\\
   &  &  &  &  &  &  &  &  &  &  &  &  &  & M2S002 & 11 & 22 & 23 & medium & ff & 37526 & 37469\\
   &  &  &  &  &  &  &  &  &  &  &  &  &  & PNS003 & 11 & 44 & 46 & thin & ff & 35934 & 35882\\
   S5 & 0601210501 & SMC survey 5 & 2009 & Sep. & 25 & 00\rahour & 58\ramin & 54\fs9 & -71\degr & 49\arcmin & 48\arcsec & 0.47 & -0.37 & M1S001 & 22 & 30 & 43 & medium & ff & 50321 & 41464\\
   &  &  &  &  &  &  &  &  &  &  &  &  &  & M2S002 & 22 & 30 & 43 & medium & ff & 50326 & 41469\\
   &  &  &  &  &  &  &  &  &  &  &  &  &  & PNS003 & 22 & 53 & 05 & thin & ff & 48734 & 39923\\
   S6 & 0601210601 & SMC survey 6 & 2009 & Sep. & 27 & 00\rahour & 55\ramin & 20\fs2 & -72\degr & 01\arcmin & 42\arcsec & -0.0 & 0.92 & M1S001 & 00 & 10 & 53 & medium & ff & 38520 & 34617\\
   &  &  &  &  &  &  &  &  &  &  &  &  &  & M2S002 & 00 & 10 & 53 & medium & ff & 38525 & 34622\\
   &  &  &  &  &  &  &  &  &  &  &  &  &  & PNS003 & 00 & 33 & 16 & thin & ff & 36933 & 34278\\
   S7 & 0601210701 & SMC survey 7 & 2009 & Sep. & 27 & 00\rahour & 49\ramin & 28\fs7 & -72\degr & 17\arcmin & 14\arcsec & 0.78 & 0.82 & M1S001 & 11 & 35 & 54 & medium & ff & 38621 & 38616\\
   &  &  &  &  &  &  &  &  &  &  &  &  &  & M2S002 & 11 & 35 & 53 & medium & ff & 38626 & 38621\\
   &  &  &  &  &  &  &  &  &  &  &  &  &  & PNS003 & 11 & 58 & 16 & thin & ff & 37034 & 37034\\
   S8 & 0601210801 & SMC survey 8 & 2009 & Oct. & 9 & 00\rahour & 56\ramin & 15\fs5 & -72\degr & 21\arcmin & 55\arcsec & 1.16 & -0.34 & M1S001 & 18 & 33 & 30 & medium & ff & 24620 & 24615\\
   &  &  &  &  &  &  &  &  &  &  &  &  &  & M2S002 & 18 & 33 & 31 & medium & ff & 24625 & 24620\\
   &  &  &  &  &  &  &  &  &  &  &  &  &  & PNS003 & 18 & 55 & 52 & thin & ff & 23034 & 23034\\
   S9 & 0601210901 & SMC survey 9 & 2009 & Sep. & 27 & 00\rahour & 48\ramin & 08\fs9 & -72\degr & 37\arcmin & 39\arcsec & 1.12 & 1.0 & M1S001 & 23 & 02 & 34 & medium & ff & 41138 & 34083\\
   &  &  &  &  &  &  &  &  &  &  &  &  &  & M2S002 & 23 & 02 & 33 & medium & ff & 41148 & 34091\\
   &  &  &  &  &  &  &  &  &  &  &  &  &  & PNS003 & 23 & 24 & 56 & thin & ff & 39685 & 32742\\
   S10 & 0601211001 & SMC survey 10 & 2009 & Nov. & 9 & 00\rahour & 43\ramin & 56\fs7 & -72\degr & 44\arcmin & 38\arcsec & -1.2 & -2.56 & M1S001 & 21 & 16 & 08 & medium & ff & 36520 & 26268\\
   &  &  &  &  &  &  &  &  &  &  &  &  &  & M2S002 & 21 & 16 & 06 & medium & ff & 36525 & 26289\\
   &  &  &  &  &  &  &  &  &  &  &  &  &  & PNS003 & 21 & 38 & 29 & thin & ff & 34934 & 24688\\
   S11 & 0601211101 & SMC survey 11 & 2009 & Oct. & 18 & 00\rahour & 42\ramin & 29\fs5 & -73\degr & 02\arcmin & 27\arcsec & -0.1 & -2.35 & M1S001 & 22 & 47 & 09 & medium & ff & 31619 & 27114\\
   &  &  &  &  &  &  &  &  &  &  &  &  &  & M2S002 & 22 & 47 & 09 & medium & ff & 31624 & 27119\\
   &  &  &  &  &  &  &  &  &  &  &  &  &  & PNS003 & 23 & 09 & 31 & thin & ff & 30034 & 25534\\
   S12 & 0601211201 & SMC survey 12 & 2009 & Oct. & 20 & 00\rahour & 42\ramin & 25\fs2 & -73\degr & 20\arcmin & 11\arcsec & 1.82 & 0.06 & M1S001 & 22 & 51 & 09 & medium & ff & 33119 & 29514\\
   &  &  &  &  &  &  &  &  &  &  &  &  &  & M2S002 & 22 & 51 & 08 & medium & ff & 33124 & 29519\\
   &  &  &  &  &  &  &  &  &  &  &  &  &  & PNS003 & 23 & 13 & 31 & thin & ff & 31534 & 27931\\
   S13 & 0601211301 & SMC survey 13 & 2009 & Oct. & 3 & 00\rahour & 46\ramin & 28\fs7 & -73\degr & 24\arcmin & 25\arcsec & -0.36 & -0.72 & M1S001 & 05 & 08 & 47 & medium & ff & 32420 & 32363\\
   &  &  &  &  &  &  &  &  &  &  &  &  &  & M2S002 & 05 & 08 & 46 & medium & ff & 32425 & 32368\\
   &  &  &  &  &  &  &  &  &  &  &  &  &  & PNS003 & 05 & 31 & 10 & thin & ff & 30833 & 30779\\
   S14 & 0601211401 & SMC survey 14 & 2009 & Nov. & 4 & 00\rahour & 52\ramin & 19\fs2 & -73\degr & 09\arcmin & 03\arcsec & 0.89 & 0.67 & M1S001 & 21 & 38 & 31 & medium & ff & 46520 & 32718\\
   &  &  &  &  &  &  &  &  &  &  &  &  &  & M2S002 & 21 & 38 & 30 & medium & ff & 46525 & 32717\\
   &  &  &  &  &  &  &  &  &  &  &  &  &  & PNS003 & 22 & 00 & 51 & thin & ff & 44290 & 31370\\
   S15 & 0601211501 & SMC survey 15 & 2009 & Oct. & 13 & 00\rahour & 56\ramin & 25\fs5 & -73\degr & 02\arcmin & 58\arcsec & 2.43 & 0.63 & M1S001 & 00 & 02 & 01 & medium & ff & 37620 & 36863\\
   &  &  &  &  &  &  &  &  &  &  &  &  &  & M2S002 & 00 & 02 & 00 & medium & ff & 37625 & 36868\\
   &  &  &  &  &  &  &  &  &  &  &  &  &  & PNS003 & 00 & 24 & 22 & thin & ff & 36034 & 35280\\
   S16 & 0601211601 & SMC survey 16 & 2009 & Oct. & 11 & 00\rahour & 58\ramin & 21\fs5 & -72\degr & 48\arcmin & 27\arcsec & 0.71 & -2.54 & M1S001 & 22 & 43 & 47 & medium & ff & 43350 & 30443\\
   &  &  &  &  &  &  &  &  &  &  &  &  &  & M2S002 & 22 & 43 & 46 & medium & ff & 43358 & 30443\\
   &  &  &  &  &  &  &  &  &  &  &  &  &  & PNS003 & 23 & 06 & 08 & thin & ff & 41769 & 29100\\
   S17 & 0601211701 & SMC survey 17 & 2009 & Oct. & 16 & 01\rahour & 02\ramin & 23\fs1 & -72\degr & 35\arcmin & 23\arcsec & 1.37 & 1.36 & M1S001 & 01 & 05 & 21 & medium & ff & 36738 & 20744\\
   &  &  &  &  &  &  &  &  &  &  &  &  &  & M2S002 & 01 & 05 & 17 & medium & ff & 36785 & 20744\\
   &  &  &  &  &  &  &  &  &  &  &  &  &  & PNS003 & 01 & 57 & 40 & thin & ff & 33664 & 17626\\
   S18 & 0601211801 & SMC survey 18 & 2009 & Nov. & 13 & 01\rahour & 04\ramin & 04\fs7 & -72\degr & 22\arcmin & 52\arcsec & 1.82 & 1.89 & M1S001 & 20 & 59 & 56 & medium & ff & 2535 & 0\\
   &  &  &  &  &  &  &  &  &  &  &  &  &  & M1U002 & 22 & 23 & 04 & medium & ff & 31531 & 22358\\
   &  &  &  &  &  &  &  &  &  &  &  &  &  & M2S002 & 20 & 59 & 56 & medium & ff & 2540 & 0\\
   &  &  &  &  &  &  &  &  &  &  &  &  &  & M2U002 & 22 & 23 & 11 & medium & ff & 31531 & 22377\\
   &  &  &  &  &  &  &  &  &  &  &  &  &  & PNS003 & 21 & 22 & 19 & thin & ff & 34934 & 21659\\
   S19 & 0601211901 & SMC survey 19 & 2009 & Nov. & 30 & 01\rahour & 08\ramin & 33\fs2 & -72\degr & 09\arcmin & 54\arcsec & 0.19 & 0.86 & M1S001 & 14 & 46 & 19 & medium & ff & 31619 & 31585\\
   &  &  &  &  &  &  &  &  &  &  &  &  &  & M2S002 & 14 & 46 & 19 & medium & ff & 31624 & 31598\\
   &  &  &  &  &  &  &  &  &  &  &  &  &  & PNS003 & 15 & 08 & 40 & thin & ff & 30034 & 30001\\
   S20 & 0601212001 & SMC survey 20 & 2009 & Nov. & 27 & 01\rahour & 12\ramin & 56\fs4 & -72\degr & 22\arcmin & 38\arcsec & 3.56 & 3.12 & M1U002 & 22 & 39 & 06 & medium & ff & 28231 & 28126\\
   &  &  &  &  &  &  &  &  &  &  &  &  &  & M2U002 & 22 & 39 & 09 & medium & ff & 28234 & 28113\\
   &  &  &  &  &  &  &  &  &  &  &  &  &  & PNU002 & 23 & 04 & 04 & thin & ff & 26490 & 26368\\
   S21 & 0601212101 & SMC survey 21 & 2009 & Nov. & 16 & 01\rahour & 11\ramin & 32\fs3 & -72\degr & 43\arcmin & 31\arcsec & 1.02 & -0.54 & M1S001 & 05 & 48 & 08 & medium & ff & 34019 & 34014\\
   &  &  &  &  &  &  &  &  &  &  &  &  &  & M2S002 & 05 & 48 & 06 & medium & ff & 34024 & 34019\\
   &  &  &  &  &  &  &  &  &  &  &  &  &  & PNS003 & 06 & 10 & 27 & thin & ff & 32434 & 32434\\
   S22 & 0601212201 & SMC survey 22 & 2009 & Nov. & 19 & 01\rahour & 13\ramin & 35\fs3 & -73\degr & 01\arcmin & 05\arcsec & 1.35 & 1.0 & M1S001 & 20 & 45 & 30 & medium & ff & 34521 & 22841\\
   &  &  &  &  &  &  &  &  &  &  &  &  &  & M2S002 & 20 & 45 & 31 & medium & ff & 34523 & 22835\\
   &  &  &  &  &  &  &  &  &  &  &  &  &  & PNS003 & 21 & 07 & 51 & thin & ff & 32934 & 22028\\
   S23 & 0601212301 & SMC survey 23 & 2009 & Sep. & 9 & 01\rahour & 17\ramin & 03\fs4 & -73\degr & 04\arcmin & 05\arcsec & 2.14 & -0.44 & M1S001 & 09 & 13 & 53 & medium & ff & 33421 & 33330\\
   &  &  &  &  &  &  &  &  &  &  &  &  &  & M2S002 & 09 & 13 & 53 & medium & ff & 33426 & 33335\\
   &  &  &  &  &  &  &  &  &  &  &  &  &  & PNS003 & 09 & 36 & 15 & thin & ff & 31834 & 31743\\
   S24 & 0601212401 & SMC survey 24 & 2009 & Jun & 29 & 01\rahour & 20\ramin & 47\fs6 & -73\degr & 15\arcmin & 35\arcsec & 2.13 & 2.05 & M1S001 & 14 & 46 & 19 & medium & ff & 36619 & 26962\\
   &  &  &  &  &  &  &  &  &  &  &  &  &  & M2S002 & 14 & 46 & 18 & medium & ff & 36624 & 26967\\
   &  &  &  &  &  &  &  &  &  &  &  &  &  & PNS003 & 15 & 08 & 41 & thin & ff & 35034 & 25372\\
   S25 & 0601212501 & SMC survey 25 & 2009 & Sep. & 9 & 01\rahour & 12\ramin & 31\fs5 & -73\degr & 18\arcmin & 24\arcsec & 1.58 & -0.12 & M1S001 & 19 & 13 & 54 & medium & ff & 33421 & 33416\\
   &  &  &  &  &  &  &  &  &  &  &  &  &  & M2S002 & 19 & 13 & 53 & medium & ff & 33426 & 33421\\
   &  &  &  &  &  &  &  &  &  &  &  &  &  & PNS003 & 19 & 36 & 15 & thin & ff & 31834 & 31831\\
   S26 & 0601212601 & SMC survey 26 & 2009 & Jun & 29 & 01\rahour & 08\ramin & 33\fs1 & -72\degr & 54\arcmin & 46\arcsec & 2.64 & -0.38 & M1S001 & 06 & 04 & 39 & medium & ff & 28720 & 17760\\
   &  &  &  &  &  &  &  &  &  &  &  &  &  & M2S002 & 06 & 04 & 38 & medium & ff & 28725 & 17770\\
   &  &  &  &  &  &  &  &  &  &  &  &  &  & PNS003 & 06 & 27 & 01 & thin & ff & 27134 & 16803\\
   S27 & 0601212701 & SMC survey 27 & 2009 & Dec. & 26 & 01\rahour & 07\ramin & 54\fs8 & -73\degr & 09\arcmin & 25\arcsec & 2.4 & 0.17 & M1S001 & 07 & 25 & 22 & medium & ff & 36622 & 36616\\
   &  &  &  &  &  &  &  &  &  &  &  &  &  & M2S002 & 07 & 25 & 20 & medium & ff & 36624 & 36619\\
   &  &  &  &  &  &  &  &  &  &  &  &  &  & PNS003 & 07 & 47 & 42 & thin & ff & 35034 & 35034\\
   S28 & 0601212801 & SMC survey 28 & 2009 & Dec. & 7 & 01\rahour & 01\ramin & 54\fs0 & -73\degr & 07\arcmin & 05\arcsec & -0.14 & 1.26 & M1U002 & 23 & 35 & 54 & medium & ff & 21352 & 17646\\
   &  &  &  &  &  &  &  &  &  &  &  &  &  & M2U002 & 23 & 36 & 22 & medium & ff & 21325 & 17620\\
   &  &  &  &  &  &  &  &  &  &  &  &  &  & PNU002 & 00 & 01 & 35 & thin & ff & 19563 & 15863\\
   S29 & 0601212901 & SMC survey 29 & 2009 & Sep. & 13 & 00\rahour & 57\ramin & 04\fs8 & -73\degr & 20\arcmin & 23\arcsec & -0.06 & 1.02 & M1S001 & 13 & 29 & 26 & medium & ff & 36120 & 34514\\
   &  &  &  &  &  &  &  &  &  &  &  &  &  & M2S002 & 13 & 29 & 24 & medium & ff & 36125 & 34520\\
   &  &  &  &  &  &  &  &  &  &  &  &  &  & PNS003 & 13 & 51 & 46 & thin & ff & 34534 & 32934\\
   S30 & 0601213001 & SMC survey 30 & 2009 & Sep. & 13 & 00\rahour & 53\ramin & 18\fs3 & -73\degr & 32\arcmin & 45\arcsec & 1.17 & -1.84 & M1S001 & 01 & 11 & 03 & medium & ff & 41720 & 40402\\
   &  &  &  &  &  &  &  &  &  &  &  &  &  & M2S002 & 01 & 11 & 03 & medium & ff & 41725 & 40407\\
   &  &  &  &  &  &  &  &  &  &  &  &  &  & PNS003 & 01 & 33 & 26 & thin & ff & 40134 & 38816\\
   S31 & 0601213201 & SMC survey 17 & 2010 & Mar. & 12 & 01\rahour & 02\ramin & 23\fs1 & -72\degr & 35\arcmin & 23\arcsec & -0.42 & 0.7 & M1S001 & 00 & 56 & 15 & medium & ff & 13619 & 10862\\
   &  &  &  &  &  &  &  &  &  &  &  &  &  & M2S002 & 00 & 56 & 15 & medium & ff & 13624 & 10867\\
   &  &  &  &  &  &  &  &  &  &  &  &  &  & PNS003 & 01 & 18 & 37 & thin & ff & 12034 & 9334\\
   S32 & 0601213301 & SMC survey 28 & 2010 & Mar. & 12 & 01\rahour & 01\ramin & 54\fs0 & -73\degr & 07\arcmin & 05\arcsec & -0.02 & 4.25 & M1S001 & 05 & 26 & 15 & medium & ff & 12621 & 8518\\
   &  &  &  &  &  &  &  &  &  &  &  &  &  & M2S002 & 05 & 26 & 15 & medium & ff & 12626 & 8523\\
   &  &  &  &  &  &  &  &  &  &  &  &  &  & PNS003 & 05 & 48 & 37 & thin & ff & 11034 & 8234\\
   S33 & 0601213401 & SMC survey 1 & 2010 & Mar. & 16 & 00\rahour & 58\ramin & 16\fs4 & -71\degr & 28\arcmin & 45\arcsec & -0.19 & 1.31 & M1S001 & 10 & 05 & 12 & medium & ff & 21521 & 15198\\
   &  &  &  &  &  &  &  &  &  &  &  &  &  & M2S002 & 10 & 05 & 14 & medium & ff & 21526 & 15198\\
   &  &  &  &  &  &  &  &  &  &  &  &  &  & PNU002 & 11 & 27 & 06 & thin & ff & 16224 & 12000\\
   C1 & 0123110201 & 1ES0102-72 & 2000 & Apr. & 16 & 01\rahour & 03\ramin & 50\fs0 & -72\degr & 01\arcmin & 55\arcsec & 0.22 & -1.41 & M1S003 & 20 & 09 & 25 & thin & lw & 18800 & 17915\\
   &  &  &  &  &  &  &  &  &  &  &  &  &  & M2S005 & 20 & 09 & 26 & thin & lw & 18799 & 17914\\
   &  &  &  &  &  &  &  &  &  &  &  &  &  & PNS001 & 19 & 56 & 32 & thin & ff & 19301 & 18418\\
   C2 & 0123110301 & 1ES0102-72 & 2000 & Apr. & 17 & 01\rahour & 03\ramin & 50\fs0 & -72\degr & 01\arcmin & 55\arcsec & 0.44 & -0.83 & M1S004 & 04 & 43 & 37 & medium & lw & 17800 & 11725\\
   &  &  &  &  &  &  &  &  &  &  &  &  &  & M2S006 & 04 & 43 & 36 & medium & lw & 17799 & 11725\\
   &  &  &  &  &  &  &  &  &  &  &  &  &  & PNS002 & 04 & 30 & 44 & medium & ff & 18300 & 12229\\
   C3 & 0135720601 & 1ES0102-72 & 2001 & Apr. & 14 & 01\rahour & 03\ramin & 50\fs0 & -72\degr & 01\arcmin & 55\arcsec & 0.47 & -0.68 & M1S003 & 20 & 47 & 25 & thin & lw & 32900 & 13352\\
   &  &  &  &  &  &  &  &  &  &  &  &  &  & M2S005 & 20 & 47 & 25 & thin & lw & 32900 & 13352\\
   &  &  &  &  &  &  &  &  &  &  &  &  &  & PNS001 & 01 & 20 & 28 & thin & ff & 16200 & 9393\\
   &  &  &  &  &  &  &  &  &  &  &  &  &  & PNS009 & 21 & 03 & 15 & thin & sw & 12600 & 0\\
   C4 & 0135720801 & 1ES0102-72 & 2001 & Dec. & 25 & 01\rahour & 04\ramin & 00\fs0 & -72\degr & 00\arcmin & 16\arcsec & 2.35 & -0.07 & M1S003 & 18 & 44 & 37 & thin & lw & 32133 & 27568\\
   &  &  &  &  &  &  &  &  &  &  &  &  &  & M2S005 & 18 & 44 & 37 & thin & lw & 32133 & 27565\\
   &  &  &  &  &  &  &  &  &  &  &  &  &  & PNS001 & 19 & 00 & 33 & thin & sw & 31317 & 26759\\
   C5 & 0135720901 & 1ES0102-72 & 2002 & Apr. & 20 & 01\rahour & 04\ramin & 01\fs7 & -72\degr & 01\arcmin & 51\arcsec & 1.39 & 0.54 & M1S003 & 22 & 28 & 33 & thin & ti & 13026 & 10077\\
   &  &  &  &  &  &  &  &  &  &  &  &  &  & M2S005 & 22 & 28 & 22 & thin & lw & 13043 & 10085\\
   &  &  &  &  &  &  &  &  &  &  &  &  &  & PNS001 & 23 & 01 & 44 & thin & ff & 13774 & 9134\\
   C6 & 0135721001 & 1ES0102-72 & 2002 & May & 18 & 01\rahour & 04\ramin & 01\fs7 & -72\degr & 01\arcmin & 51\arcsec & 0.02 & 1.61 & M1S003 & 11 & 09 & 57 & thin & ti & 13089 & 9393\\
   &  &  &  &  &  &  &  &  &  &  &  &  &  & M1S018 & 15 & 01 & 36 & thin & lw & 16691 & 7459\\
   &  &  &  &  &  &  &  &  &  &  &  &  &  & M2S005 & 10 & 39 & 47 & thin & lw & 14890 & 9385\\
   &  &  &  &  &  &  &  &  &  &  &  &  &  & M2S019 & 14 & 57 & 17 & thin & ti & 16687 & 7467\\
   &  &  &  &  &  &  &  &  &  &  &  &  &  & PNS001 & 11 & 43 & 08 & thin & ff & 12216 & 10516\\
   &  &  &  &  &  &  &  &  &  &  &  &  &  & PNS017 & 15 & 55 & 51 & thin & lw & 13000 & 4256\\
   C7 & 0135721101 & 1ES0102-72 & 2002 & Oct. & 13 & 01\rahour & 03\ramin & 43\fs5 & -72\degr & 01\arcmin & 31\arcsec & 0.17 & 2.07 & M1S003 & 03 & 19 & 38 & thin & lw & 23676 & 23064\\
   &  &  &  &  &  &  &  &  &  &  &  &  &  & M2S005 & 03 & 19 & 38 & thin & lw & 23677 & 23066\\
   &  &  &  &  &  &  &  &  &  &  &  &  &  & PNS001 & 03 & 24 & 22 & thin & sw & 10240 & 0\\
   &  &  &  &  &  &  &  &  &  &  &  &  &  & PNS017 & 06 & 53 & 28 & thin & lw & 10237 & 9837\\
   C8 & 0135721301 & 1ES0102-72 & 2002 & Dec. & 14 & 01\rahour & 03\ramin & 56\fs4 & -72\degr & 00\arcmin & 28\arcsec & 0.64 & 0.73 & M1S003 & 03 & 53 & 55 & thin & lw & 28678 & 28372\\
   &  &  &  &  &  &  &  &  &  &  &  &  &  & M2S005 & 03 & 53 & 55 & thin & lw & 28677 & 28367\\
   &  &  &  &  &  &  &  &  &  &  &  &  &  & PNS001 & 03 & 58 & 40 & thin & sw & 10940 & 0\\
   &  &  &  &  &  &  &  &  &  &  &  &  &  & PNS017 & 07 & 39 & 23 & thin & lw & 14538 & 14222\\
   C9 & 0135721401 & 1ES0102-72 & 2003 & Apr. & 20 & 01\rahour & 04\ramin & 18\fs1 & -72\degr & 02\arcmin & 32\arcsec & 1.43 & 0.81 & M1S003 & 12 & 04 & 59 & calc & ff & 9176 & 0\\
   &  &  &  &  &  &  &  &  &  &  &  &  &  & M1U002 & 15 & 03 & 09 & thin & lw & 34532 & 30351\\
   &  &  &  &  &  &  &  &  &  &  &  &  &  & M2S005 & 12 & 04 & 58 & calc & ff & 9181 & 0\\
   &  &  &  &  &  &  &  &  &  &  &  &  &  & M2U002 & 15 & 03 & 14 & thin & lw & 34533 & 30365\\
   &  &  &  &  &  &  &  &  &  &  &  &  &  & PNS001 & 12 & 27 & 08 & calc & ff & 7278 & 0\\
   &  &  &  &  &  &  &  &  &  &  &  &  &  & PNS017 & 19 & 21 & 26 & medium & lw & 18741 & 18725\\
   &  &  &  &  &  &  &  &  &  &  &  &  &  & PNU002 & 14 & 56 & 20 & medium & sw & 13604 & 0\\
   C10 & 0135721501 & 1ES0102-72 & 2003 & Oct. & 27 & 01\rahour & 03\ramin & 45\fs6 & -72\degr & 01\arcmin & 07\arcsec & 0.09 & 1.22 & M1S003 & 07 & 55 & 06 & thin & lw & 30177 & 24241\\
   &  &  &  &  &  &  &  &  &  &  &  &  &  & M2S005 & 07 & 55 & 06 & thin & lw & 30181 & 24249\\
   &  &  &  &  &  &  &  &  &  &  &  &  &  & PNS001 & 08 & 17 & 30 & thick & ff & 28535 & 22597\\
   C11 & 0135721701 & 1ES0102-72 & 2003 & Nov. & 16 & 01\rahour & 03\ramin & 45\fs6 & -72\degr & 01\arcmin & 07\arcsec & 0.31 & 1.64 & M1S003 & 06 & 12 & 02 & thin & lw & 27322 & 25158\\
   &  &  &  &  &  &  &  &  &  &  &  &  &  & M2S005 & 06 & 12 & 03 & thin & lw & 27322 & 25170\\
   &  &  &  &  &  &  &  &  &  &  &  &  &  & PNS001 & 06 & 34 & 29 & thick & ff & 31218 & 24492\\
   C12 & 0135721901 & 1ES0102-72 & 2004 & Apr. & 28 & 01\rahour & 04\ramin & 17\fs3 & -72\degr & 02\arcmin & 38\arcsec & 0.0 & 1.13 & M1S003 & 07 & 09 & 57 & thin & lw & 33280 & 32977\\
   &  &  &  &  &  &  &  &  &  &  &  &  &  & M2S005 & 07 & 09 & 57 & thin & lw & 33285 & 32971\\
   &  &  &  &  &  &  &  &  &  &  &  &  &  & PNS001 & 07 & 29 & 42 & thick & lw & 31800 & 31484\\
   C13 & 0135722401 & 1ES0102-72 & 2004 & Oct. & 14 & 01\rahour & 03\ramin & 45\fs6 & -72\degr & 01\arcmin & 07\arcsec & 0.1 & -0.46 & M1S003 & 09 & 04 & 02 & thick & lw & 30878 & 30419\\
   &  &  &  &  &  &  &  &  &  &  &  &  &  & M2S005 & 09 & 04 & 02 & thick & lw & 30883 & 30406\\
   &  &  &  &  &  &  &  &  &  &  &  &  &  & PNS001 & 09 & 09 & 28 & thick & sw & 30672 & 30191\\
   C14 & 0135722001 & 1ES0102-72 & 2004 & Oct. & 26 & 01\rahour & 04\ramin & 03\fs6 & -72\degr & 01\arcmin & 44\arcsec & 0.63 & 0.56 & M1S003 & 06 & 56 & 39 & thin & lw & 31679 & 31599\\
   &  &  &  &  &  &  &  &  &  &  &  &  &  & M2S005 & 06 & 56 & 39 & thin & lw & 31683 & 31601\\
   &  &  &  &  &  &  &  &  &  &  &  &  &  & PNS001 & 07 & 25 & 23 & thick & lw & 29659 & 29578\\
   C15 & 0135722101 & 1ES0102-72 & 2004 & Nov. & 7 & 01\rahour & 03\ramin & 59\fs6 & -72\degr & 01\arcmin & 44\arcsec & 0.37 & 1.89 & M1S003 & 03 & 38 & 05 & thin & lw & 31577 & 24008\\
   &  &  &  &  &  &  &  &  &  &  &  &  &  & M1S019 & 22 & 38 & 09 & calc & ff & 17441 & 0\\
   &  &  &  &  &  &  &  &  &  &  &  &  &  & M2S005 & 03 & 38 & 04 & thin & lw & 31583 & 23998\\
   &  &  &  &  &  &  &  &  &  &  &  &  &  & M2S020 & 22 & 38 & 08 & calc & ff & 17444 & 0\\
   &  &  &  &  &  &  &  &  &  &  &  &  &  & PNS001 & 04 & 00 & 29 & thin & ff & 29937 & 22638\\
   &  &  &  &  &  &  &  &  &  &  &  &  &  & PNS021 & 23 & 00 & 29 & calc & ff & 15122 & 0\\
   C16 & 0135722201 & 1ES0102-72 & 2004 & Nov. & 7 & 01\rahour & 04\ramin & 03\fs6 & -72\degr & 01\arcmin & 57\arcsec & 0.64 & 1.01 & M1U002 & 13 & 09 & 50 & thin & lw & 22643 & 0\\
   &  &  &  &  &  &  &  &  &  &  &  &  &  & M1U003 & 21 & 15 & 30 & thin & lw & 2294 & 0\\
   &  &  &  &  &  &  &  &  &  &  &  &  &  & M2S005 & 13 & 06 & 24 & thin & lw & 22858 & 0\\
   &  &  &  &  &  &  &  &  &  &  &  &  &  & M2U002 & 21 & 18 & 05 & thin & lw & 2144 & 0\\
   &  &  &  &  &  &  &  &  &  &  &  &  &  & PNS001 & 13 & 28 & 49 & thin2 & ff & 10565 & 0\\
   &  &  &  &  &  &  &  &  &  &  &  &  &  & PNU014 & 16 & 36 & 13 & thin & ff & 5383 & 0\\
   &  &  &  &  &  &  &  &  &  &  &  &  &  & PNU027 & 18 & 15 & 48 & thin2 & ff & 12781 & 0\\
   C17 & 0135722301 & 1ES0102-72 & 2004 & Nov. & 7 & 01\rahour & 03\ramin & 59\fs6 & -72\degr & 01\arcmin & 57\arcsec & -0.31 & 1.74 & M1S003 & 22 & 35 & 47 & thin & lw & 31636 & 14035\\
   &  &  &  &  &  &  &  &  &  &  &  &  &  & M2U002 & 22 & 44 & 58 & thin & lw & 31094 & 14038\\
   &  &  &  &  &  &  &  &  &  &  &  &  &  & PNS001 & 22 & 58 & 12 & thin & ff & 30000 & 13768\\
   C18 & 0135722501 & 1ES0102-72 & 2005 & Apr. & 17 & 01\rahour & 04\ramin & 17\fs3 & -72\degr & 02\arcmin & 38\arcsec & 0.25 & 1.64 & M1S003 & 22 & 15 & 58 & thin & lw & 36878 & 23753\\
   &  &  &  &  &  &  &  &  &  &  &  &  &  & M2S005 & 22 & 15 & 58 & thin & lw & 36882 & 23769\\
   &  &  &  &  &  &  &  &  &  &  &  &  &  & PNS001 & 22 & 44 & 29 & thin & lw & 34872 & 21804\\
   C19 & 0135722601 & 1ES0102-72 & 2005 & Nov. & 5 & 01\rahour & 03\ramin & 47\fs1 & -72\degr & 00\arcmin & 57\arcsec & 0.14 & 1.07 & M1S003 & 06 & 45 & 03 & thin & lw & 30207 & 28998\\
   &  &  &  &  &  &  &  &  &  &  &  &  &  & M2S005 & 06 & 45 & 04 & thin & lw & 30211 & 29009\\
   &  &  &  &  &  &  &  &  &  &  &  &  &  & PNS001 & 06 & 50 & 29 & medium & sw & 30001 & 28796\\
   C20 & 0135722701 & 1ES0102-72 & 2006 & Apr. & 20 & 01\rahour & 04\ramin & 01\fs7 & -72\degr & 01\arcmin & 51\arcsec & 0.02 & 2.32 & M1S003 & 02 & 25 & 04 & thin & lw & 30207 & 30203\\
   &  &  &  &  &  &  &  &  &  &  &  &  &  & M2S005 & 02 & 25 & 03 & thin & lw & 30212 & 30208\\
   &  &  &  &  &  &  &  &  &  &  &  &  &  & PNS001 & 02 & 30 & 30 & thin & sw & 30001 & 30000\\
   C21 & 0412980101 & 1ES0102-72 & 2006 & Nov. & 5 & 01\rahour & 03\ramin & 47\fs1 & -72\degr & 00\arcmin & 57\arcsec & 1.34 & 2.64 & M1S002 & 00 & 55 & 20 & thin & lw & 32127 & 30986\\
   &  &  &  &  &  &  &  &  &  &  &  &  &  & M2S003 & 00 & 55 & 20 & thin & lw & 32131 & 31000\\
   &  &  &  &  &  &  &  &  &  &  &  &  &  & PNS001 & 01 & 00 & 46 & medium & sw & 31971 & 30889\\
   C22 & 0412980201 & 1ES0102-72 & 2007 & Apr. & 25 & 01\rahour & 04\ramin & 01\fs7 & -72\degr & 01\arcmin & 51\arcsec & 0.04 & 2.21 & M1S002 & 12 & 36 & 26 & thin & lw & 36128 & 19571\\
   &  &  &  &  &  &  &  &  &  &  &  &  &  & M2S003 & 12 & 36 & 26 & thin & lw & 36133 & 19586\\
   &  &  &  &  &  &  &  &  &  &  &  &  &  & PNS001 & 12 & 41 & 52 & thin & sw & 35972 & 19435\\
   C23 & 0412980301 & 1ES0102-72 & 2007 & Oct. & 26 & 01\rahour & 03\ramin & 47\fs1 & -72\degr & 00\arcmin & 57\arcsec & 1.29 & 1.49 & M1S002 & 09 & 49 & 09 & thin & lw & 36828 & 34966\\
   &  &  &  &  &  &  &  &  &  &  &  &  &  & M2S003 & 09 & 49 & 09 & thin & lw & 36833 & 34972\\
   &  &  &  &  &  &  &  &  &  &  &  &  &  & PNS001 & 09 & 54 & 35 & medium & sw & 36671 & 34811\\
   C24 & 0412980501 & 1ES0102-72 & 2008 & Apr. & 19 & 01\rahour & 04\ramin & 01\fs7 & -72\degr & 01\arcmin & 51\arcsec & -0.61 & 1.78 & M1S002 & 09 & 22 & 50 & thin & lw & 29627 & 21997\\
   &  &  &  &  &  &  &  &  &  &  &  &  &  & M2S003 & 09 & 22 & 50 & thin & lw & 29633 & 22005\\
   &  &  &  &  &  &  &  &  &  &  &  &  &  & PNS001 & 09 & 28 & 17 & thin & sw & 29471 & 21738\\
   C25 & 0412980701 & 1ES0102-72 & 2008 & Nov. & 14 & 01\rahour & 04\ramin & 01\fs7 & -72\degr & 01\arcmin & 51\arcsec & 1.97 & 2.54 & M1S002 & 19 & 49 & 11 & thin & lw & 28628 & 28604\\
   &  &  &  &  &  &  &  &  &  &  &  &  &  & M2S003 & 19 & 49 & 11 & thin & lw & 28633 & 28607\\
   &  &  &  &  &  &  &  &  &  &  &  &  &  & PNS001 & 19 & 54 & 36 & medium & sw & 28471 & 28438\\
   C26 & 0412980801 & 1ES0102-72 & 2009 & Apr. & 13 & 01\rahour & 04\ramin & 01\fs7 & -72\degr & 01\arcmin & 51\arcsec & 1.09 & 1.31 & M1S002 & 00 & 04 & 42 & thin & lw & 28628 & 6522\\
   &  &  &  &  &  &  &  &  &  &  &  &  &  & M2S003 & 00 & 04 & 42 & thin & lw & 28633 & 6527\\
   &  &  &  &  &  &  &  &  &  &  &  &  &  & PNS001 & 00 & 10 & 08 & thin & sw & 28471 & 6369\\
   C27 & 0412980901 & 1ES0102-72 & 2009 & Oct. & 21 & 01\rahour & 04\ramin & 01\fs7 & -72\degr & 01\arcmin & 51\arcsec & 1.63 & 2.05 & M1S002 & 09 & 03 & 03 & thin & lw & 28628 & 28624\\
   &  &  &  &  &  &  &  &  &  &  &  &  &  & M2S003 & 09 & 03 & 03 & thin & lw & 28633 & 28629\\
   &  &  &  &  &  &  &  &  &  &  &  &  &  & PNS001 & 09 & 08 & 28 & medium & sw & 28471 & 28471\\
   C28 & 0412981001 & 1ES0102-72 & 2010 & Apr. & 21 & 01\rahour & 04\ramin & 01\fs7 & -72\degr & 01\arcmin & 51\arcsec & -0.15 & 0.93 & M1S002 & 01 & 36 & 37 & thin & lw & 30228 & 27254\\
   &  &  &  &  &  &  &  &  &  &  &  &  &  & M2S003 & 01 & 36 & 37 & thin & lw & 30233 & 27260\\
   &  &  &  &  &  &  &  &  &  &  &  &  &  & PNS001 & 01 & 42 & 03 & thin & sw & 30072 & 27101\\
   A1 & 0112780201 & RX J0059.2-7138 & 2000 & Sep. & 19 & 00\rahour & 59\ramin & 13\fs0 & -71\degr & 38\arcmin & 50\arcsec & -0.36 & -4.78 & M1S002 & 02 & 05 & 37 & medium & ti & 4116 & 2425\\
   &  &  &  &  &  &  &  &  &  &  &  &  &  & M2S003 & 02 & 05 & 34 & thin & ti & 4163 & 2428\\
   &  &  &  &  &  &  &  &  &  &  &  &  &  & PNS001 & 01 & 24 & 24 & medium & sw & 8598 & 4900\\
   A2 & 0110000101 & IKT 5 & 2000 & Oct. & 15 & 00\rahour & 49\ramin & 07\fs0 & -73\degr & 14\arcmin & 06\arcsec & -0.64 & -1.21 & M1S003 & 15 & 18 & 28 & medium & ff & 26897 & 25388\\
   &  &  &  &  &  &  &  &  &  &  &  &  &  & M2S004 & 15 & 18 & 22 & medium & ff & 26897 & 25388\\
   &  &  &  &  &  &  &  &  &  &  &  &  &  & PNS005 & 16 & 24 & 25 & medium & eff & 23000 & 21599\\
   A3 & 0110000201 & IKT 18 & 2000 & Oct. & 17 & 00\rahour & 59\ramin & 26\fs0 & -72\degr & 10\arcmin & 11\arcsec & -0.07 & 1.87 & M1S003 & 15 & 10 & 44 & medium & ff & 19747 & 16656\\
   &  &  &  &  &  &  &  &  &  &  &  &  &  & M2S004 & 15 & 10 & 35 & medium & ff & 19747 & 16656\\
   &  &  &  &  &  &  &  &  &  &  &  &  &  & PNS005 & 16 & 16 & 36 & medium & eff & 15850 & 13418\\
   A4 & 0110000301 & IKT 23 & 2000 & Oct. & 17 & 01\rahour & 04\ramin & 52\fs0 & -72\degr & 23\arcmin & 10\arcsec & 1.74 & 1.49 & M1S003 & 21 & 45 & 54 & medium & ff & 32630 & 11854\\
   &  &  &  &  &  &  &  &  &  &  &  &  &  & M2S004 & 21 & 45 & 53 & medium & ff & 32659 & 11854\\
   &  &  &  &  &  &  &  &  &  &  &  &  &  & PNS005 & 22 & 51 & 56 & medium & eff & 30950 & 8148\\
   A5 & 0112780601 & RX J0059.2-7138 & 2001 & Apr. & 29 & 00\rahour & 59\ramin & 13\fs0 & -71\degr & 38\arcmin & 50\arcsec & 0.0 & 0.0 & M1S002 & 21 & 07 & 59 & medium & ti & 4966 & 52\\
   &  &  &  &  &  &  &  &  &  &  &  &  &  & M2S003 & 20 & 58 & 31 & thin & sw & 5795 & 52\\
   &  &  &  &  &  &  &  &  &  &  &  &  &  & PNS001 & 21 & 14 & 07 & medium & sw & 5000 & 0\\
   A6 & 0011450101 & SMC X-1 & 2001 & May & 31 & 01\rahour & 17\ramin & 05\fs1 & -73\degr & 26\arcmin & 35\arcsec & 0.77 & -0.23 & M1S001 & 02 & 20 & 10 & thin & ff & 59018 & 43157\\
   &  &  &  &  &  &  &  &  &  &  &  &  &  & M2S002 & 02 & 20 & 10 & thin & ff & 59018 & 43157\\
   &  &  &  &  &  &  &  &  &  &  &  &  &  & PNS003 & 02 & 59 & 37 & thin & ff & 56390 & 40790\\
   A7 & 0084200801 & SMC Pointing 8 & 2001 & Oct. & 17 & 00\rahour & 54\ramin & 31\fs7 & -73\degr & 40\arcmin & 56\arcsec & -0.14 & -0.13 & M1S001 & 10 & 07 & 40 & medium & ff & 21115 & 11491\\
   &  &  &  &  &  &  &  &  &  &  &  &  &  & M2S002 & 10 & 07 & 40 & medium & ff & 21115 & 11491\\
   &  &  &  &  &  &  &  &  &  &  &  &  &  & PNS003 & 10 & 46 & 50 & thin & ff & 18500 & 8866\\
   A8 & 0011450201 & SMC X-1 & 2001 & Nov. & 16 & 01\rahour & 17\ramin & 05\fs1 & -73\degr & 26\arcmin & 35\arcsec & -0.41 & 0.21 & M1S001 & 03 & 23 & 32 & thin & ti & 40761 & 40035\\
   &  &  &  &  &  &  &  &  &  &  &  &  &  & M2S002 & 03 & 23 & 32 & thin & ti & 40761 & 40027\\
   &  &  &  &  &  &  &  &  &  &  &  &  &  & PNS003 & 03 & 39 & 16 & thin & sw & 40218 & 0\\
   A9 & 0018540101 & HD 5980 & 2001 & Nov. & 20 & 00\rahour & 59\ramin & 26\fs8 & -72\degr & 09\arcmin & 55\arcsec & 0.67 & 0.84 & M1S001 & 23 & 42 & 37 & medium & ff & 27025 & 1469\\
   &  &  &  &  &  &  &  &  &  &  &  &  &  & M2S003 & 23 & 42 & 37 & medium & ff & 27025 & 1471\\
   &  &  &  &  &  &  &  &  &  &  &  &  &  & PNS002 & 00 & 21 & 48 & medium & ff & 24410 & 1347\\
   A10 & 0084200101 & SMC Pointing 1 & 2002 & Mar. & 30 & 00\rahour & 56\ramin & 41\fs7 & -72\degr & 20\arcmin & 24\arcsec & 0.62 & 0.18 & M1S001 & 13 & 48 & 28 & medium & ff & 21279 & 9197\\
   &  &  &  &  &  &  &  &  &  &  &  &  &  & M2S002 & 13 & 48 & 29 & medium & ff & 21279 & 9197\\
   &  &  &  &  &  &  &  &  &  &  &  &  &  & PNS003 & 14 & 21 & 45 & thin & ff & 18999 & 7927\\
   A11 & 0142660801 & RXJ0059.4-7118 & 2003 & Nov. & 17 & 00\rahour & 59\ramin & 26\fs4 & -71\degr & 18\arcmin & 48\arcsec & 0.04 & -0.78 & M1S001 & 03 & 55 & 54 & thin & ff & 12397 & 7338\\
   &  &  &  &  &  &  &  &  &  &  &  &  &  & M2S002 & 03 & 55 & 56 & thin & ff & 12345 & 7337\\
   &  &  &  &  &  &  &  &  &  &  &  &  &  & PNS003 & 04 & 18 & 12 & thin & ff & 11785 & 5996\\
   A12 & 0157960201 & XTE J0055-727 & 2003 & Dec. & 18 & 00\rahour & 55\ramin & 22\fs0 & -72\degr & 42\arcmin & 00\arcsec & 2.37 & 0.79 & M1S004 & 14 & 32 & 45 & medium & lw & 3198 & 0\\
   &  &  &  &  &  &  &  &  &  &  &  &  &  & M1U002 & 15 & 45 & 40 & medium & ff & 14760 & 14346\\
   &  &  &  &  &  &  &  &  &  &  &  &  &  & M2S005 & 14 & 32 & 54 & medium & ff & 18673 & 14309\\
   &  &  &  &  &  &  &  &  &  &  &  &  &  & PNS003 & 14 & 52 & 23 & medium & lw & 17198 & 14017\\
   A13 & 0164560401 & XTE J0051-727 & 2004 & Apr. & 28 & 00\rahour & 51\ramin & 15\fs0 & -72\degr & 44\arcmin & 24\arcsec & 0.0 & 0.0 & M1S001 & 22 & 07 & 38 & medium & ff & 739 & 0\\
   &  &  &  &  &  &  &  &  &  &  &  &  &  & M2S002 & 22 & 07 & 38 & medium & ff & 791 & 0\\
   &  &  &  &  &  &  &  &  &  &  &  &  &  & PNS003 & 22 & 27 & 43 & medium & lw & 777 & 0\\
   A14 & 0212282601 & HD 5980 & 2005 & Mar. & 27 & 00\rahour & 59\ramin & 26\fs8 & -72\degr & 09\arcmin & 54\arcsec & 0.94 & 0.8 & M1S001 & 16 & 20 & 35 & medium & ff & 26406 & 4495\\
   &  &  &  &  &  &  &  &  &  &  &  &  &  & M2S002 & 16 & 20 & 35 & medium & ff & 26398 & 4501\\
   &  &  &  &  &  &  &  &  &  &  &  &  &  & PNS003 & 16 & 42 & 56 & closed & ff & 34402 & 0\\
   A15 & 0304250401 & HD 5980 & 2005 & Nov. & 27 & 00\rahour & 59\ramin & 26\fs8 & -72\degr & 09\arcmin & 54\arcsec & 0.65 & 1.36 & M1S009 & 06 & 22 & 46 & medium & ff & 17571 & 17503\\
   &  &  &  &  &  &  &  &  &  &  &  &  &  & M2S010 & 06 & 22 & 48 & medium & ff & 17574 & 17511\\
   &  &  &  &  &  &  &  &  &  &  &  &  &  & PNS011 & 06 & 45 & 04 & medium & ff & 15937 & 15866\\
   A16 & 0304250501 & HD 5980 & 2005 & Nov. & 29 & 00\rahour & 59\ramin & 26\fs8 & -72\degr & 09\arcmin & 54\arcsec & 2.29 & -0.08 & M1S001 & 05 & 11 & 36 & medium & ff & 16570 & 16565\\
   &  &  &  &  &  &  &  &  &  &  &  &  &  & M2S002 & 05 & 11 & 33 & medium & ff & 16575 & 16570\\
   &  &  &  &  &  &  &  &  &  &  &  &  &  & PNS003 & 05 & 33 & 52 & medium & ff & 14937 & 14933\\
   A17 & 0304250601 & HD 5980 & 2005 & Dec. & 11 & 00\rahour & 59\ramin & 26\fs8 & -72\degr & 09\arcmin & 54\arcsec & 0.8 & -0.03 & M1S001 & 12 & 48 & 45 & medium & ff & 16669 & 16664\\
   &  &  &  &  &  &  &  &  &  &  &  &  &  & M2S002 & 12 & 48 & 42 & medium & ff & 16674 & 16669\\
   &  &  &  &  &  &  &  &  &  &  &  &  &  & PNU002 & 14 & 25 & 31 & medium & ff & 10566 & 10566\\
   A18 & 0311590601 & Nova SMC 2005 & 2006 & Mar. & 13 & 01\rahour & 14\ramin & 59\fs9 & -73\degr & 25\arcmin & 36\arcsec & 2.41 & 1.98 & M1S002 & 15 & 17 & 13 & thin & ff & 11370 & 6765\\
   &  &  &  &  &  &  &  &  &  &  &  &  &  & M2S003 & 15 & 17 & 13 & thin & ff & 11375 & 6770\\
   &  &  &  &  &  &  &  &  &  &  &  &  &  & PNS001 & 15 & 39 & 31 & thin & ff & 9737 & 5126\\
   A19 & 0301170501 & SMC Field-5 & 2006 & Mar. & 19 & 00\rahour & 48\ramin & 23\fs4 & -73\degr & 41\arcmin & 00\arcsec & 0.0 & 0.0 & M1S001 & 14 & 23 & 19 & medium & ff & 13198 & 621\\
   &  &  &  &  &  &  &  &  &  &  &  &  &  & M1U002 & 19 & 38 & 30 & medium & ff & 2155 & 0\\
   &  &  &  &  &  &  &  &  &  &  &  &  &  & M2S002 & 14 & 23 & 17 & medium & ff & 13198 & 621\\
   &  &  &  &  &  &  &  &  &  &  &  &  &  & M2U002 & 19 & 38 & 36 & medium & ff & 2153 & 0\\
   &  &  &  &  &  &  &  &  &  &  &  &  &  & PNS003 & 14 & 45 & 35 & medium & ff & 19437 & 0\\
   A20 & 0301170101 & SMC Field-1 & 2006 & Mar. & 22 & 01\rahour & 08\ramin & 06\fs4 & -72\degr & 52\arcmin & 23\arcsec & -0.24 & 1.81 & M1S001 & 21 & 39 & 54 & medium & ff & 23172 & 17960\\
   &  &  &  &  &  &  &  &  &  &  &  &  &  & M2S002 & 21 & 39 & 54 & medium & ff & 23174 & 17961\\
   &  &  &  &  &  &  &  &  &  &  &  &  &  & PNS003 & 22 & 02 & 14 & medium & ff & 21537 & 17107\\
   A21 & 0301170201 & SMC Field-2 & 2006 & Mar. & 23 & 00\rahour & 52\ramin & 12\fs1 & -72\degr & 01\arcmin & 42\arcsec & 0.49 & 3.01 & M1S001 & 04 & 48 & 17 & medium & ff & 23569 & 18364\\
   &  &  &  &  &  &  &  &  &  &  &  &  &  & M2S002 & 04 & 48 & 18 & medium & ff & 23574 & 18369\\
   &  &  &  &  &  &  &  &  &  &  &  &  &  & PNS003 & 05 & 10 & 34 & medium & ff & 21937 & 16737\\
   A22 & 0301170601 & SMC Field-6 & 2006 & Mar. & 27 & 00\rahour & 40\ramin & 23\fs8 & -72\degr & 46\arcmin & 50\arcsec & -0.95 & 0.29 & M1S001 & 12 & 21 & 01 & thin & ff & 24872 & 14035\\
   &  &  &  &  &  &  &  &  &  &  &  &  &  & M2S002 & 12 & 20 & 59 & thin & ff & 24875 & 14038\\
   &  &  &  &  &  &  &  &  &  &  &  &  &  & PNS003 & 12 & 43 & 17 & thin & ff & 23238 & 12677\\
   A23 & 0301170301 & SMC Field-3 & 2006 & Apr. & 6 & 00\rahour & 42\ramin & 46\fs6 & -73\degr & 35\arcmin & 38\arcsec & -0.51 & -0.51 & M1S001 & 04 & 32 & 35 & medium & ff & 21570 & 17065\\
   &  &  &  &  &  &  &  &  &  &  &  &  &  & M2S002 & 04 & 32 & 32 & medium & ff & 21575 & 17070\\
   &  &  &  &  &  &  &  &  &  &  &  &  &  & PNS003 & 04 & 54 & 50 & medium & ff & 19937 & 15437\\
   A24 & 0402000101 & RXJ0103.8-7254 & 2006 & Oct. & 3 & 01\rahour & 03\ramin & 52\fs2 & -72\degr & 54\arcmin & 28\arcsec & 1.35 & -2.35 & M1S001 & 00 & 09 & 09 & thin & ff & 21619 & 20288\\
   &  &  &  &  &  &  &  &  &  &  &  &  &  & M2S002 & 00 & 09 & 11 & thin & ff & 21625 & 20291\\
   &  &  &  &  &  &  &  &  &  &  &  &  &  & PNS003 & 00 & 31 & 26 & thin & ff & 20037 & 18705\\
   A25 & 0404680101 & SMC Pointing 5\_1 & 2006 & Oct. & 5 & 00\rahour & 47\ramin & 36\fs0 & -73\degr & 08\arcmin & 24\arcsec & 0.21 & -2.78 & M1S001 & 00 & 22 & 33 & medium & ff & 23099 & 9294\\
   &  &  &  &  &  &  &  &  &  &  &  &  &  & M2S002 & 00 & 22 & 36 & medium & ff & 23101 & 9296\\
   &  &  &  &  &  &  &  &  &  &  &  &  &  & PNS003 & 00 & 44 & 52 & thin & ff & 21514 & 7714\\
   A26 & 0404680201 & SMC Pointing 5\_2 & 2006 & Nov. & 1 & 00\rahour & 52\ramin & 26\fs4 & -72\degr & 52\arcmin & 12\arcsec & 1.61 & 0.26 & M1S001 & 00 & 56 & 29 & medium & ff & 32318 & 32282\\
   &  &  &  &  &  &  &  &  &  &  &  &  &  & M2S002 & 00 & 56 & 29 & medium & ff & 32324 & 32290\\
   &  &  &  &  &  &  &  &  &  &  &  &  &  & PNS003 & 01 & 18 & 46 & thin & ff & 30737 & 30703\\
   A27 & 0403970301 & SMC01: N19 & 2007 & Mar. & 12 & 00\rahour & 47\ramin & 39\fs4 & -72\degr & 59\arcmin & 31\arcsec & -0.02 & 1.56 & M1S001 & 20 & 02 & 20 & thin & ff & 38826 & 24662\\
   &  &  &  &  &  &  &  &  &  &  &  &  &  & M2S002 & 20 & 02 & 18 & thin & ff & 38834 & 24674\\
   &  &  &  &  &  &  &  &  &  &  &  &  &  & PNS003 & 21 & 02 & 54 & thin & eff & 34949 & 20927\\
   A28 & 0404680301 & SMC Pointing 5\_3 & 2007 & Apr. & 11 & 00\rahour & 51\ramin & 00\fs7 & -73\degr & 24\arcmin & 17\arcsec & 0.38 & -1.15 & M1S001 & 19 & 38 & 25 & medium & ff & 23606 & 17618\\
   &  &  &  &  &  &  &  &  &  &  &  &  &  & M2S002 & 19 & 38 & 25 & medium & ff & 23613 & 17626\\
   &  &  &  &  &  &  &  &  &  &  &  &  &  & PNS003 & 20 & 00 & 45 & thin & ff & 22024 & 16039\\
   A29 & 0404680501 & SMC Pointing 5\_5 & 2007 & Apr. & 12 & 01\rahour & 07\ramin & 42\fs3 & -72\degr & 30\arcmin & 11\arcsec & 0.29 & 1.37 & M1S001 & 03 & 07 & 23 & medium & ff & 23621 & 23216\\
   &  &  &  &  &  &  &  &  &  &  &  &  &  & M2S002 & 03 & 07 & 22 & medium & ff & 23626 & 23221\\
   &  &  &  &  &  &  &  &  &  &  &  &  &  & PNS003 & 03 & 29 & 42 & thin & ff & 22037 & 21637\\
   A30 & 0501470101 & RX J0059.6-7138 & 2007 & Jun & 4 & 00\rahour & 59\ramin & 41\fs8 & -71\degr & 38\arcmin & 15\arcsec & 1.76 & 1.41 & M1S001 & 08 & 59 & 50 & thin & ff & 33405 & 10560\\
   &  &  &  &  &  &  &  &  &  &  &  &  &  & M2S002 & 08 & 59 & 50 & thin & ff & 33410 & 10565\\
   &  &  &  &  &  &  &  &  &  &  &  &  &  & PNS003 & 09 & 22 & 09 & thin & ff & 31822 & 10318\\
   A31 & 0500980201 & SMC Pointing 6\_2 & 2007 & Jun & 6 & 01\rahour & 00\ramin & 00\fs0 & -72\degr & 27\arcmin & 00\arcsec & 2.72 & 1.57 & M1S001 & 08 & 52 & 16 & medium & ff & 28621 & 12905\\
   &  &  &  &  &  &  &  &  &  &  &  &  &  & M2S002 & 08 & 52 & 11 & medium & ff & 28626 & 12913\\
   &  &  &  &  &  &  &  &  &  &  &  &  &  & PNS003 & 09 & 14 & 30 & thin & ff & 27038 & 11421\\
   A32 & 0500980101 & SMC Pointing 6\_1 & 2007 & Jun & 23 & 00\rahour & 53\ramin & 02\fs4 & -72\degr & 26\arcmin & 17\arcsec & 4.25 & 0.9 & M1S001 & 05 & 51 & 39 & medium & ff & 25673 & 24164\\
   &  &  &  &  &  &  &  &  &  &  &  &  &  & M2S002 & 05 & 51 & 39 & medium & ff & 25678 & 24169\\
   &  &  &  &  &  &  &  &  &  &  &  &  &  & PNS003 & 06 & 13 & 58 & thin & ff & 24088 & 22688\\
   A33 & 0503000201 & SMC Field-5 & 2007 & Oct. & 28 & 00\rahour & 48\ramin & 23\fs4 & -73\degr & 41\arcmin & 00\arcsec & 1.55 & -0.29 & M1S001 & 05 & 49 & 58 & medium & ff & 21531 & 20301\\
   &  &  &  &  &  &  &  &  &  &  &  &  &  & M2S002 & 05 & 49 & 58 & medium & ff & 21536 & 20301\\
   &  &  &  &  &  &  &  &  &  &  &  &  &  & PNS003 & 06 & 11 & 55 & medium & ff & 19969 & 19164\\
   A34 & 0503000301 & SMC Field-6 & 2008 & Mar. & 16 & 00\rahour & 40\ramin & 23\fs8 & -72\degr & 46\arcmin & 50\arcsec & -2.35 & 0.98 & M1S001 & 15 & 25 & 16 & thin & ff & 9129 & 5353\\
   &  &  &  &  &  &  &  &  &  &  &  &  &  & M1U002 & 18 & 39 & 00 & thin & ff & 14711 & 0\\
   &  &  &  &  &  &  &  &  &  &  &  &  &  & M2S002 & 15 & 25 & 15 & thin & ff & 9147 & 5353\\
   &  &  &  &  &  &  &  &  &  &  &  &  &  & M2U002 & 18 & 39 & 04 & thin & ff & 14703 & 0\\
   &  &  &  &  &  &  &  &  &  &  &  &  &  & PNS003 & 15 & 47 & 13 & thin & ff & 28865 & 0\\
   O1 & 0112880901 & CF Tuc & 2000 & Nov. & 30 & 00\rahour & 53\ramin & 04\fs8 & -74\degr & 39\arcmin & 07\arcsec & 1.5 & 1.73 & M1S001 & 12 & 12 & 57 & medium & ff & 40548 & 39542\\
   &  &  &  &  &  &  &  &  &  &  &  &  &  & M2S002 & 12 & 12 & 54 & medium & ff & 40547 & 39542\\
   &  &  &  &  &  &  &  &  &  &  &  &  &  & PNS003 & 12 & 54 & 23 & medium & ff & 38150 & 37150\\
   O2 & 0142661001 & RXJ0050.5-7455 & 2003 & Nov. & 16 & 00\rahour & 50\ramin & 35\fs1 & -74\degr & 55\arcmin & 44\arcsec & -1.57 & -2.42 & M1U002 & 16 & 55 & 51 & thin & ff & 15119 & 12106\\
   &  &  &  &  &  &  &  &  &  &  &  &  &  & M2U002 & 16 & 55 & 54 & thin & ff & 15122 & 12109\\
   &  &  &  &  &  &  &  &  &  &  &  &  &  & PNU002 & 17 & 19 & 00 & thin & ff & 13437 & 10429\\
   O3 & 0142660401 & RXJ0059.1-7505 & 2003 & Nov. & 16 & 00\rahour & 59\ramin & 10\fs7 & -75\degr & 05\arcmin & 23\arcsec & 0.45 & -0.9 & M1S001 & 21 & 50 & 00 & thin & ff & 18671 & 16445\\
   &  &  &  &  &  &  &  &  &  &  &  &  &  & M2S002 & 21 & 50 & 02 & thin & ff & 18676 & 16471\\
   &  &  &  &  &  &  &  &  &  &  &  &  &  & PNS003 & 22 & 12 & 21 & thin & ff & 17036 & 14806\\
   O4 & 0301150101 & F00521-7054 & 2006 & Mar. & 22 & 00\rahour & 53\ramin & 56\fs1 & -70\degr & 38\arcmin & 04\arcsec & -1.26 & 2.1 & M1U002 & 17 & 50 & 55 & thin & ff & 10460 & 8270\\
   &  &  &  &  &  &  &  &  &  &  &  &  &  & M2U002 & 18 & 05 & 32 & thin & ff & 9589 & 8228\\
   &  &  &  &  &  &  &  &  &  &  &  &  &  & PNU002 & 18 & 08 & 59 & thin & ff & 9081 & 7778\\
   O5 & 0301151601 & F00521-7054 & 2006 & Apr. & 22 & 00\rahour & 53\ramin & 56\fs1 & -70\degr & 38\arcmin & 04\arcsec & 1.11 & 1.48 & M1S001 & 00 & 55 & 16 & thin & ff & 13970 & 12165\\
   &  &  &  &  &  &  &  &  &  &  &  &  &  & M2S002 & 00 & 55 & 14 & thin & ff & 13975 & 12170\\
   &  &  &  &  &  &  &  &  &  &  &  &  &  & PNS003 & 01 & 17 & 32 & thin & ff & 12338 & 10538\\
\end{longtable}
\end{landscape}
}

\end{appendix}

\end{document}